%% file: paper.tex
\documentclass[11pt]{article}

\usepackage{algorithm}
\usepackage{algorithmic}
\usepackage{paralist}
\usepackage{framed}
\usepackage{amssymb}
\usepackage{amsfonts}
\usepackage{amsmath}
\usepackage{mathtools}
\usepackage{graphicx}
\usepackage{caption}
\usepackage{url}
\usepackage{rotating}
\usepackage{multirow}
\usepackage{subfigure}
\usepackage{color}
\usepackage{xcolor}
\usepackage{enumitem}
\usepackage{listings}
\usepackage{varwidth}
\usepackage{tikz}
\usepackage{footnote}
\usetikzlibrary{shapes.gates.logic.US,trees,positioning,arrows}
\usepackage{booktabs}
\newcommand{\bigcell}[2]{\begin{tabular}{@{}#1@{}}#2\end{tabular}}

\newlength{\defbaselineskip}
\definecolor{darkgreen}{rgb}{0,0.6,0}

\newif\ifwithcomments
\withcommentstrue
\ifwithcomments
\newcommand{\Jiyan}[1]{{\color{blue} Jiyan: #1}}
\else
\newcommand{\Jiyan}[1]{}
\fi

 \newtheorem{definition}{Definition}
 \newtheorem{lemma}{Lemma}
 \newtheorem{theorem}{Theorem}
 \newtheorem{corollary}{Corollary}
 \newtheorem{remark}{Remark}

\newcommand{\T}{^{\raisebox{0.09em}{\hbox{\tiny $T$}}\hspace{-0.1em}}}

\DeclareMathOperator{\poly}{poly}
\DeclareMathOperator{\bigO}{\mathcal{O}}
\DeclareMathOperator{\nnz}{nnz}
\DeclareMathOperator{\A}{\mathcal{A}}
\DeclareMathOperator{\R}{\mathbb{R}}

\DeclareMathOperator{\E}{\mathcal{E}}
\DeclareMathOperator{\rank}{rank}
\DeclareMathOperator{\X}{\mathcal{X}}
\DeclareMathOperator{\N}{\mathcal{N}}

\def\norm#1{{\|#1\|}}

\begin{document}

\input{tit_abst}

\newpage

\input{intro}

\input{rla_ram}

\input{background}

\input{basic_theory}

\input{implementation}

\input{conclusion}

\vspace{5mm}
\textbf{Acknowledgments.}
We would like to thank Michael Saunders for advice and helpful discussions.
We would also like to acknowledge 
the Army Research Office, 
the Defense Advanced Research Projects Agency, 
and 
the Department of Energy 
for providing partial support for this~work.
\vspace{5mm}

\begin{small}

\bibliographystyle{unsrt}

\bibliography{pardist}

\end{small}

\end{document}

%% file: tit_abst.tex
\title{%
Implementing Randomized Matrix Algorithms in Parallel and Distributed Environments
}

\author{
Jiyan Yang
\thanks{
Institute for Computational and Mathematical Engineering,
Stanford University,
Stanford, CA 94305.
Email: jiyan@stanford.edu
}
\and
Xiangrui Meng
\thanks{
Databricks,
160 Spear Street,
Floor 13,
San Francisco, CA 94105.
Email: meng@databricks.com
}
\and
Michael W. Mahoney
\thanks{
International Computer Science Institute 
and Department of Statistics,
University of California at Berkeley,
Berkeley, CA 94720.
Email:  mmahoney@stat.berkeley.edu
}
}

\date{}
\maketitle


\begin{abstract}
\noindent
In this era of large-scale data, distributed systems built on top of clusters 
of commodity hardware provide cheap and reliable storage and scalable 
processing of massive data.
With cheap storage, instead of storing only currently-relevant data, it is 
common to store as much data as possible, hoping that its value can be 
extracted later.
In this way, exabytes ($10^{18}$ bytes) of data are being created on a daily 
basis.
Extracting value from these data however, requires scalable implementations
of advanced analytical algorithms beyond simple data processing, e.g., 
statistical regression methods, linear algebra, and optimization algorithms.
Many traditional methods are designed to minimize floating-point 
operations, which is the dominant cost of in-memory computation on a single 
machine.
In parallel and distributed environments, however, load balancing and 
communication, including disk and network I/O, can easily dominate 
computation.
These factors greatly increase the complexity of algorithm design and 
challenge traditional ways of thinking about the design of parallel and 
distributed algorithms.

Here, we review recent work on developing and implementing randomized matrix 
algorithms in large-scale parallel and distributed environments.
Randomized algorithms for matrix problems have received a great deal of 
attention in recent years, thus far typically either in theory or in machine 
learning applications or with implementations on a single machine.
Our main focus is on the underlying theory and practical implementation of 
random projection and random sampling algorithms for very large very 
overdetermined (\emph{i.e.}, overconstrained) $\ell_1$ and $\ell_2$ 
regression problems. 
Randomization can be used in one of two related ways: 
either to construct sub-sampled problems that can be solved, exactly or 
approximately, with traditional numerical methods; or  
to construct preconditioned versions of the original full problem that are 
easier to solve with traditional iterative algorithms. 
Theoretical results demonstrate that in near input-sparsity time and with 
only a few passes through the data one can obtain very strong relative-error 
approximate solutions, with high probability.
Empirical results highlight the importance of various trade-offs (\emph{e.g.}, 
between the time to construct an embedding and the conditioning quality of the 
embedding, between the relative importance of computation versus 
communication, etc.) and demonstrate that $\ell_1$ and $\ell_2$ regression 
problems can be solved to low, medium, or high precision in existing 
distributed systems on up to terabyte-sized data. 
\end{abstract}

%% file: intro.tex
\section{Introduction}
\label{sxn:intro}

Matrix algorithms lie at the heart of many applications, both historically in 
areas such as signal processing and scientific computing as well as more 
recently in areas such as machine learning and data analysis.
Essentially, the reason is that matrices provide a convenient mathematical 
structure with which to model data arising in a broad range of applications:
an $m \times n$ real-valued matrix $A$ provides a natural structure for 
encoding information about $m$ objects, each of which is described by $n$ 
features. 
Alternatively, an $n \times n$ real-valued matrix $A$ can be used to describe
the correlations between all pairs of $n$ data points, or the weighted 
edge-edge adjacency matrix structure of an $n$-node graph.
In astronomy, for example, very small angular regions of the sky imaged at 
a range of electromagnetic frequency bands can be represented as a 
matrix---in that case, an object is a region and the features are the 
elements of the frequency bands. 
Similarly, in genetics, DNA SNP (Single Nucleotide Polymorphism) or DNA 
microarray expression data can be represented in such a framework, with 
$A_{ij}$ representing the expression level of the $i^{th}$ gene or SNP 
in the $j^{th}$ experimental condition or individual. 
Similarly, term-document matrices can be constructed in many Internet 
applications, with $A_{ij}$ indicating the frequency of the $j^{th}$ term 
in the $i^{th}$ document.

Most traditional algorithms for matrix problems are designed to run on a 
single machine, focusing on minimizing the number of floating-point 
operations per second (FLOPS).
On the other hand, motivated by the ability to generate very large quantities 
of data in relatively automated ways, analyzing data sets of billions or 
more of records has now become a regular task in many companies and 
institutions.
In a distributed computational environment, which is typical in these 
applications, communication costs, \emph{e.g.}, between different machines, 
are often much more important than computational costs. 
What is more, if the data cannot fit into memory on a single machine, then 
one must scan the records from secondary storage, \emph{e.g.}, hard disk, 
which makes each pass through the data associated with enormous I/O costs.
Given that, in many of these large-scale applications, regression, low-rank 
approximation, and related matrix problems are ubiquitous, the fast 
computation of their solutions on large-scale data platforms is of interest.

In this paper, we will provide an overview of recent work in Randomized Numerical Linear 
Algebra (RandNLA) on implementing randomized matrix algorithms in 
large-scale parallel and distributed computational environments.
RandNLA is a large area that applies randomization as an algorithmic 
resource to develop improved algorithms for regression, low-rank matrix 
approximation, and related problems~\cite{mahoney2011randomized}.
To limit the presentation, here we will be most interested in very large very 
rectangular linear regression problems on up to terabyte-sized data: in 
particular, in the $\ell_2$ regression (also known as least squares, or 
LS) problem and its robust alternative, the $\ell_1$ regression (also known 
as least absolute deviations, LAD, or least absolute errors, LAE) problem, 
with strongly rectangular ``tall'' data.
Although our main focus is on $\ell_2$ and $\ell_1$ regression, much of the
underlying theory holds for $\ell_p$ regression, either for $p\in[1,2]$ or 
for all $p\in[1,\infty)$, and thus for simplicity we formulate many of our 
results in $\ell_p$.

Several important conclusions will emerge from our presentation.
\begin{itemize}
\item
First, many of the basic ideas from RandNLA in RAM extend to RandNLA in 
parallel/distributed environments in a relatively straightforward manner, 
assuming that one is more concerned about communication than computation.
This is important from an algorithm design perspective, as it highlights 
which aspects of these RandNLA algorithms are peculiar to the use of 
randomization and which aspects are peculiar to parallel/distributed 
environments.
\item
Second, with appropriate engineering of random sampling and random projection 
algorithms, it is possible to compute good approximate solutions---to low 
precision (\emph{e.g.}, $1$ or $2$ digits of precision), medium precision 
(\emph{e.g.}, $3$ or $4$ digits of precision), or high precision 
(\emph{e.g.}, up to machine precision)---to several common matrix problems 
in only a few passes over the original matrix on up to terabyte-sized data.
While low precision is certainly appropriate for many data analysis and 
machine learning applications involving noisy input data, the appropriate 
level of precision is a choice for user of an algorithm to make; and 
there are obvious advantages to having the developer of an algorithm provide
control to the user on the quality of the answer returned by the algorithm.
\item
Third, the design principles for developing high-quality RandNLA matrix 
algorithms depend strongly on whether one is interested in low, medium,
or high precision. 
(An example of this is whether to solve the randomized subproblem with a 
traditional method or to use the randomized subproblem to create a 
preconditioned version of the original problem.)
Understanding these principles, the connections between them, and how they 
relate to traditional principles of NLA algorithm design is important for 
providing high-quality implementations of recent theoretical developments 
in the RandNLA literature.
\end{itemize}

\begin{table}
  \centering
  \begin{tabular}{c|c|c}
    & $m$ & $n$ \\
    \hline
    SNP & number of SNPs ($10^7$) & number of subjects ($10^3$)\\
    TinyImages & number of images ($10^8$) & number of pixels in each image ($10^3$)\\
    PDE & number of degrees of freedom  & number of time steps\\
    sensor network & size of sensing data & number of sensors\\
    NLP & number of words and $n$-grams & number of documents\\
    tick data & number of ticks & number of stocks
  \end{tabular}
  \caption{Examples of strongly rectangular datasets}
  \label{tab:strongly_rectangular_data}
\end{table}

Although many of the ideas we will discuss can be extended to related matrix 
problems such as low-rank matrix approximation, there are two main reasons 
for restricting attention to strongly rectangular data.
The first, most obvious, reason is that strongly rectangular data arises in 
many fields to which machine learning and data analysis methods are 
routinely applied.
Consider, \emph{e.g.}, Table~\ref{tab:strongly_rectangular_data}, which 
lists a few examples.
\begin{itemize}
\item 
In genetics, single nucleotide polymorphisms (SNPs) are important in the study 
of human health.
There are roughly $10$ million SNPs in the human genome.
However, there are typically at most a few thousand subjects for a study of a 
certain type of disease, due to the high cost of determination of genotypes and
limited number of target subjects.
\item 
In Internet applications, strongly rectangular datasets are common. 
For example, the image dataset called TinyImages~\cite{torralba2008tiny} which 
contains $80$ million images of size $32 \times 32$ collected from Internet.
\item 
In spatial discretization of high-dimensional partial differential
equations~(PDEs), the number of degrees of freedom grows exponentially as
dimension increases.
For 3D problems, it is common that the number of degrees of freedom reaches
$10^9$, for example, by having a $1000 \times 1000 \times 1000$ discretization
of a cubic domain.
However, for a time-dependent problem, time stays one-dimensional.
Though depending on spatial discretization (\emph{e.g.}, the 
Courant-Friedrichs-Lewy condition for hyperbolic PDEs), the number of time 
steps is usually much less than the number of degrees of freedoms in spatial 
discretization.
\item In geophysical applications, especially in seismology, the number of
  sensors is much less than the number of data points each sensor collects.
  For example, Werner-Allen \emph{et al.}~\cite{werner2005monitoring} deployed three
  wireless sensors to monitor volcanic eruptions.
  In 54 hours, each sensor sent back approximately $20$ million packets.
\item In natural language processing (NLP), the number of documents is much less
  than the number of $n$-grams, which grows geometrically as $n$ increases.
  For example, the
  webspam\footnote{\url{http://www.csie.ntu.edu.tw/~cjlin/libsvmtools/datasets/binary.html}} dataset
  contains 350,000 documents and 254 unigrams, but 680,715 trigrams.
\item In high-frequency trading, the number of relevant stocks is much less than
  the number of ticks, changes to the best bid and ask.
  For example, in 2012 ISE Historical Options Tick Data\footnote{\url{http://www.ise.com/hotdata}} has
  daily files with average size greater than 100GB uncompressed.
\end{itemize}
A second, less obvious, reason for restricting attention to strongly 
rectangular data is that many of the algorithmic methods that are developed 
for them (both the RandNLA methods we will review as well as deterministic 
NLA methods that have been used traditionally) have extensions to low-rank 
matrix approximation and to related problems on more general ``fat'' matrices.
For example, many of the methods for SVD-based low-rank approximation and 
related rank-revealing QR decompositions of 
general matrices have strong connections to QR decomposition methods for 
rectangular matrices; and, similarly, many of the methods for more general 
linear and convex programming arise in special (\emph{e.g.}, $\ell_1$ 
regression) linear programming problems.
Thus, they are a good problem class to consider the development of matrix 
algorithms (either in general or for RandNLA algorithms) in parallel and 
distributed environments.

It is worth emphasizing that the phrase ``parallel and distributed'' can 
mean quite different things to different research communities, in particular 
to what might be termed HPC (high performance computing) or scientific 
computing researchers versus data analytics or database or distributed data 
systems researchers.
There are important technical and cultural differences here, but there are also 
some important similarities.
For example, to achieve parallelism, one can use multi-threading on a 
shared-memory machine, or one can use message passing on a multi-node cluster.
Alternatively, to process massive data on large commodity clusters, 
Google's MapReduce \cite{dean2004mapreduce} describes a computational
framework for distributed computation with fault tolerance.
For computation not requiring any internode communication, one can achieve 
even better parallelism.
We don't want to dwell on many of these important details here: this is a 
complicated and evolving space; and no doubt the details of the implementation 
of many widely-used algorithms will evolve as the space evolves.
To give the interested reader a quick sense of some of these issues, though, 
here we provide a very high-level representative description of parallel 
environments and how they scale.
As one goes down this list, one tends to get larger and larger.
\begin{savenotes}
\begin{table}[H]
\centering
\begin{tabular}{c|ccc}
name & cores & memory & notes \\
\hline
Shared memory & $[10, 10^3]$\footnote{\url{http://www.sgi.com/pdfs/4358.pdf}} & $[100\text{GB}, 100\text{TB}]$ \\
\multirow{2}{*}{Message passing} & \multirow{2}{*}{$[200, 10^5]$\footnote{\url{http://www.top500.org/list/2011/11/100}}} & \multirow{2}{*}{$[1\text{TB}, 1000\text{TB}]$} & CUDA cores: $[5 \times 10^4, 3 \times 10^6]$\footnote{\url{http://i.top500.org/site/50310}} \\
& & & GPU memory: $[500\text{GB}, 20\text{TB}]$ \\
MapReduce & $[40, 10^5]$\footnote{\url{http://www.cloudera.com/blog/2010/04/pushing-the-limits-of-distributed-processing/}} & $[240\text{GB}, 100\text{TB}]$ & storage: $[100\text{TB}, 100\text{PB}]$\footnote{\url{http://hortonworks.com/blog/an-introduction-to-hdfs-federation/}} \\
Distributed computing & $[-, 3 \times 10^5]$\footnote{\url{http://folding.stanford.edu/}}
\end{tabular}
\caption{High-level representative description of parallel 
environments.}
\end{table}
\end{savenotes}


In addition, it is also worth emphasizing that there is a great deal of 
related work in parallel and distributed computing, both in numerical 
linear algebra as well as more generally in scientific computing.
For example, 
Valiant has provided a widely-used model for parallel 
computation~\cite{Val90}; 
Aggarwal \emph{et al.} have analyzed the communication complexity of 
PRAMs~\cite{ACS90};
Lint and Agerwala have highlighted communication issues that arise in the 
design of parallel algorithms~\cite{LA81};
Heller has surveyed parallel algorithms in numerical linear 
algebra~\cite{Hel76};
Toledo has provided a survey of out-of-core algorithms in numerical linear 
algebra~\cite{Tol99};
Ballard \emph{et al.} have focused on developing algorithms for minimizing 
communication in numerical linear algebra~\cite{BJHS11}; and
Bertsekas and Tsitsiklis have surveyed parallel and distributed iterative 
algorithms~\cite{bertsekas1991some}. 
We expect that some of the most interesting developments in upcoming years 
will involve coupling the ideas for implementing RandNLA algorithms in 
parallel and distributed environments that we describe in this review with 
these more traditional ideas for performing parallel and distributed 
computation.

In the next section, Section~\ref{sxn:rla_ram}, we will review the basic 
ideas underlying RandNLA methods, as they have been developed in the special 
case of $\ell_2$ regression in the RAM model.
Then, in Section~\ref{sxn:background}, we will provide notation, some
background and preliminaries on $\ell_2$ and more general $\ell_p$ 
regression problems, as well as traditional methods for their solution.
Then, in Section~\ref{sxn:round_embed}, we will describe rounding and 
embedding methods that are used in a critical manner by RandNLA algorithms; 
and in Section~\ref{sxn:implementations}, we will review recent empirical 
results on implementing these ideas to solve up to terabyte-sized $\ell_2$ 
and $\ell_1$ regression problems.
Finally, in Section~\ref{sxn:conc}, we will provide a brief discussion and 
conclusion.
An overview of the general RandNLA area has been 
provided~\cite{mahoney2011randomized}, and 
we refer the interested reader to this overview.
In addition, two other reviews are available to the interested reader: 
an overview of how RandNLA methods can be coupled with traditional NLA 
algorithms for low-rank matrix approximation~\cite{halko2011finding}; and
an overview of how data-oblivious subspace embedding methods are used in 
RandNLA~\cite{Woodruff_sketching_NOW}.

%% file: rla_ram.tex
\section{RandNLA in RAM}
\label{sxn:rla_ram}

In this section, we will highlight several core ideas that have been central 
to prior work in RandNLA in (theory and/or practice in) RAM that we will see 
are also important as design principles for extending RandNLA methods to 
larger-scale parallel and distributed environments.
We start in Section~\ref{sxn:rla_ram-meta_alg} by describing a prototypical 
example of a RandNLA algorithm for the very overdetermined LS problem;
then we describe in Section~\ref{sxn:rla_ram-lev_cond} two problem-specific 
complexity measures that are important for low-precision and high-precision 
solutions to matrix problems, respectively, as well as two complementary ways 
in which randomization can be used by RandNLA algorithms; and 
we conclude in Section~\ref{sxn:rla_ram-run_time} with a brief discussion of 
running time considerations.

\subsection{A meta-algorithm for RandNLA}
\label{sxn:rla_ram-meta_alg}

A prototypical example of the RandNLA approach is given by the following 
meta-algorithm for very overdetermined LS 
problems~\cite{drineas2006sampling,mahoney2011randomized,DMMW12_JMLR,MMY14_JMLR}.
In particular, the problem of interest is to solve:
\begin{equation}
  \min_x \|Ax - b\|_2.
\label{eq:ls_min_length}
\end{equation}
The following meta-algorithm takes as input an $m \times n$ matrix $A$, where 
$m \gg n$, a vector $b$, and a probability distribution 
$\{\pi_i\}_{i=1}^{m}$, and it returns as output an approximate solution 
$\hat{x}$, which is an estimate of the exact answer $x^*$ of
Problem~(\ref{eq:ls_min_length}).

\begin{itemize}
\item
\textbf{Randomly sampling.}
Randomly sample $r > n$ constraints, \emph{i.e.}, rows of $A$ and the 
corresponding elements of $b$, using $\{\pi_i\}_{i=1}^{m}$ as an importance 
sampling distribution.
\item
\textbf{Subproblem construction.}
Rescale each sampled row/element by $1/(r\pi_{i})$ to form a weighted LS 
subproblem.
\item
\textbf{Solving the subproblem.}
Solve the weighted LS subproblem, formally given in~(\ref{lsq-sample}) 
below, and then return the solution $\hat{x}$.
\end{itemize}

\noindent
It is convenient to describe this meta-algorithm in terms of a random 
``sampling matrix'' $S$, in the following manner.
If we draw $r$ samples (rows or constraints or data points) with 
replacement, then define an $r \times m$ sampling matrix, $S$, where each of 
the $r$ rows of $S$ has one non-zero element indicating which row of $A$ 
(and element of $b$) is chosen in a given random trial.
In this case, the $(i,k)^{th}$ element of $S$ equals $1/\sqrt{r\pi_k}$ if the 
$k^{th}$ data point is chosen in the $i^{th}$ random trial (meaning, in
particular, that every non-zero element of $S$ equals $\sqrt{n/r}$ for 
sampling uniformly at random).
With this notation, this meta-algorithm constructs and solves the weighted 
LS estimator:
\begin{eqnarray}
  \label{lsq-sample}
 \hat x = \arg\min_x \|SAx - Sb\|_2.
\end{eqnarray}

Since this meta-algorithm samples constraints and not variables, the 
dimensionality of the vector $\hat{x}$ that solves the (still 
overconstrained, but smaller) weighted LS subproblem is the same as that 
of the vector $x^*$ that solves the original LS problem.
The former may thus be taken as an approximation of the latter, where,
of course, the quality of the approximation depends critically on the 
choice of $\{\pi_i\}_{i=1}^{n}$. 
Although uniform subsampling (with or without replacement) is very simple to 
implement, it is easy to construct examples where it will perform very 
poorly~\cite{drineas2006sampling,mahoney2011randomized,MMY14_JMLR}.
On the other hand, it has been shown that, for a parameter 
$\gamma \in (0,1]$ that can be tuned, if
\begin{equation}
\label{eqn:approx-lev-score-probs}
\pi_i \ge \gamma \frac{h_{ii}}{p}, \;\;\mbox{and }
r=\bigO(p \log(p) / (\gamma \epsilon^2) ),
\end{equation}
where the so-called statistical leverage scores $h_{ii}$ are defined in~(\ref{eqn:statlev1}) below, \emph{i.e.}, if one draws the sample 
according to an importance sampling distribution that is proportional to the 
leverage scores of $A$, then with constant probability (that can be easily 
boosted to probability $1-\delta$, for any $\delta>0$) the following 
relative-error bounds hold:
\begin{eqnarray}
\label{eq:ls-bound-eq1}
||b-A\hat{x}||_2 
   &\leq& (1+\epsilon) ||b-Ax^*||_2  \hspace{2mm} \mbox{ and } \\
\label{eq:ls-bound-eq2}
|| x^* - \hat{x} ||_2
   &\leq& \sqrt{\epsilon} \left( \kappa(A)\sqrt{\xi^{-2}-1} \right) ||x^*||_2  ,
\end{eqnarray}
where $\kappa(A)$ is the condition number of $A$ and where 
$\xi = ||UU^Tb||_2/||b||_2$ is a parameter defining the amount of the mass 
of $b$ inside the column space of 
$A$~\cite{drineas2006sampling,mahoney2011randomized,DMMW12_JMLR}.

Due to the crucial role of the statistical leverage scores in~(\ref{eqn:approx-lev-score-probs}), this canonical RandNLA procedure has 
been referred to as the \emph{algorithmic leveraging} approach to 
approximating LS approximation~\cite{MMY14_JMLR}.
In addition, although this meta-algorithm has been described here only for 
very overdetermined LS problems, it generalizes to other linear regression 
problems and low-rank matrix approximation problems on less rectangular 
matrices\footnote{Let $A$ be a matrix with dimension $m$ by $n$ where $m > n$. A less rectangular matrix is a matrix that has smaller $m/n$.}~\cite{DMM08_CURtheory_JRNL,mahoney2009cur,clarkson2013fast,CW13sparse_STOC,GM13_TR}.

\subsection{Leveraging, conditioning, and using randomization}
\label{sxn:rla_ram-lev_cond}

Leveraging and conditioning refer to two types of problem-specific complexity
measures, \emph{i.e.}, quantities that can be computed for any problem 
instance that characterize how difficult that problem instance is for a 
particular class of algorithms.
Understanding these, as well as different uses of randomization in algorithm 
design, is important for designing RandNLA algorithms, both in theory and/or
practice in RAM as well as in larger parallel and distributed environments.
For now, we describe these in the context of very overdetermined LS problems.

\begin{itemize}
\item
\textbf{Statistical leverage.}
(Related to eigenvectors; important for obtaining low-precision solutions.)
If we let $H=A(A^{T}A)^{-1}A^{T}$, where the inverse can be replaced with 
the Moore-Penrose pseudoinverse if $A$ is rank deficient, be the projection
matrix onto the column span of $A$, then the $i^{th}$ diagonal element of 
$H$, 
\begin{equation}
\label{eqn:statlev1}
h_{ii}=A_{(i)}(A^{T}A)^{-1}A_{(i)}^{T} , 
\end{equation}
where $A_{(i)}$ is the $i^{th}$ row of $A$, is the \emph{statistical 
leverage} of $i^{th}$ observation or sample.
Since $H$ can alternatively be expressed as $H=UU^{T}$, where $U$ is any 
orthogonal basis for the column space of $X$, \emph{e.g.}, the $Q$ matrix 
from a QR decomposition or the matrix of left singular vectors from the 
thin SVD, the leverage of the $i^{th}$ observation can also be expressed~as
\begin{equation}
\label{eqn:statlev2}
h_{ii}=\sum_{j=1}^{n}U_{ij}^2=||U_{(i)}||^2 ,
\end{equation}
where $U_{(i)}$ is the $i^{th}$ row of $U$.
Leverage scores provide a notion of ``coherence'' or ``outlierness,'' in 
that they measure how well-correlated the singular vectors are with the 
canonical basis~\cite{mahoney2009cur,DMMW12_JMLR,CR12} as well as which 
rows/constraints have largest ``influence'' on the LS 
fit~\cite{HW78,CH86,VW81,ChatterjeeHadi88}.
Computing the leverage scores $\{h_{ii}\}_{i=1}^{m}$ \emph{exactly} is 
generally as hard as solving the original LS problem (but $1\pm\epsilon$
approximations to them can be computed more quickly, for arbitrary input 
matrices~\cite{DMMW12_JMLR}).

Leverage scores are important from an algorithm design perspective since they 
define the key nonuniformity structure needed to control the complexity of 
high-quality random sampling algorithms.
In particular, na\"{\i}ve uniform random sampling algorithms perform poorly 
when the leverage scores are very nonuniform, while randomly sampling in a
manner that depends on the leverage scores leads to high-quality solutions.
\emph{Thus, in designing RandNLA algorithms, whether in RAM or in 
parallel-distributed environments, one must either quickly compute 
approximations to the leverage scores or quickly preprocess the input 
matrix so they are nearly uniformized---in which case uniform random sampling 
on the preprocessed matrix performs well.}

Informally, the leverage scores characterize where in the high-dimensional
Euclidean space the (singular value) information in $A$ is being sent, 
\emph{i.e.}, how the quadratic well (with aspect ratio $\kappa(A)$ that is 
implicitly defined by the matrix $A$) ``sits'' with respect to the canonical 
axes of the high-dimensional Euclidean space. 
\emph{If one is interested in obtaining \emph{low-precision solutions}, 
\emph{e.g.}, $\epsilon=10^{-1}$, that can be obtained by an algorithm that 
provides $1\pm\epsilon$ relative-error approximations for a fixed value of 
$\epsilon$ but whose $\epsilon$ dependence is polynomial in $1/\epsilon$, 
then the key quantities that must be dealt with are statistical leverage 
scores of the input~data.}
\item
\textbf{Condition number.}
(Related to eigenvalues; important for obtaining high-precision solutions.)
If we let $\sigma_{\max}(A)$ and $\sigma_{\min}(A)$ denote the largest and 
smallest nonzero singular values of $A$, respectively, then 
$\kappa(A) = \sigma_{\max}(A) / \sigma^+_{\min}(A)$ is the $\ell_2$-norm 
condition number of $A$ which is formally defined in Definition~\ref{def:l2normcond}.
Computing $\kappa(A)$ \emph{exactly} is generally as hard as solving the 
original LS problem.
The condition number $\kappa(A)$ is important from an algorithm design 
perspective since $\kappa(A)$ defines the key nonuniformity structure 
needed to control the complexity of high-precision iterative algorithms, 
\emph{i.e.}, it bounds the number of iterations needed for iterative 
methods to converge.
In particular, for ill-conditioned problems, \emph{e.g.}, if 
$\kappa(A) \approx 10^6 \gg 1$, then the convergence speed of iterative 
methods is very slow, while if $\kappa \gtrsim 1$ then iterative 
algorithms converge very quickly.
Informally, $\kappa(A)$ defines the aspect ratio of the quadratic well 
implicitly defined by $A$ in the high-dimensional Euclidean space. 
\emph{If one is interested in obtaining \emph{high-precision solutions}, 
\emph{e.g.}, $\epsilon=10^{-10}$, that can be obtained by iterating a 
low-precision solution to high precision with an iterative algorithm that 
converges as $\log(1/\epsilon)$, then the key quantity that must be dealt 
with is the condition number of the input~data.}
\item
\textbf{Monte Carlo versus Las Vegas uses of randomization.}
Note that the guarantee provided by the meta-algorithm, as stated above, 
is of the following form: the algorithm runs in no more than a 
specified time $T$, and with probability at least $1-\delta$ it returns 
a solution that is an $\epsilon$-good approximation to the exact solution. 
Randomized algorithms that provide guarantees of this form, \emph{i.e.}, 
with running time that is is deterministic, but whose output may be 
incorrect with a certain small probability, are known as Monte Carlo 
algorithms~\cite{MotwaniRaghavan95}.
A related class of randomized algorithms, known as Las Vegas algorithms, 
provide a different type of guaranatee: they always produce the correct 
answer, but the amount of time they take varies 
randomly~\cite{MotwaniRaghavan95}.
In many applications of RandNLA algorithms, guarantees of this latter 
form are preferable.
\end{itemize}

\noindent
The notions of condition number and leverage scores have been described here only for very overdetermined $\ell_2$ regression problems.
However, as discussed in Section~\ref{sxn:background} below (as well as 
previously~\cite{DMM08_CURtheory_JRNL,clarkson2013fast}), these notions generalize 
to very overdetermined $\ell_p$, for $p \ne 2$, regression 
problems~\cite{clarkson2013fast} as well as to $p=2$ for less rectangular 
matrices, as long as one specifies a rank parameter 
$k$~\cite{DMM08_CURtheory_JRNL}.
Understanding these generalizations, as well as the associated tradeoffs, 
will be important for developing RandNLA algorithms in parallel and 
distributed environments.

\subsection{Running Time Considerations in RAM}
\label{sxn:rla_ram-run_time}

As presented, the meta-algorithm of the previous subsection has a running 
time that depends on both the time to construct the probability 
distribution, $\{\pi_i\}_{i=1}^{n}$, and the time to solve the subsampled 
problem.
For uniform sampling, the former is trivial and the latter depends on the 
size of the subproblem.
For estimators that depend on the exact or approximate (recall the 
flexibility in~(\ref{eqn:approx-lev-score-probs}) provided by $\gamma$) 
leverage scores, the running time is dominated by the exact or approximate 
computation of those scores.
A na\"{i}ve algorithm involves using a QR decomposition or the thin SVD of 
$A$ to obtain the exact leverage scores.  
This na\"{\i}ve implementation of the meta-algorithm takes roughly 
$\bigO(mn^2/\epsilon)$ time and is thus no faster (in the RAM model) than 
solving the original LS problem 
exactly~\cite{drineas2006sampling,DMM08_CURtheory_JRNL}. 
There are two other potential problems with practical implementations of 
the meta-algorithm:
the running time dependence of roughly $\bigO(mn^2/\epsilon)$ time scales 
polynomially with $1/\epsilon$, which is prohibitive if one is interested 
in moderately small (\emph{e.g.}, $10^{-4}$) to very small (\emph{e.g.}, 
$10^{-10}$) values of $\epsilon$; and, since this is a randomized Monte 
Carlo algorithm, with some probability $\delta$ the algorithm might 
completely fail.

Importantly, all three of these potential problems can be solved to yield 
improved variants of the meta-algorithm.

\begin{itemize}
\item
\textbf{Making the algorithm fast: improving the dependence on $m$ and $n$.}
We can make this meta-algorithm ``fast'' in worst-case theory in 
RAM~\cite{sarlos2006improved,drineas2011faster,drineas2006sampling,DMMW12_JMLR,CW13sparse_STOC}.
In particular, this meta-algorithm runs in $\bigO(mn \log n /\epsilon)$ time in 
RAM if one does either of the following:
if one performs a Hadamard-based random random projection and then performs 
uniform sampling in the randomly rotated 
basis~\cite{sarlos2006improved,drineas2011faster} (which, recall, is 
basically what random projection algorithms do when applied to vectors in a 
Euclidean space~\cite{mahoney2011randomized}); or
if one quickly computes approximations to the statistical leverage scores 
(using the algorithm of~\cite{DMMW12_JMLR}, the running time bottleneck of 
which is applying a random projection to the input data) and then uses those 
approximate scores as an importance sampling 
distribution \cite{drineas2006sampling,DMMW12_JMLR}.
In addition, by using carefully-constructed extremely-sparse random 
projections, both of these two approaches can be made to run in so-called
``input sparsity time,'' \emph{i.e.}, in time proportional to the number of 
nonzeros in the input data, plus lower-order terms that depend on the lower 
dimension of the input matrix~\cite{CW13sparse_STOC}.
\item
\textbf{Making the algorithm high-precision: improving the dependence on $\epsilon$.}
We can make this meta-algorithm ``fast'' in practice, \emph{e.g.}, in 
``high precision'' numerical implementation in 
RAM~\cite{rokhlin2008fast,avron2010blendenpik,meng2011lsrn,coakley2011fast}.
In particular, this meta-algorithm runs in $\bigO(mn \log n \log(1/\epsilon))$ 
time in RAM if one uses the subsampled problem constructed by the random 
projection/sampling process to construct a preconditioner, using it as a 
preconditioner for a traditional iterative algorithm on the original full 
problem~\cite{rokhlin2008fast,avron2010blendenpik,meng2011lsrn}.
This is important since, although the worst-case theory holds for any fixed 
$\epsilon$, it is quite coarse in the sense that the sampling complexity 
depends on $\epsilon$ as $1/\epsilon$ and not $\log(1/\epsilon)$.
In particular, this means that obtaining high-precision with (say) 
$\epsilon=10^{-10}$ is not practically possible.
In this iterative use case, there are several tradeoffs: \emph{e.g.}, one 
could construct a very high-quality preconditioner (\emph{e.g.}, using a 
number of samples that would yield a $1+\epsilon$ error approximation if one 
solved the LS problem on the subproblem) and perform fewer iterations, or 
one could construct a lower quality preconditioner by drawing many fewer 
samples and perform a few extra iterations.
Here too, the input sparsity time algorithm of~\cite{CW13sparse_STOC} 
could be used to improve the running time still further.
\item
\textbf{Dealing with the $\delta$ failure probability.}
Although fixing a failure probability $\delta$ is convenient for
theoretical analysis, in certain applications having even a very small 
probability that the algorithm might return a completely meaningless 
answer is undesirable.
In this case, one is interested in converting a Monte Carlo algorithm 
into a Las Vegas algorithm.
Fortuitously, those application areas, \emph{e.g.}, scientific computing,
are often more interested in moderate to high precision solutions than in 
low precision solutions.
In these case, using the subsampled problem to create a preconditioner 
for iterative algorithms on the original problem has the side effect that 
one changes a ``fixed running time but might fail'' algorithm to an 
``expected running time but will never fail'' algorithm.
\end{itemize}

\noindent
From above, we can make the following conclusions.
The ``fast in worst-case theory'' variants of our meta-algorithm
(\cite{sarlos2006improved,drineas2011faster,drineas2006sampling,DMMW12_JMLR,CW13sparse_STOC})
represent qualitative improvements to the $\bigO(mn^2)$ worst-case asymptotic 
running time of traditional algorithms for the LS problem going back to 
Gaussian elimination.
The ``fast in numerical implementation'' variants of the meta-algorithm
(\cite{rokhlin2008fast,avron2010blendenpik,meng2011lsrn}) have been shown to 
beat \textsc{Lapack}'s direct dense least-squares solver by a large margin 
on essentially any dense tall matrix, illustrating that the worst-case 
asymptotic theory holds for matrices as small as several thousand by several 
hundred~\cite{avron2010blendenpik}.

While these results are a remarkable success for RandNLA in RAM, they leave 
open the question of how these RandNLA methods perform in larger-scale 
parallel/distributed environments, and they raise the question of whether 
the same RandNLA principles can be extended to other common regression 
problems.
In the remainder of this paper, we will review recent work showing that if
one wants to solve $\ell_2$ regression problems in parallel/distributed
environments, and if one wants to solve $\ell_1$ regression problems in 
theory or in RAM or in parallel/distributed environments, then one can use 
the same RandNLA meta-algorithm and design principles.
Importantly, though, depending on the exact situation, one must instantiate 
the same algorithmic principles in different ways, \emph{e.g.}, one must 
worry much more about communication rather than FLOPS.

%% file: background.tex
\section{Preliminaries on $\ell_p$ regression problems}
\label{sxn:background}

In this section, we will start in Section~\ref{sxn:background-notation} with 
a brief review of notation that we will use in the remainder of the paper.
Then, in Sections~\ref{sxn:background-lp_problems}, 
\ref{sxn:background-condition_number}, 
and~\ref{sxn:background-preconditioning},
we will review $\ell_p$ regression problems and the notions of condition number 
and preconditioning for these problems.
Finally, in Sections~\ref{sxn:background-ls_solvers}
and~\ref{sxn:background-lp_solvers}, we will review traditional deterministic 
solvers for $\ell_2$ as well as $\ell_1$ and more general $\ell_p$ regression 
problems.

\subsection{Notation conventions}
\label{sxn:background-notation}

We briefly list the notation conventions we follow in this work:
\begin{itemize}
\item We use uppercase letters to denote matrices and constants, e.g., $A$, $R$,
  $C$, etc.
\item We use lowercase letters to denote vectors and scalars, e.g., $x$, $b$,
  $p$, $m$, $n$, etc.
\item We use $\|\cdot\|_p$ to denote the $\ell_p$ norm of a vector,
  $\|\cdot\|_2$ the spectral norm of a matrix, $\|\cdot\|_F$ the Frobenius norm
  of a matrix, and $|\cdot|_p$ the element-wise $\ell_p$ norm of a matrix.
\item We use uppercase calligraphic letters to denote point sets, e.g.,
  $\mathcal{A}$ for the linear subspace spanned by $A$'s columns, $\mathcal{C}$
  for a convex set, and $\mathcal{E}$ for an ellipsoid, except that $\bigO$ is
  used for big O-notation.
\item The ``$\tilde{\ }$'' accent is used for sketches of matrices, e.g.,
  $\tilde{A}$, the ``$^*$'' superscript is used for indicating optimal
  solutions, e.g., $x^*$, and the ``$\hat{\ }$'' accent is used for estimates of
  solutions, e.g., $\hat{x}$.
\end{itemize}

\subsection{$\ell_p$ regression problems}
\label{sxn:background-lp_problems}

In this work, a parameterized family of linear regression problems that is of
particular interest is the \emph{$\ell_p$ regression} problem.
\begin{definition}[$\ell_p$ regression]
  Given a matrix $A \in \R^{m \times n}$, a vector $b \in \R^m$, and
  $p\in[1,\infty]$, the \emph{$\ell_p$~regression problem} specified by $A$,
  $b$, and $p$ is the following optimization problem:
  \begin{equation}
    \label{eq:lp_reg}
    \mathrm{minimize}_{x \in \R^n}~\|A x - b\|_p,
  \end{equation}
  where the $\ell_p$ norm of a vector $x$ is $
  \|x\|_p=\left(\sum_i|x_i|^p\right)^{1/p} $, defined to be $\max_i |x_i|$ for
  $p=\infty$.
  We call the problem \emph{strongly over-determined} if $m \gg n$, and
  \emph{strongly under-determined} if $m \ll n$.
\end{definition}
Important special cases include the $\ell_2$ regression problem, also known 
as linear least squares (LS), and the $\ell_1$ regression problem, also 
known as least absolute deviations (LAD) or least absolute errors (LAE).
The former is ubiquitous; and the latter is of particular interest as a 
robust regression technique, in that it is less sensitive to the presence of 
outliers than the former.

For general $p \in [1, \infty]$, denote $\X^*$ the set of optimal solutions 
to \eqref{eq:lp_reg}.
Let $x^* \in \X^*$ be an arbitrary optimal solution, and let 
$f^* = \|A x^* - b\|_p$ be the optimal objective value.
We will be particularly interested in finding a \emph{relative-error 
approximation}, in terms of the objective value, to the general $\ell_p$
regression problem \eqref{eq:lp_reg}.

\begin{definition}[Relative-error approximation]
  Given an error parameter $\epsilon > 0$, $\hat{x} \in \R^n$ is a
  \emph{$(1+\epsilon)$-approximate solution} to the $\ell_p$ regression problem
  \eqref{eq:lp_reg} if and only if
  \begin{equation*}
    \hat{f} = \|A \hat{x} - b\|_p \leq (1+\epsilon) f^*.
  \end{equation*}
\end{definition}

In order to make our theory and our algorithms for general $\ell_p$ 
regression simpler more concise, we can use an equivalent formulation of 
\eqref{eq:lp_reg} in our discussion.
\begin{equation}
  \label{eq:lp_reg_homo}
  \begin{aligned}
    & \mathrm{minimize}_{x \in \R^n} && \|A x\|_p \\
    & \mathrm{subject~to} && c^T x = 1.
  \end{aligned}
\end{equation}
Above, the ``new'' $A$ is $A$ concatenated with $-b$, i.e., $\begin{pmatrix} A & -b \end{pmatrix}$ and $c$ is a
vector with a $1$ at the last coordinate and zeros elsewhere, i.e., $c \in \R^{d+1}$ and $c = \begin{pmatrix} 0 & \ldots & 0 & 1 \end{pmatrix}$, to force the last element of any feasible solution to be $1$.
We note that the same formulation is also used by
\cite{nesterov2009unconstrained} for solving unconstrained convex problems in
relative scale.
This formulation of $\ell_p$ regression, which consists of a homogeneous
objective and an affine constraint, can be shown to be equivalent to the
formulation of \eqref{eq:lp_reg}.

Consider, next, the special case $p=2$.
If, in the LS problem
\begin{equation}
  \label{eq:ls}
  \text{minimize}_{x \in \mathbb{R}^n}  \quad \| A x - b \|_2  ,
\end{equation}
we let $r = \rank(A) \leq \min(m,n)$, then recall that if $r < n$ (the LS 
problem is under-determined or rank-deficient), then \eqref{eq:ls} has an 
infinite number of minimizers.
In that case, the set of all minimizers is convex and hence has a unique 
element having minimum length.
On the other hand, if $r = n$ so the problem has full rank, there exists 
only one minimizer to \eqref{eq:ls} and hence it must have the minimum 
length.
In either case, we denote this unique min-length solution to \eqref{eq:ls} 
by $x^*$, and we are interested in computing $x^*$ in this work.
This was defined in Problem \eqref{eq:ls_min_length} above.
In this case, we will also be interested in bounding $\| x^* - \hat{x} \|_2$, 
for arbitrary or worst-case input, where $\hat{x}$ was defined in 
Problem~(\ref{lsq-sample}) above and is an approximation to $x^*$.

\subsection{$\ell_p$-norm condition number}
\label{sxn:background-condition_number}

An important concept in $\ell_2$ and more general $\ell_p$ regression problems, 
and in developing efficient algorithms for their solution, is the concept of 
\emph{condition number}.
For linear systems and LS problems, the $\ell_2$-norm condition number is 
already a well-established term.

\begin{definition}[$\ell_2$-norm condition number]
\label{def:l2normcond}
  Given a matrix $A \in \R^{m \times n}$ with full column rank, let
  $\sigma^{\max}_2(A)$ be the largest singular value and $\sigma^{\min}_2(A)$ be
  the smallest singular value of $A$.
  The \emph{$\ell_2$-norm condition number} of $A$ is defined as $\kappa_2(A) =
  \sigma^{\max}_2(A)/\sigma^{\min}_2(A)$.
  For simplicity, we use $\kappa_2$, $\sigma_2^{\min}$, and
  $\sigma_2^{\max}$ when the underlying matrix is clear from context.
\end{definition}

\noindent
For general $\ell_p$ norm and general $\ell_p$ regression problems, here we 
state here two related notions of condition number and then a lemma that 
characterizes the relationship between them.

\begin{definition}[$\ell_p$-norm condition number
  (Clarkson~et~al.~\cite{clarkson2013fast})]
  \label{def:lpnormcond}
  Given a matrix $A \in \R^{m \times n}$ and $p \in [1, \infty]$, let
  \begin{equation*}
    \sigma_p^{\max}(A) = \max_{\|x\|_2 = 1} \|A x\|_p \text{ and } \sigma_p^{\min}(A) = \min_{\|x\|_2 = 1} \|A x\|_p.
  \end{equation*}
  Then, we denote by $\kappa_p(A)$ the \emph{$\ell_p$-norm condition number of
    $A$}, defined to be:
  \begin{equation*}
    \kappa_p(A) = \sigma_p^{\max}(A) / \sigma_p^{\min}(A).
  \end{equation*}
  For simplicity, we use $\kappa_p$, $\sigma_p^{\min}$, and
  $\sigma_p^{\max}$ when the underlying matrix is clear.
\end{definition}

\begin{definition}[$(\alpha, \beta, p)$-conditioning
  (Dasgupta~et~al.~\cite{dasgupta2009sampling})]
  \label{def:lpbasis}
  Given a matrix $A \in \R^{m \times n}$ and $p\in[1,\infty]$, let $\|\cdot\|_q$
  be the dual norm of $\|\cdot\|_p$.
  Then $A$ is \emph{$(\alpha,\beta,p)$-conditioned} if (1) $|A|_p \leq \alpha$,
  and (2) for all $z \in \R^n$, $\|z\|_q \leq \beta \|A z\|_p$.
  Define $\bar{\kappa}_p(A)$, the $(\alpha, \beta, p)$-condition number of $A$,
  as the minimum value of $\alpha \beta$ such that $A$ is $(\alpha, \beta,
  p)$-conditioned.
  We use $\bar{\kappa}_p$ for simplicity if the underlying matrix is clear.
\end{definition}

\begin{lemma}[Equivalence of $\kappa_p$ and $\bar{\kappa}_p$
  (Clarkson~et~al.~\cite{clarkson2013fast})]
  \label{lemma:kappa_equiv}
  Given a matrix $A \in \R^{m \times n}$ and $p \in [1, \infty]$, we always have
  \begin{equation*}
    n^{-|1/2-1/p|} \kappa_p(A) \leq \bar{\kappa}_p(A) \leq n^{\max \{1/2, 1/p\}} \kappa_p(A).
  \end{equation*}
\end{lemma}

\noindent
That is, by Lemma~\ref{lemma:kappa_equiv}, if $m \gg n$, then the notions of 
condition number provided by Definition~\ref{def:lpnormcond} and 
Definition~\ref{def:lpbasis} are equivalent, up to low-dimensional 
factors.
These low-dimensional factors typically do not matter in theoretical 
formulations of the problem, but they can matter in practical 
implementations.

The $\ell_p$-norm condition number of a matrix can be arbitrarily large.
Given the equivalence established by Lemma~\ref{lemma:kappa_equiv}, we say
that a matrix $A$ is \emph{well-conditioned in the $\ell_p$ norm} if 
$\kappa_p$ or $\bar{\kappa}_p = \bigO(\poly(n))$, independent of the high
dimension $m$.
We see in the following sections that the condition number plays a very
important part in the analysis of traditional algorithms.

\subsection{Preconditioning $\ell_p$ regression problems}
\label{sxn:background-preconditioning}

Preconditioning refers to the application of a transformation, called the 
preconditioner, to a given problem instance such that the transformed 
instance is more-easily solved by a given class of algorithms.
Most commonly, the preconditioned problem is solved with an iterative 
algorithm, the complexity of which depends on the condition number of the 
preconditioned problem. 

To start, consider $p=2$, and recall that for a \emph{square} linear system 
$A x = b$ of full rank, this \emph{preconditioning} usually takes one of 
the following forms:
\begin{align*}
  \label{eq:linear_precond}
  \text{left preconditioning} &\quad M\T A x = M\T b, \\
  \text{right preconditioning} &\quad A N y = b, ~ x = N y, \\
  \text{left and right preconditioning} &\quad M\T A N y = M\T b, ~ x = N y.
\end{align*}
Clearly, the preconditioned system is consistent with the original one, 
\emph{i.e.}, has the same $x^*$ as the unique solution, if the preconditioners 
$M$ and $N$ are nonsingular.

For the general LS Problem \eqref{eq:ls_min_length}, more care should be taken
so that the preconditioned system has the same min-length solution as the
original one.
In particular, if we apply left preconditioning to the LS problem 
$\min_x \| A x - b \|_2$, then the preconditioned system becomes 
$\min_x \| M\T A x - M\T b \|_2$, and its min-length solution is given by
\begin{equation*}
  x^*_{\text{left}} = (M\T A)^\dagger M\T b.
\end{equation*}
Similarly, the min-length solution to the right preconditioned system is given
by
\begin{equation*}
  x^*_{\text{right}} = N ( A N )^\dagger b.
\end{equation*}
The following lemma states the necessary and sufficient conditions for
$A^\dagger = N ( A N )^\dagger$ or $A^\dagger = ( M\T A )^\dagger M\T$ to hold.
Note that these conditions holding certainly imply that $x_{\text{right}}^* =
x^*$ and $x_{\text{left}}^* = x^*$, respectively.

\begin{lemma}[Left and right preconditioning (Meng \emph{et al.} \cite{meng2011lsrn}]
  \label{lemma:ls_precond}
  Given $A \in \mathbb{R}^{m \times n}$, $N \in \mathbb{R}^{n \times p}$ and $M
  \in \mathbb{R}^{m \times q}$, we have
  \begin{enumerate}
  \item $A^\dagger = N ( A N )^\dagger$ if and only if $\text{range}(N N\T A\T )
    = \text{range} ( A\T )$,
  \item $A^\dagger = ( M\T A )^\dagger M\T$ if and only if $\text{range}(M M\T
    A) = \text{range}(A)$.
  \end{enumerate}
\end{lemma}

\noindent
Given this preconditioned problem, \eqref{eq:cg_convergence_rate} (see 
below) bounds the number of itrations for certain iterative algorithms for the 
LS problem.

Just as with $p=2$, for more general $\ell_p$ regression problems with matrix 
$A \in \R^{m \times n}$ with full column rank, although its condition numbers 
$\kappa_p(A)$ and $\bar{\kappa}_p(A)$ can be arbitrarily large, we can often 
find a matrix $R \in \R^{n \times n}$ such that $A R^{-1}$ is well-conditioned.
(This is \emph{not} the $R$ from a QR decomposition of $A$, unless $p=2$, 
but some other matrix $R$.)
In this case, the $\ell_p$ regression Problem \eqref{eq:lp_reg_homo} is 
equivalent to the following well-conditioned problem:
\begin{equation}
  \label{eq:lp_reg_precond}
  \begin{aligned}
    & \text{minimize}_{y \in \R^n} && \|A R^{-1} y\|_p, \\
    & \text{subject to} && c^T R^{-1} y = 1. \\
  \end{aligned}
\end{equation}
Clearly, if $y^*$ is an optimal solution to \eqref{eq:lp_reg_precond}, then 
$x^* = R^{-1} y$ is an optimal solution to \eqref{eq:lp_reg_homo}, and vice 
versa; however, \eqref{eq:lp_reg_precond} may be easier to solve than
\eqref{eq:lp_reg_homo} because of better conditioning.


Since we want to reduce the condition number of a problem instance via 
preconditioning, it is natural to ask what the best possible outcome would 
be in theory.
For $p=2$, an orthogonal matrix, \emph{e.g.}, the matrix $Q$ computed from a QR
decomposition, has $\kappa_2(Q) = 1$.
More generally, for the $\ell_p$-norm condition number $\kappa_p$, we have the 
following existence result.
\begin{lemma}
  \label{lemma:best_kappa}
  Given a matrix $A \in \R^{m \times n}$ with full column rank and $p \in [1,
  \infty]$, there exist a matrix $R \in \R^{n \times n}$ such that $\kappa_p(A
  R^{-1}) \leq n^{1/2}$.
\end{lemma}
This is a direct consequence of John's theorem~\cite{john1948extremum} on
ellipsoidal rounding of centrally symmetric convex sets.
For the $(\alpha, \beta, p)$-condition number $\bar{\kappa}_p$, we have the
following lemma.
\begin{lemma}
  \label{lemma:best_kappa_bar}
  Given a matrix $A \in \R^{m \times n}$ with full column rank and $p \in [1,
  \infty]$, there exist a matrix $R \in \R^{n \times n}$ such that
  $\bar{\kappa}_p(A R^{-1}) \leq n$.
\end{lemma}
Note that Lemmas \ref{lemma:best_kappa} and \ref{lemma:best_kappa_bar} are both 
existential results.
Unfortunately, except the case when $p = 2$, no polynomial-time algorithm is 
known that can provide such preconditioning for general matrices.
Below, in Section~\ref{sxn:round_embed}, we will discuss two practical 
approaches for $\ell_p$-norm preconditioning: via ellipsoidal rounding and 
via subspace embedding, as well as subspace-preserving sampling algorithms 
built on top of them.

\subsection{Traditional solvers for $\ell_2$ regression}
\label{sxn:background-ls_solvers}

Least squares is a classic problem in linear algebra.
It has a long history, tracing back to Gauss, and it arises in numerous
applications.
A detailed survey of numerical algorithms for least squares is certainly beyond
the scope of this work.
In this section, we briefly describe some well-known direct methods and
iterative methods that compute the min-length solution to a possibly
rank-deficient least squares problem, and we refer readers to
Bj{\"o}rck~\cite{bjorck1996numerical} for additional details.

\subsubsection*{Direct methods}

It is well known that the min-length solution of a least squares problem can be
computed using the singular value decomposition (SVD).
Let $A = U \Sigma V\T$ be the compact SVD, where $U \in \R^{m \times r}$,
$\Sigma \in \R^{r \times r}$, and $V \in \R^{n \times r}$, \emph{i.e.}, only singular
vectors corresponding to the non-zero singular values are calculated.
We have $x^* = V \Sigma^{-1} U\T b$.
The matrix $V \Sigma^{-1} U\T$ is the Moore-Penrose pseudoinverse of $A$,
denoted by $A^\dagger$, which is defined and unique for any matrix.
Hence we can simply write $x^* = A^\dagger b$.
The SVD approach is accurate and robust to rank-deficiency.

Another way to solve a least squares problem is using complete orthogonal
factorization.
If we can find orthonormal matrices $Q \in \R^{m \times r}$ and $Z \in \R^{n
  \times r}$, and a matrix $T \in \R^{r \times r}$, such that $A = Q T Z\T$,
then the min-length solution is given by $x^* = Z T^{-1} Q\T b$.
We can treat SVD as a special case of complete orthogonal factorization.
In practice, complete orthogonal factorization is usually computed via
rank-revealing QR factorizations, making $T$ a triangular matrix.
The QR approach is less expensive than SVD, but it is slightly less robust at
determining the rank.

A third way to solve a least squares problem is by computing the min-length
solution to the normal equation $A\T A x = A\T b$, namely
\begin{equation}
  \label{eq:ls_ne_min_length}
  x^* = (A\T A)^\dagger A\T b = A\T (A A\T)^\dagger b.
\end{equation}
It is easy to verify the correctness of \eqref{eq:ls_ne_min_length} by replacing
$A$ by its compact SVD $U \Sigma V\T$.
If $r = \min(m,n)$, a Cholesky factorization of either $A\T A$ (if $m \geq n$)
or $AA\T$ (if $m \leq n$) solves \eqref{eq:ls_ne_min_length}.
If $r < \min(m,n)$, we need the eigensystem of $A\T A$ or $A A\T$ to compute
$x^*$.
The normal equation approach is the least expensive among the three direct
approaches we have mentioned, but it is
also the least accurate one, especially on ill-conditioned problems.
See Chapter~5 of Golub and Van Loan \cite{golub1996matrix} for a detailed
analysis.
A closely related direct solver is the semi-normal equation method. It is often useful when the $R$-factor of the QR decomposition is known; see~\cite{semi-normal} for more details.

For sparse least squares problems, by pivoting $A$'s columns and rows, we may
find a sparse factorization of $A$, which is preferred to a dense factorization
for more efficient storage.
For sparse direct methods, we refer readers to Davis~\cite{davis2006direct}.

\subsubsection*{Iterative methods}

Instead of direct methods, we can use iterative methods to solve \eqref{eq:ls}.
If all the iterates $\{ x^{(k)} \}$ are in $\text{range}(A\T)$ and if 
$\{ x^{(k)} \}$ converges to a minimizer, it must be the minimizer having 
minimum length, \emph{i.e.}, the solution to Problem \eqref{eq:ls_min_length}.
This is the case when we use a Krylov subspace method starting with a zero
vector.
For example, the conjugate gradient (CG) method on the normal equation leads to
the min-length solution (see Paige and Saunders~\cite{paige1975solution}).
In practice, CGLS~\cite{hestenesmethods}, LSQR~\cite{paige1982lsqr} are
preferable because they are equivalent to applying CG to the normal equation in
exact arithmetic but they are numerically more stable.
Other Krylov subspace methods such as LSMR~\cite{fong2011lsmr} can also solve
\eqref{eq:ls} as well.
The Chebyshev semi-iterative method \cite{golub1961chebyshev} can also be
modified to solve LS problems.

Importantly, however, it is in general hard to predict the number of iterations
for CG-like methods.
The convergence rate is affected by the condition number of $A\T A$.
A classical result \cite[p.187]{luenberger1973introduction} states that
\begin{equation}
  \label{eq:cg_convergence_rate}
  \frac{\| x^{(k)} - x^* \|_{A\T A}}{\|x^{(0)} - x^*\|_{A\T A}} 
  \leq 2 \left( \frac{\sqrt{\kappa(A\T A)} - 1}{\sqrt{\kappa(A\T A)} + 1} \right)^k,
\end{equation}
where $\| z \|_{A\T A} = z\T A\T A z = \|A z\|^2$ for any $z \in \R^n$,
and where $\kappa(A\T A)$ is the condition number of $A\T A$ under the $2$-norm.
Estimating $\kappa(A\T A)$ is generally as hard as solving the LS problem
itself, and in practice the bound does not hold in any case unless
reorthogonalization is used.
Thus, the computational cost of CG-like methods remains unpredictable in
general, except when $A\T A$ is very well-conditioned and the condition number
can be well estimated.

\subsection{Traditional solvers for $\ell_1$ and more general $\ell_p$ regression}
\label{sxn:background-lp_solvers}

While $\ell_2$ regression can be solved with direct methods such as SVD and QR,
the solution of general $\ell_p$ regression has to rely on iterative methods due
to the lack of analytical solution.
In particular, 
$\ell_1$ and $\ell_{\infty}$ regression problems can be formulated as linear
programs and solved by linear programming solvers, and general $\ell_p$
regression problems can be formulated as convex programs and hence solved by
general convex solvers. 
This, however, comes at the cost of increased complexity, compared to the
$\ell_2$ case.
For example, it is easy to see that all $\ell_p$ regression problems are convex
due to the convexity of vector norms.
Therefore, standard convex solvers, \emph{e.g.}, gradient-based
methods~\cite{nesterov2004introductory}, interior-point
methods~(IPMs)~\cite{ye2011interior}, and interior-point cutting-plane
methods~(IPCPMs)\cite{mitchell2003polynomial} can be used to solve $\ell_p$
regression problems.
Discussing those convex solvers is beyond the scope of the work.
We refer readers to the monographs mentioned above or Boyd and
Vandenberghe~\cite{boyd2004convex} for a general introduction.

When $p=1$ or $\infty$, the problem is still convex but not smooth.
Subgradient methods~\cite{clarkson2005subgradient} or gradient methods with
smoothing~\cite{nesterov2005smooth} can be used to handle non-smoothness, while
another solution is via linear programming.
In particular, an $\ell_1$ regression problem specified by $A \in \R^{m \times n}$ and $b \in
\R^m$ is equivalent to the following linear program:
\begin{equation*}
  \begin{aligned}
    & \text{minimize} && \mathbf{1}_m^Ty_+ + \mathbf{1}_m^Ty_- \\
    & \text{subject to} &&  A x - b = y_+ - y_-, \\
    & && y_+, y_- \geq 0, \quad y_+, y_- \in \R^m, \quad x \in \R^n,
  \end{aligned}
\end{equation*}
and an $\ell_\infty$ regression problem specified by $A$ and $b$ is equivalent
to the following:
\begin{equation*}
  \begin{aligned}
    & \text{minimize} && y \\
    & \text{subject to} &&  -y \leq A x - b \leq y, \\
    & && y \in \R, \quad x \in \R^n,
  \end{aligned}
\end{equation*}
where $\mathbf{1}_m \in \R^m$ indicates a vector of length $m$ with all ones.
As a linear programming problem, an $\ell_1$ or $\ell_\infty$ regression problem
can be solved by any linear programming solver, using the simplex
method~\cite{dantzig1998linear} or IPMs.
Similar to the case for least squares, the $\ell_p$ condition number affects the
performance of $\ell_p$ regression solvers, \emph{e.g.}, on the convergence rate 
for subgradient~\cite{clarkson2005subgradient} or gradient
method~\cite{nesterov2008rounding}, on the search of an initial feasible point
for IPMs~\cite{vavasis1995condition}, and on the initial search region for
ellipsoid methods and IPCPMs~\cite{mitchell2003polynomial}.
Generally speaking, a smaller $\ell_p$ condition number makes the problem easier to
solve.

Another popular way to solve $\ell_p$ regression problems is via iteratively
re-weighted least squares (IRLS)~\cite{holland1977robust}, which solves a
sequence of weighted least squares problems and makes the solutions converge to
an optimal solution of the original $\ell_p$ regression problem.
At step $k$, it solves the following weighted least squares problem:
\begin{equation*}
  x^{(k+1)} = \arg \min_{x \in \R^n} \| W^{(k)} (A x - b) \|_2,
\end{equation*}
where $W^{(k)}$ is a diagonal matrix with positive diagonals $w^{(k)}_i$,
$i=1,\ldots,m$.
Let $W^{(0)}$ be an identity matrix and choose
\begin{equation*}
  w^{(k)}_i = |a_i^T x^{(k)} - b_i|^{p-2}, \quad i = 1,\ldots,m, \quad k=1,\ldots.
\end{equation*}
until $\{x^{(k)}\}$ converges.
The choice of $w^{(k)}_i$ is often smoothed to avoid dividing by zero in
practice.
It is not hard to show that if $\{x^{(k)}\}$ converges, it converges to an
optimal solution of the $\ell_p$ regression problem.
However, the convergence theory of IRLS only exists under certain assumptions
and the convergence rate is much harder to derive.
See Burrus~\cite{burrus2013iterative} for a survey of related work.

%% file: basic_theory.tex
\section{Rounding, embedding, and sampling $\ell_p$ regression problems}
\label{sxn:round_embed}

Preconditioning, ellipsoidal rounding, and low-distortion subspace embedding 
are three core technical tools underlying RandNLA regression algorithms.
In this section, we will describe in detail how these methods are used for
$\ell_p$ regression problems, with an emphasis on tradeoffs that arise when 
applying these methods in parallel and distributed environments.
Recall that, for any matrix $A \in \R^{m \times n}$ with full column rank, 
Lemmas~\ref{lemma:best_kappa} and~\ref{lemma:best_kappa_bar} above show that 
there always exists a preconditioner matrix $R \in \R^{n \times n}$ such that 
$A R^{-1}$ is well-conditioned, for $\ell_p$ regression, for general 
$p\in[1,\infty]$.
For $p=2$, such a matrix $R$ can be computed in $O(mn^2)$ time as the ``R'' 
matrix from a QR decomposition, although it is of interest to compute other
such preconditioner matrices $R$ that are nearly as good more quickly; and
for $p=1$ and other values of $p$, it is of interest to compute a
preconditioner matrix $R$ in time that is linear in $m$ and low-degree 
polynomial in $n$.
In this section, we will discuss these and related issues.

In particular, in Sections~\ref{sxn:round_embed-round} 
and~\ref{sxn:round_embed-embed}, we discuss practical algorithms to find 
such $R$ matrices, and we describe the trade-offs between speed 
(\emph{e.g.}, FLOPS, number of passes, additional space/time, etc.) and 
conditioning quality.
The algorithms fall into two general families: ellipsoidal rounding 
(Section~\ref{sxn:round_embed-round}) and subspace embedding 
(Section~\ref{sxn:round_embed-embed}).
We present them roughly in the order of speed (in the RAM model), from 
slower ones to faster ones.  
We will discuss practical tradeoffs in Section~\ref{sxn:implementations}.
For simplicity, here we assume $m \gg \poly(n)$, and hence 
$m n^2 \gg m n + \poly(n)$; and if $A$ is sparse, we assume that 
$m n \gg \nnz(A)$. Hereby, the degree of $\poly(n)$ depends on the underlying algorithm, which may range from $\bigO(n)$ to $\bigO(n^7)$.

Before diving into the details, it is worth mentioning a few high-level 
considerations about subspace embedding methods.
(Similar considerations apply to ellipsoidal rounding methods.)
Subspace embedding algorithms involve mapping data points, 
\emph{e.g.}, the columns of an $m \times n$ matrix, where $m \gg n$ to a 
lower-dimensional space such that some property of the data, \emph{e.g.}, 
geometric properties of the point set, is approximately preserved; see Definition~\ref{def:low_dist} for definition for low-distortion subspace embedding matrix. 
As such, they are critical building blocks for developing improved random 
sampling and random projection algorithms for common linear algebra 
problems more generally, and they are one of the main technical tools for 
RandNLA algorithms.
There are several properties of subspace embedding algorithms that are 
important in order to optimize their performance in theory and/or in 
practice.
For example, given a subspace embedding algorithm, we may want to know:
\begin{itemize}
\item whether it is \emph{data-oblivious} (\emph{i.e.}, independent of the 
input subspace) or \emph{data-aware} (\emph{i.e.}, dependent on some property 
of the input matrix or input space),
\item the time and storage it needs to \emph{construct} an embedding,
\item the time and storage to \emph{apply} the embedding to an input matrix,
\item the \emph{failure rate}, if the construction of the embedding is 
randomized,
\item the \emph{dimension} of the embedding, \emph{i.e.}, the number of 
dimensions being sampled by sampling algorithms or being projected onto by 
projection algorithms,
\item the \emph{distortion} of the embedding, and
\item how to balance the \emph{trade-offs} among those properties.
\end{itemize}
Some of these considerations may not be important for typical theoretical 
analysis but still affect the practical performance of implementations of 
these algorithms.

After the discussion of rounding and embedding methods, we will then show in 
Section~\ref{sxn:app_to_lp} that ellipsoidal rounding and subspace embedding 
methods (that show that the $\ell_p$ norms of the entire subspace of vectors 
can be well-preserved) can be used in one of two complementary ways: one can 
solve an $\ell_p$ regression problem on the rounded/embedded subproblem; or 
one can use the rounding/embedding to construct a preconditioner for the 
original problem.
(We loosely refer to these two complementary types of approaches as 
low-precision methods and high-precision methods, respectively.  The reason 
is that the running time complexity with respect to the error parameter 
$\epsilon$ for the former is $\mbox{poly}(1/\epsilon)$, while the running 
time complexity with respect to $\epsilon$ for the latter is 
$\log(1/\epsilon)$.) 
We also discuss various ways to combine these two types of approaches to 
improve their performance in practice.

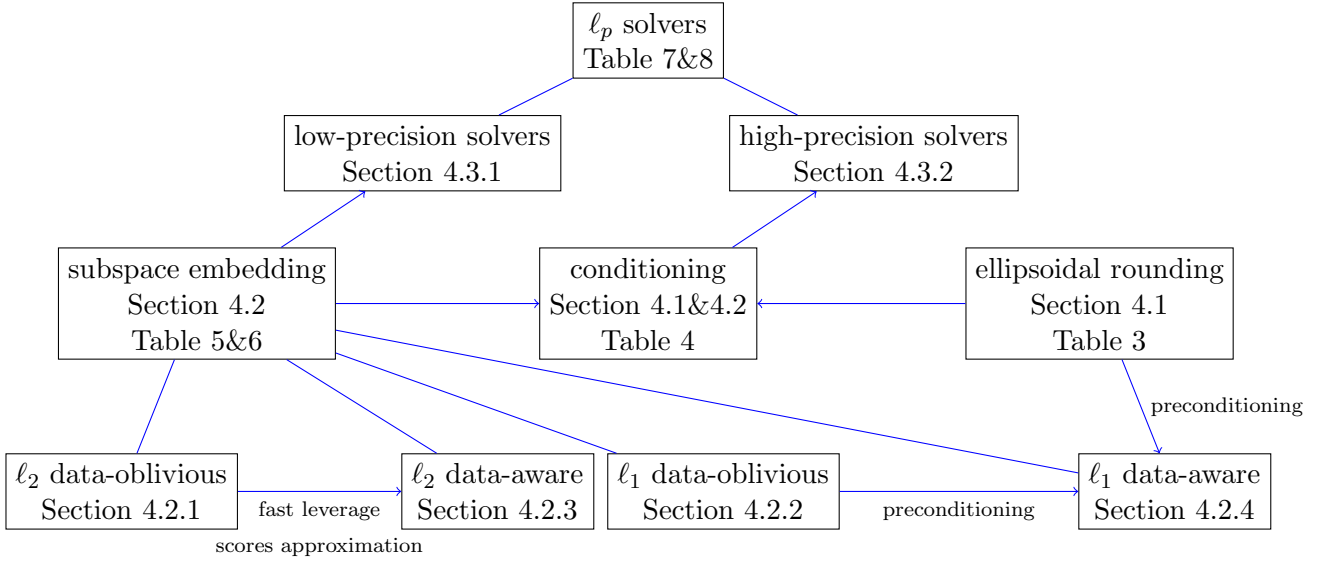
\begin{figure*}
\centering
\begin{tikzpicture}

\node[draw,align=center] (lp) at (20,10) {$\ell_p$ solvers\\
Table~\ref{tab:l2_solve}\&\ref{tab:l1_solve}};
\node[draw,align=center] (low) at (17,8.5) {low-precision solvers\\
Section~\ref{sxn:low-precision-solvers}};
\node[draw,align=center] (high) at (23,8.5) {high-precision solvers\\
Section~\ref{sxn:high-precision-solvers}};

\draw[-,draw=blue] (lp) -- (low);
\draw[-,draw=blue] (lp) -- (high);

\node[draw,align=center] (embed) at (14,6.5) {subspace embedding\\
Section~\ref{sxn:round_embed-embed}\\
Table~\ref{tab:l2_embed}\&\ref{tab:l1_embed}};
\node[draw,align=center] (cond) at (20,6.5) {conditioning\\
Section~\ref{sxn:round_embed-round}\&\ref{sxn:round_embed-embed}\\
Table~\ref{tab:cond}};
\node[draw,align=center] (round) at (26,6.5) {ellipsoidal rounding\\
Section~\ref{sxn:round_embed-round}\\
Table~\ref{tab:rounding}};

\draw[->,draw=blue] (embed) -- (low);
\draw[->,draw=blue] (cond) -- (high);
\draw[->,draw=blue] (embed) -- (cond);
\draw[->,draw=blue] (round) -- (cond);

\node[draw,align=center] (l1_obli) at (21,4) {$\ell_1$ data-oblivious\\
Section~\ref{sxn:data-oblivious-l1-embed}};
\node[draw,align=center] (l2_obli) at (13,4) {$\ell_2$ data-oblivious\\
Section~\ref{sxn:data-oblivious-l2-embed}};
\node[draw,align=center] (l1_awa) at (27,4) {$\ell_1$ data-aware\\
Section~\ref{sxn:data-aware-l1-embed}};
\node[draw,align=center] (l2_awa) at (18,4) {$\ell_2$ data-aware\\
Section~\ref{sxn:data-aware-l2-embed}};

\draw[-,draw=blue] (embed) -- (l1_obli);
\draw[-,draw=blue] (embed) -- (l2_obli);
\draw[-,draw=blue] (embed) -- (l1_awa);
\draw[-,draw=blue] (embed) -- (l2_awa);

\draw[->,draw=blue] (round) -- (l1_awa) node [midway, right, align=center] {\scriptsize preconditioning};
\draw[->,draw=blue] (l1_obli) -- (l1_awa) node [midway, below, sloped, align=center] {\scriptsize preconditioning};
\draw[->,draw=blue] (l2_obli) -- (l2_awa) node [midway, below, sloped, align=center] {\scriptsize fast leverage \\ \scriptsize scores approximation};

    
 \end{tikzpicture}
 \caption{Overview of relationships between several core technical components in RandNLA algorithms for solving $\ell_p$ regression. Relevant subsection and tables in this section are also shown. A directed edge implies the tail component contributes to the head component. }
  \label{fig:flow}
\end{figure*}

Since we will introduce several important and distinct but closely-related 
concepts in this long section, in Figure~\ref{fig:flow} we provide an 
overview of these relations as well as of the structure of this~section.

\subsection{Ellipsoidal rounding and fast ellipsoid rounding}
\label{sxn:round_embed-round}

In this subsection, we will describe \emph{ellipsoidal rounding} methods.
In particular, we are interested in the ellipsoidal rounding of a centrally 
symmetric convex set and its application to $\ell_p$-norm preconditioning.
We start with a definition.

\begin{definition}[Ellipsoidal rounding]
  Let $\mathcal{C} \subseteq \R^n$ be a convex set that is full-dimensional,
  closed, bounded, and centrally symmetric with respect to the origin. 
  An ellipsoid $\mathcal{E}(0, E) = \{ x \in \R^n \,|\, \|E x\|_2 \leq 1 \}$ is
  a \emph{$\kappa$-rounding} of $\mathcal{C}$ if it satisfies $
  \mathcal{E}/\kappa \subseteq \mathcal{C} \subseteq \mathcal{E} $, for some
  $\kappa \geq 1$, where $\mathcal{E}/\kappa$ means shrinking $\mathcal{E}$ by a
  factor of $1/\kappa$. 
\end{definition}

Finding an ellipsoidal rounding with a small $\kappa$ factor for a given convex
set has many applications such as in computational
geometry~\cite{barequet2001efficiently}, convex
optimization~\cite{lovasz1986algorithmic}, and computer
graphics~\cite{bouville1985bounding}.
In addition, 
the $\ell_p$-norm condition number $\kappa_p$ naturally connects to ellipsoidal
rounding.
To see this, let $\mathcal{C} = \{ x \in \R^n \,|\, \|A x\|_p \leq 1\} $ and 
assume that we have a $\kappa$-rounding of 
$\mathcal{C}$: $\mathcal{E} = \{ x \,|\, \|R x\|_2 \leq 1 \}$.
This implies
\begin{equation*}
  \|R x\|_2 \leq \|A x\|_p \leq \kappa \|R x\|_2, \quad \forall x \in \R^n.
\end{equation*}
If we let $y = R x$, then we get
\begin{equation*}
  \|y\|_2 \leq \|A R^{-1} y\|_p \leq \kappa \|y\|_2, \quad \forall y \in \R^n.
\end{equation*}
Therefore, we have $\kappa_p(A R^{-1}) \leq \kappa$.
So a $\kappa$-rounding of $\mathcal{C}$ leads to a $\kappa$-preconditioning of $A$.

Recall the well-known result due to John~\cite{john1948extremum} that for a
centrally symmetric convex set $\mathcal{C}$ there exists a 
$n^{1/2}$-rounding.
It is known that this result is sharp and that such rounding is given by the
L{\"o}wner-John (LJ) ellipsoid of $\mathcal{C}$, \emph{i.e.}, the 
minimal-volume ellipsoid containing $\mathcal{C}$.
This leads to Lemma~\ref{lemma:best_kappa} above.
Unfortunately, finding an $n^{1/2}$-rounding is a hard problem.
No constant-factor approximation in polynomial time is known for general
centrally symmetric convex sets, and hardness results have been
shown~\cite{lovasz1986algorithmic}.

To state algorithmic results, suppose that $\mathcal{C}$ is described by a
separation oracle and that we are provided an ellipsoid $\mathcal{E}_0$ that
gives an $L$-rounding for some $L \geq 1$.
In this case, we can find a $(n(n+1))^{1/2}$-rounding in polynomial time, in
particular, in $\bigO(n^4 \log L)$ calls to the oracle; see 
Lov{\'a}sz~\cite[Theorem 2.4.1]{lovasz1986algorithmic}.
(Polynomial time algorithms with better $\kappa$ have been proposed for 
special convex sets, \emph{e.g.}, the convex hull of a finite point 
set~\cite{khachiyan1993complexity} and the convex set specified by the 
matrix $\ell_\infty$ norm~\cite{nesterov2008rounding}.)
This algorithmic result was used by Clarkson~\cite{clarkson2005subgradient} 
and then by Dasgupta \emph{et al.}~\cite{dasgupta2009sampling} for $\ell_p$ 
regression.
Note that, in these works, only $\bigO(n)$-rounding is actually needed, 
instead of $(n(n+1))^{1/2}$-rounding.

\begin{table}[t]
  \centering
  \begin{tabular}{c|c|c|c|c}
    & $\kappa$ & time & \# passes & \# calls to oracle\\
    \hline
    ER~\cite{clarkson2005subgradient,dasgupta2009sampling} & $(n (n+1))^{1/2}$ & $\bigO(m n^5 \log m)$ & $n^3\log m$  & $\bigO(n^4\log m)$ \\
    Fast ER~\cite{clarkson2013fast} & $2 n$ & $\bigO(m n^3 \log m)$ & $n\log m$ & $\bigO(n^2 \log m)$\\
    Single-pass ER~\cite{meng2013robust} & $2 n^{|2/p-1|+1}$ & $\bigO(m n^2 \log m)$ & $1$ & $\bigO(n^2 \log m)^\ast$ \\
  \end{tabular}
  \caption{
   Summary of several ellipsoidal rounding for $\ell_p$ conditioning.
           Above, the $\ast$ superscript denotes that the oracles are described and called through a smaller matrix with size $m/n$ by $n$.
          }
  \label{tab:rounding}
\end{table}

Recent work has focused on constructing ellipsoid rounding methods that are 
much faster than these more classical techniques but that lead to only slight 
degredation in preconditioning quality.
See Table~\ref{tab:rounding} for a summary of these results.
In particular, Clarkson \emph{et al.}~\cite{clarkson2013fast} follow the 
same construction as in the proof of Lov{\'a}sz~\cite{lovasz1986algorithmic} 
but show that it is much faster (in $\bigO(n^2 \log L)$ calls to the oracle) 
to find a (slightly worse) $2n$-rounding of a centrally symmetric convex set 
in $\R^n$ that is described by a separation oracle.

\begin{lemma}[Fast ellipsoidal rounding (Clarkson \emph{et al.}~\cite{clarkson2013fast})]
  \label{thm:rounding}
  Given a centrally symmetric convex set $\mathcal{C} \subseteq \R^n$, which is
  centered at the origin and described by a separation oracle, and an ellipsoid
  $\mathcal{E}_0$ centered at the origin such that $\mathcal{E}_0/L \subseteq
  \mathcal{C} \subseteq \mathcal{E}_0$ for some $L \geq 1$, it takes at most
  $3.15 n^2 \log L$ calls to the oracle and additional $\bigO(n^4 \log L)$ time to
  find a $2n$-rounding of $\mathcal{C}$.
\end{lemma}

\noindent
By applying Lemma \ref{thm:rounding} to the convex set 
$\mathcal{C} = \{ x \,|\, \|A x\|_p \leq 1\}$, with the separation oracle 
described via a subgradient of $\|A x\|_p$ and the initial rounding provided 
by the ``R'' matrix from the QR decomposition of $A$, one immediately 
improves the running time of the algorithm used by
Clarkson~\cite{clarkson2005subgradient} and by Dasgupta 
\emph{et al.}~\cite{dasgupta2009sampling} from $\bigO(m n^5 \log m)$ to 
$\bigO(m n^3 \log m)$ while maintaining an $\bigO(n)$-conditioning.

\begin{corollary}
  \label{coro:lp_cond_2d}
  Given a matrix $A \in \R^{m \times n}$ with full column rank, it takes at most
  $\bigO(m n^3 \log m)$ time to find a matrix $R \in \R^{n \times
    n}$ such that $\kappa_p(A R^{-1}) \leq 2 n$.
\end{corollary}

Unfortunately, even this improvement for computing a $2n$-conditioning is 
not immediately applicable to very large matrices.
The reason is that such matrices are usually distributively stored on 
secondary storage and each call to the oracle requires a pass through the 
data.
We could group $n$ calls together within a single pass, but this would still 
need $\bigO(n \log m)$ passes. 
Instead, Meng and Mahoney~\cite{meng2013robust} present a deterministic 
single-pass conditioning algorithm that balances the cost-performance 
trade-off to provide a $2 n^{|2/p-1|+1}$-conditioning of 
$A$~\cite{meng2013robust}.
This algorithm essentially invoke the fast ellipsoidal rounding 
(Lemma~\ref{thm:rounding}) method on a smaller problem which is constructed 
via a single-pass on the original dataset.
Their main algorithm is stated in  Algorithm~\ref{alg:l1_cond}, and the main 
result for Algorithm~\ref{alg:l1_cond} is the following.

\begin{lemma}[One-pass conditioning (Meng and Mahoney~\cite{meng2013robust})]
\label{lem:one-pass-cond}
Algorithm~\ref{alg:l1_cond} is a $2 n^{|2/p-1|+1}$-conditioning algorithm, 
and it runs in $\bigO((m n^2 + n^4) \log m)$ time.
 It needs to compute a $2n$-rounding on a problem with size $m/n$ by $n$ which needs $\bigO(n^2 \log m)$ calls to the separation oracle on the smaller problem.
\end{lemma}

\begin{algorithm}[t]
  \caption{A single-pass conditioning algorithm.}
  \label{alg:l1_cond}
  \begin{algorithmic}[1]
    \REQUIRE $A \in \mathbb{R}^{m \times n}$ with full column rank and $p \in
    [1, \infty]$.  

    \ENSURE A non-singular matrix $E \in \mathbb{R}^{n \times n}$ such
    that $$\|y\|_2 \leq \|A E y\|_p \leq 2 n^{|2/p-1|+1} \|y\|_2,\ \forall y \in
    \mathbb{R}^n.$$

    \STATE Partition $A$ along its rows into sub-matrices of size $n^2 \times
    n$, denoted by $A_1, \ldots, A_M$.

    \STATE For each $A_i$, compute its economy-sized SVD: $A_i = U_i \Sigma_i
    V_i^T$.

    \STATE Let $\tilde{A}_i = \Sigma_i V_i^T$ for $i=1,\ldots,M$,
    \begin{equation*}
      \tilde{\mathcal{C}} = \left\{ x \in \R^n \,\left|\, \left(\sum_{i=1}^M
            \|\tilde{A}_i x\|_2^p \right)^{1/p} \leq
          1\right.\right\}, \text{ and } \tilde{A} = {\tiny \begin{pmatrix} \tilde{A}_1 \\
          \vdots \\ \tilde{A}_M
          \end{pmatrix}}.
    \end{equation*}

    \STATE Compute $\tilde{A}$'s SVD: $\tilde{A} = \tilde{U} \tilde{\Sigma}
    \tilde{V}^T$.
    
    \STATE Let $\mathcal{E}_0 = \mathcal{E}(0, E_0)$ where $E_0 =
    n^{\max\{1/p-1/2, 0\}} \tilde{V} \tilde{\Sigma}^{-1}$.

    \STATE Compute an ellipsoid $\mathcal{E} = \mathcal{E}(0, E)$ that gives a
    $2n$-rounding of~$\tilde{\mathcal{C}}$ starting from $\E_0$ that gives an
    $(M n^2)^{|1/p - 1/2|}$-rounding of $\tilde{\mathcal{C}}$.

    \STATE Return $n^{\min \{1/p-1/2, 0\}} E$.
  \end{algorithmic}
\end{algorithm}

\noindent
\begin{remark}
Solving the rounding problem of size $m/n \times n$ in 
Algorithm~\ref{alg:l1_cond} requires $\bigO(m)$ RAM, which might be too much 
for very large-scale problems.
In such cases, one can increase the block size from $n^2$ to, \emph{e.g.}, 
$n^3$.
A modification to the proof of Lemma~\ref{lem:one-pass-cond} shows that this 
gives us a $2 n^{|3/p - 3/2| + 1}$-conditioning algorithm that only needs
$\bigO(m/n)$ RAM and $\bigO((m n + n^4) \log m)$ FLOPS for the rounding
problem.
\end{remark}
\begin{remark}
One can replace SVD computation in 
Algorithm~\ref{alg:l1_cond} by a fast randomized $\ell_2$ subspace embedding 
(\emph{i.e.}, a fast low-rank approximation algorithm as described 
in~\cite{halko2011finding,mahoney2011randomized} and that we describe 
below).
This reduces the overall running time to $\bigO((m n + n^4) \log(m n)$), and 
this is an improvement in terms of FLOPS; but this would lead to a 
non-deterministic result with additional variability due to the 
randomization (that in our experience substantially degrades the
embedding/conditioning quality in practice).
How to balance those trade-offs in real applications and implementations 
depends on the underlying application and problem details.
\end{remark}

\subsection{Low-distortion subspace embedding and subspace-preserving embedding}
\label{sxn:round_embed-embed}

In this subsection, we will describe in detail \emph{subspace embedding} 
methods.
Subspace embedding methods were first used in RandNLA by Drineas 
\emph{et al.} in their relative-error approximation algorithm for $\ell_2$ 
regression (basically, the meta-algorithm described in 
Section~\ref{sxn:rla_ram-meta_alg})~\cite{drineas2006sampling}; they were
first used in a data-oblivious manner in RandNLA by 
Sarl{\'o}s~\cite{sarlos2006improved}; and an overview of data-oblivious 
subspace embedding methods as used in RandNLA has been provided by 
Woodruff~\cite{Woodruff_sketching_NOW}.
Based on the properties of the subspace embedding methods, we will present 
them in the following four categories.
In Section~\ref{sxn:data-oblivious-l2-embed} 
and~\ref{sxn:data-oblivious-l1-embed}, we will discuss the data-oblivious 
subspace embedding methods for $\ell_2$ and $\ell_1$ norms, respectively; 
and then in Section~\ref{sxn:data-aware-l2-embed} 
and~\ref{sxn:data-aware-l1-embed}, we will discuss the data-aware subspace 
embedding methods for $\ell_2$ and $\ell_1$ norms, respectively. 
Before getting into the details of these methods, we first provide some 
background and definitions.

Let us denote by $\A \subset \R^m$ the subspace spanned by the columns of $A$.
A subspace embedding of $\A$ into $\R^{s}$ with $s > 0$ is a
structure-preserving mapping $\phi: \A \hookrightarrow \R^{s}$, where the
meaning of ``structure-preserving'' varies depending on the application.
Here, we are interested in low-distortion linear embeddings of the normed 
vector space $\A_p = (\A, \|\cdot\|_p)$, the subspace $\A$ paired with the 
$\ell_p$ norm $\|\cdot\|_p$.
(Again, although we are most interested in $\ell_1$ and $\ell_2$, some of 
the results hold more generally than for just $p=2$ and $p=1$, and so we 
formulate some of these results for general $p$.) 
We start with the following definition.

\begin{definition}[Low-distortion $\ell_p$ subspace embedding]
 \label{def:low_dist}
  Given a matrix $A \in \R^{m \times n}$ and $p \in [1, \infty]$, $\Phi \in \R^{s
    \times m}$ is an embedding of $\A_p$ if $s = \bigO(\poly(n))$,
  independent of $m$, and there exist $\sigma_\Phi > 0$ and $\kappa_\Phi >
  0$ such that
  \begin{equation*}
    \sigma_\Phi \cdot \|y\|_p \leq \|\Phi y\|_p 
    \leq \kappa_\Phi \sigma_\Phi \cdot \|y\|_p, \quad \forall y \in \A_p.
  \end{equation*}
  We call $\Phi$ a low-distortion subspace embedding of $\A_p$ if the distortion
  of the embedding $\kappa_\Phi = \bigO(\poly(n))$, independent of $m$.
\end{definition}

\noindent
We remind the reader that low-distortion subspace embeddings can be used in 
one of two related ways:
for $\ell_p$-norm preconditioning and/or for solving directly $\ell_p$ 
regression subproblems.
We will start by establishing some terminology for their use for 
preconditioning.

Given a low-distortion embedding matrix $\Phi$ of $\A_p$ with distortion 
$\kappa_\Phi$, let $R$ be the ``R'' matrix from the QR decomposition of 
$\Phi A$. 
Then, the matrix $A R^{-1}$ is well-conditioned in the $\ell_p$ norm. 
To see this, note that we~have
\begin{align*}
  \|A R^{-1} x\|_p &\leq \sigma_\Phi \kappa_\Phi \|\Phi A R^{-1} x\|_p \leq
   \sigma_\Phi \kappa_\Phi s^{\max\{0, 1/p-1/2\}} \cdot \|\Phi A R^{-1}\|_2 \cdot \|x\|_2 \\
  &= \sigma_\Phi \kappa_\Phi s^{\max\{0, 1/p-1/2\}} \cdot \|x\|_2, \quad \forall x \in \R^n,
\end{align*}
where the first inequality is due to low distortion and the second 
inequality is due to the equivalence of vector norms.
By similar arguments, we can show that
\begin{align*}
  \|A R^{-1} x\|_p &\geq \sigma_\Phi \cdot \|\Phi A R^{-1}\|_p \geq
  \sigma_\Phi s^{\min\{0,1/p-1/2\}} \cdot \|\Phi A R^{-1} x\|_2 \\
  &= \sigma_\Phi s^{\min\{0,1/p-1/2\}} \cdot \|x\|_2, \quad \forall x \in \R^n.
\end{align*}
Hence, by combining these results, we have 
$$\kappa_p(A R^{-1}) \leq \kappa_\Phi s^{|1/p-1/2|} = \bigO(\poly(n)),$$
\emph{i.e.}, the matrix $A R^{-1}$ is well-conditioned in the $\ell_p$ norm.
\emph{We call a conditioning method that is obtained via computing the QR 
factorization of a low-distortion embedding a QR-type method; and we call a
conditioning method that is obtained via an ellipsoid rounding of a 
low-distortion embedding an ER-type method.}

Furthermore, one can construct a well-conditioned basis by combining QR-like 
and ER-like methods.  To see this, let $R$ be the matrix obtained by applying
Corollary~\ref{coro:lp_cond_2d} to $\Phi A$.
We have
\begin{equation*}
  \|A R^{-1} x\|_p \leq \sigma_\Phi \kappa_\Phi \cdot \|\Phi A R^{-1} x\|_p 
  \leq 2 n \sigma_\Phi \kappa_\Phi \|x\|_2, \quad \forall x \in \R^n,
\end{equation*}
where the second inequality is due to the ellipsoidal rounding result, and 
\begin{equation*}
  \|A R^{-1} x\|_p \geq \sigma_\Phi \|\Phi A R^{-1} x\|_p \geq \sigma_\Phi \|x\|_2, \quad \forall x \in\R^n.
\end{equation*}
Hence $$\kappa_p(A R^{-1}) \leq 2 n \kappa_\Phi = \bigO(\poly(n))$$ and $A R^{-1}$
is well-conditioned.
\emph{Following our previous conventions, we call this combined type of 
conditioning method a QR+ER-type method.}

 \begin{table}[t]
  \centering
   \begin{tabular}{c|ccccc}
     name &  $\kappa$ & running time & \# passes & type & norm \\
    \hline
     ER~\cite{clarkson2005subgradient,dasgupta2009sampling} & $(n (n+1))^{1/2}$ & $\bigO(m n^5 \log m)$ & $\bigO(n^3 L)$ & ER & $\ell_1$ \\
    Fast ER~\cite{clarkson2013fast} & $2 n$ & $\bigO(m n^3 \log m)$ & $\bigO(n L)$ & ER & $\ell_1$ \\
    Single-pass ER~\cite{meng2013robust} & $2 n^2$ & $\bigO(m n^2 \log m)$ & $1$ & ER & $\ell_1$ \\
    CT~\cite{sohler2011subspace} & $\bigO(n^{3/2} \log^{3/2} n)$ & $\bigO(mn^2 \log n)$  & $1$ &  QR & $\ell_1$ \\
      FCT~\cite{clarkson2013fast}   & $\bigO(n^{9/2} \log^{9/2} n)$ & $\bigO(mn \log n)$  & $1$ & QR & $\ell_1$ \\ 
      SPCT~\cite{meng2013robust} &  $\bigO(n^{11/2} \log^{11/2} n)$  &  $\bigO(\textup{nnz}(A))$  & $1$ & QR & $\ell_1$ \\
      SPCT2~\cite{meng2013robust}   & $6n$ & $\bigO(\textup{nnz}(A) \cdot \log n)$ & $2$ & QR+ER & $\ell_1$ \\
      RET~\cite{WZ13} &  $\bigO(n^{5/2} \log^{5/2} n)$  &  $\bigO(\textup{nnz}(A))$  & $1$ & QR & $\ell_1$ \\
      Gaussian & $ \bigO(1) $ & $\bigO(mn^2)$ & 1 & QR  & $\ell_2$ \\
      SRHT~\cite{tropp2011improved,drineas2011faster,DMMW12_JMLR} & $ \bigO(1)$ & $\bigO(mn\log m)$ & 1 & QR & $\ell_2$ \\
      CW~\cite{CW13sparse_STOC,meng2013low,nelson2012osnap}  & $ \bigO(1)$ & $\bigO(\textup{nnz}(A))$ & 1 & QR & $\ell_2$ 
   \end{tabular}
   \caption{Summary of of $\ell_1$ and $\ell_2$ norm conditioning methods.
            QR and ER refer, respectively, to methods based on the QR 
            factorization and methods based on Ellipsoid Rounding, as 
            discussed in the text.
            }
    \label{tab:cond}
 \end{table}

In Table~\ref{tab:cond}, we summarize several different types of 
conditioning methods for $\ell_1$ and $\ell_2$ conditioning.
Comparing the QR-type approach and the ER-type approach to obtaining the 
preconditioner matrix $R$, we see there are trade-offs between running times 
and conditioning quality.
Performing the QR decomposition takes $\bigO(s n^2)$ time, which is faster 
than fast ellipsoidal rounding that takes $\bigO(s n^3 \log s)$ time.
However, the latter approach might provide a better conditioning quality 
when $2 n < s^{|1/p-1/2|}$.
We note that those trade-offs are not important in most theoretical 
formulations, as long as both take $\bigO(\poly(n))$ time and provide 
$\bigO(\poly(n))$ conditioning, independent of $m$, but they certainly do 
affect the performance in practice.

A special family of low-distortion subspace embedding that has very low 
distortion factor is called subspace-preserving embedding.

\begin{definition}[Subspace-preserving embedding]
Given a matrix $A \in \R^{m\times n}$, $p \in [1, \infty]$
  and $\epsilon \in (0, 1)$, $\Phi \in \R^{s
    \times m}$ is a
  subspace-preserving embedding of $\A_p$ if
  $s = \bigO(\poly(n))$, independent of $m$, and
  \begin{equation*}
    (1-\epsilon) \cdot \|y\|_p \leq \|\Phi y\|_p \leq (1+\epsilon) \cdot \|y\|_p, \quad \forall y \in \A_p.
  \end{equation*}
\end{definition}


\subsubsection{Data-oblivious low-distortion $\ell_2$ subspace embeddings}
\label{sxn:data-oblivious-l2-embed}

An $\ell_2$ subspace embedding is distinct from but closely related to the 
embedding provided by the Johnson-Lindenstrauss (J-L) lemma.

\begin{lemma}[Johnson-Lindenstrauss lemma~\cite{johnson1984extensions}]
  \label{lemma:jl}
  Given $\epsilon \in (0, 1)$, a point set $\X$ of $N$ points in $\R^m$,
  there is a linear map $\phi: \R^m \hookrightarrow \R^s$ with $s = C \log N /
  \epsilon^2$, where $C > 0$ is a global constant, such that
  \begin{equation*}
    (1-\epsilon) \|x-y\|^2 \leq \|\phi(x) - \phi(y)\|^2 
    \leq (1+\epsilon)\|x-y\|^2, \quad \forall x, y \in \X.
  \end{equation*}
  We say a mapping has J-L property if it satisfies the above condition with a
  constant probability.
\end{lemma}

\noindent
The original proof of the J-L lemma is done by constructing a projection 
from $\R^m$ to a randomly chosen $s$-dimensional subspace.
The projection can be represented by a random orthonormal matrix in
$\R^{s \times m}$.
Indyk and Motwani~\cite{indyk1998approximate} show that a matrix whose
entries are independent random variables drawn from the standard normal
distribution scaled by $s^{-1/2}$ also satisfies the J-L property.
This simplifies the construction of a J-L transform, and it has improved 
algorithmic properties.
Later, Achlioptas~\cite{achlioptas2001database} show that the random normal
variables can be replaced by random signs, and moreover, we can zero out
approximately $2/3$ of the entries with proper scaling, while still 
maintaining the J-L property.
The latter approach allows faster construction and projection with less 
storage, although still at the same order as the random normal projection.

The original J-L lemma applies to an arbitrary set of $N$ vectors in $\R^m$. 
By using an $\epsilon$-net argument and triangle inequality,
Sarl\'os~\cite{sarlos2006improved} shows that a J-L transform can also 
preserve the Euclidean geometry of an entire $n$-dimensional subspace of 
vectors in $\R^m$, with embedding dimension 
$\bigO(n \log (n/\epsilon) / \epsilon^2)$.
\begin{lemma}[Sarl\'os~\cite{sarlos2006improved}]
  \label{lemma:jl_subspace}
  Let $\A_2$ be an arbitrary $n$-dimensional subspace of $\R^m$ and $0 \leq
  \epsilon, \delta < 1$.
  If $\Phi$ is a J-L transform from $\R^m$ to $\bigO(n \log(n / \epsilon) /
  \epsilon^2 \cdot f(\delta))$ dimensions for some function $f$. Then
  \begin{equation*}
    \Pr(\forall x \in \A_2 : |\|x\|_2 - \|\Phi x\|_2 | \leq \epsilon \|x\|_2) \geq 1 - \delta.
  \end{equation*}
\end{lemma}
The result of Lemma~\ref{lemma:jl_subspace} applies to any J-L transform, 
\emph{i.e.}, to any transform (including those with better or worse 
asymptotic FLOPS behavior) that satisfies the J-L distortion property.

It is important to note, however, that for some J-L transforms, we are able 
to obtain more refined results.
In particular, these can be obtained  by bounding the spectral norm of 
$(\Phi U)^T(\Phi U) - I$, where $U$ is an orthonormal basis of $\A_2$.
If $\|(\Phi U)^T(\Phi U) - I\| \leq \epsilon$, for any $x \in
\A_2$, we have
\begin{equation*}
  |\|\Phi x\|_2^2 - \|x\|_2^2 |= |(U x)^T ( (\Phi U)^T (\Phi U) - I ) ( U x ) |
  \leq \epsilon \|U x\|_2^2 = \epsilon \|x\|_2^2,
\end{equation*}
and hence
\begin{equation*}
  |\|\Phi x\|_2 - \|x\|_2| \leq \frac{\epsilon \|x\|_2^2}{\|\Phi x\|_2 + \|x\|_2} \leq \epsilon \|x\|_2.
\end{equation*}


We show some results following this approach.
First consider the a random normal matrix, which has the following 
concentration result on its extreme singular values.

\begin{lemma}[Davidson and Szarek \cite{davidson2001local}] 
  \label{lemma:concentration}
Consider an $s \times n$ random matrix
  $G$ with $s > n$, whose entries are independent random variables following the
  standard normal distribution.
  Let the singular values be $\sigma_1 \geq \cdots \geq \sigma_n$.
  Then for any $t > 0$,
  \begin{equation}
    \label{eq:concentration}
    \max \left\{ \Pr(\sigma_1 \geq \sqrt{s} + \sqrt{n} + t), \Pr( \sigma_n 
      \leq \sqrt{s} - \sqrt{n} - t) \right\} < e^{- t^2 / 2}.
  \end{equation}
\end{lemma}

\noindent
Using this concentration result, we can easily present a better analysis of
random normal projection than in Lemma~\ref{lemma:jl_subspace}.

\begin{corollary}
  \label{lemma:gaussian_jl_subspace}
  Given an $n$-dimensional subspace $\A_2 \subset \R^m$ and $\epsilon, \delta \in
  (0, 1)$, let $G \in \R^{s \times m}$ be a random matrix whose entries are
  independently drawn from the standard normal distribution.
  There exist $s = \bigO((\sqrt{n} + \log(1/\delta))^2/\epsilon^2)$ such that,
  with probability at least $1-\delta$, we have
  \begin{equation*}
    (1-\epsilon) \|x\|_2 \leq \|s^{-1/2} G x\|_2 \leq (1+\epsilon) \|x\|_2, \quad \forall x \in \A_2.
  \end{equation*}
\end{corollary}

Dense J-L transforms, \emph{e.g.}, a random normal projection and its 
variants, use matrix-vector multiplication for the embedding.
Given a matrix $A \in \R^{m \times n}$, computing $\tilde{A} = \Phi A$ takes
$\bigO(\nnz(A) \cdot s)$ time when $\Phi$ is a dense matrix of size 
$s \times m$ and $\nnz(A)$ is the number of non-zero elements in $A$.
There is also a line of research work on ``fast'' J-L transforms that 
started with~\cite{ailono2006approximate,AC06-JRNL09}.
These use FFT-like algorithms for the embedding, and thus they lead to 
$\bigO(m \log m)$ time for each projection.
Hence, computing $\tilde{A} = \Phi A$ takes $\bigO(m n \log m)$ time when 
$\Phi$ is a fast J-L transform.
Before stating these results, we borrow the notion of FJLT 
from~\cite{ailono2006approximate,AC06-JRNL09} and use that to define a 
stronger and faster version of the simple J-L transform.
\begin{definition}[FJLT]
	Given an $n$-dimensional subspace $\A_2 \subset \R^m$, we say $\Phi \in \R^{r\times m}$ is an FJLT for $\A_2$ if
	 $\Phi$ satisfies the following two properties:
 \begin{compactitem}
	\item  $ \|(\Phi U)^T(\Phi U) - I_n\|_2 \leq \epsilon $, where $U$ is an orthonormal basis of $\A_2$.
	\item  Given any $x \in \R^n$,
	 $\Phi x$ can be computed in at most $\bigO(m \log m)$ time.
 \end{compactitem}
\end{definition}

Ailon and Chazelle construct the so-called fast Johnson-Lindenstrauss transform 
(FJLT)~\cite{AC06-JRNL09}, which is a product of three 
matrices $\Phi = P H D$, where $P \in \R^{s \times m}$ is a sparse J-L 
transform with approximately $\bigO(s \log^2 N)$ nonzeros, 
$H \in \R^{m \times m}$ is a normalized Walsh-Hadamard matrix, and 
$D \in \R^{m \times m}$ is a diagonal matrix with its diagonals drawn 
independently from $\{-1,1\}$ with probability $1/2$.
Because multiplying $H$ with a vector can be done in $\bigO(m \log m)$ time
using an FFT-like algorithm, it reduces the projection time from 
$\bigO(s m)$ to $\bigO(m \log m)$.
This FJLT construction is further simplified by Ailon and
Liberty~\cite{ailon2009fast, ailon2011almost}.

A subsequently-refined FJLT was analyzed by Tropp~\cite{tropp2011improved}, 
and it is named the subsampled randomized Hadamard transform (SRHT). 
As with other FJLT methods, the SRHT preserves the geometry of an entire 
$\ell_2$ subspace of vectors by using a matrix Chernoff inequality to bound 
$\|(\Phi U)^T (\Phi U) - I\|_2$.
We describe this particular FJLT in more detail.

\begin{definition}
  An SRHT is an $s \times m$ matrix of the form 
  \begin{equation*}
    \Phi = \sqrt{\frac{m}{s}} R H D,
  \end{equation*}
  where
  \begin{itemize}
  \item $D \in \R^{m \times m}$ is a diagonal matrix whose entries are
    independent random signs,
  \item $H \in \R^{m \times m}$ is a Walsh-Hadamard matrix scaled by $m^{-1/2}$,
  \item $R \in \R^{s \times m}$ restricts an $n$-dimensional vector to $s$
    coordinates, chosen uniformly at random.
  \end{itemize}
\end{definition}

\noindent
Below we present the main results for SRHT from~\cite{DMMW12_JMLR} since it 
has a better characterization of the subspace-preserving properties. 
We note that its proof is essentially a combination of the results 
in~\cite{tropp2011improved,drineas2011faster}.

\begin{lemma}[SRHT~\cite{tropp2011improved,drineas2011faster,DMMW12_JMLR}]
  \label{lemma:srht}
  Given an $n$-dimensional subspace $\A_2 \subset \R^m$ and $\epsilon, \delta \in
  (0, 1)$, let $\Phi \in \R^{s \times m}$ be a random SRHT
  with embedding dimension
   $s \geq \frac{14n\ln(40mn)}{\epsilon^2}\ln\left(\frac{30^2n\ln(40mn)}{\epsilon^2}\right)$. Then,
  with probability at least $0.9$, we have
  \begin{equation*}
    (1-\epsilon) \|x\|_2 \leq \|\Phi x\|_2 \leq (1+\epsilon) \|x\|_2, \quad \forall x \in \A_2.
  \end{equation*}
\end{lemma}

\noindent
Note that besides Walsh-Hardamard transform, other FFT-based transform, e.g., discrete Hartley transform (DHT), discrete cosine transform (DCT) which have more practical advantages can be also be used; see~\cite{avron2010blendenpik} for an details of other choices.
Another important point to keep in mind (in particular, for parallel and 
distributed applications) is that, although called ``fast,'' a fast 
transform might be slower than a dense transform: 
when $\nnz(A) = \bigO(m)$ (since machines are optimized for matrix-vector 
multiplies);
when $A$'s columns are distributively stored (since this slows down FFT-like 
algorithms, due to communication issues); or 
for other machine-related issues.

More recently, Clarkson and Woodruff~\cite{CW13sparse_STOC} developed an 
algorithm for the $\ell_2$ subspace embedding that runs in so-called 
\emph{input-sparsity} time, \emph{i.e.}, in $\bigO(\nnz(A))$ time, plus 
lower-order terms that depend polynomially on the low dimension of the input.
Their construction is exactly the CountSketch matrix in the data
stream literature~\cite{charikar2002finding}, which is an extremely simple and
sparse matrix.
It can be written as the product of two matrices $\Phi = S D \in \R^{s \times
  m}$, where $S \in \R^{s \times m}$ has each column chosen independently and
uniformly from the $s$ standard basis vectors of $\R^s$ and $D \in \R^{m \times
  m}$ is a diagonal matrix with diagonal entries chosen independently and
uniformly from $\pm 1$.
By decoupling $\A$ into two orthogonal subspaces, called ``heavy'' and 
``light'' based on the row norms of $U$, an orthonormal basis of $\A$, 
\emph{i.e.}, based on the statistical leverage scores of $A$, they proved 
that with an embedding dimension $\bigO(n^2/\epsilon^2)$, the above 
construction gives an $\ell_2$ subspace embedding matrix.
Improved bounds and simpler proofs (that have much more linear algebraic 
flavor) were subsequently provided by Mahoney and Meng \cite{meng2013low} 
and Nelson and Nguyen \cite{nelson2012osnap}.
In rest of this paper, we refer to this method as CW.
Below, we present the main results
from~\cite{CW13sparse_STOC,meng2013low,nelson2012osnap}.

\begin{lemma}[Input-sparsity time embedding for $\ell_2$ \cite{CW13sparse_STOC,meng2013low,nelson2012osnap}]
  \label{thm:sparse_l2}
  Given an $n$-dimensional subspace $\A_2 \subset \R^m$ and any $\delta \in (0,
  1)$, let $s = (n^2 + n) / (\epsilon^2 \delta)$.
  Then, with probability at least $1-\delta$,
  \begin{equation*}
    (1-\epsilon) \|x\|_2 \leq \| \Phi x \|_2 \leq (1+\epsilon) \|x\|_2, \quad \forall x \in \A_2,
  \end{equation*}
  where $\Phi \in \R^{s\times m}$ is the CountSketch matrix described above.  
\end{lemma}

\begin{remark}
It is easy to see that computing $\Phi A$, \emph{i.e.}, computing the 
subspace embedding, takes $\bigO(\nnz(A))$ time.
The $\bigO(\nnz(A))$ running time is indeed optimal, up to constant factors, 
for general inputs.
Consider the case when $A$ has an important row $a_i$ such that $A$ becomes
rank-deficient without it.
Thus, we have to observe $a_i$ in order to compute a low-distortion 
embedding.
However, without any prior knowledge, we have to scan at least a constant
portion of the input to guarantee that $a_i$ is observed with a constant
probability, which takes $\bigO(\nnz(A))$ time.
Also note that this optimality result applies to general $\ell_p$ norms.
\end{remark}

To summarize, in Table~\ref{tab:l2_embed}, we provide a summary of the basic 
properties of several data-oblivious $\ell_2$ subspace embeddings discussed 
here (as well as of several data-aware $\ell_2$ subspace-preserving 
embeddings that will be discussed in Section~\ref{sxn:data-aware-l2-embed}).

\begin{remark}
With these low-distortion $\ell_2$ subspace embeddings, one can use the QR-type method to compute an $\ell_2$ preconditioner. That is,
one can compute the QR factorization of the low-distortion subspace embeddings in Table~\ref{tab:l2_embed} and use $R^{-1}$ as the preconditioner; see Table~\ref{tab:cond} for more details.
We note that the tradeoffs in running time are implicit although they have the same conditioning quality.
This is because the running time for computing the QR factorization depends on the embedding dimension which is varied from method to method.
However, normally this is absorbed by the time for forming $\Phi A$ (theoretically, and it is in practice not the dominant~effect).
\end{remark}

 \begin{table}[t]
  \centering
   \begin{tabular}{c|ccc}
     name  &  running time & $s$ & $\kappa_\Phi$ \\
    \hline
    Gaussian~(REF) & $\bigO(mns)$ & $\bigO(n/\epsilon^2)$ & $1+\epsilon$  \\
    SRHT~\cite{tropp2011improved,drineas2011faster,DMMW12_JMLR} & $\bigO(mn\log m)$ & $\bigO(n \log(mn) \log(n/\epsilon^2) /\epsilon^2)$ & $1+\epsilon$\\
    CW~\cite{CW13sparse_STOC,meng2013low,nelson2012osnap} & $\bigO(\texttt{nnz}(A))$ & $(n^2+n)/\epsilon^2$ & $1+\epsilon$  \\
    Exact lev. scores sampling~\cite{drineas2006sampling} & $\bigO(mn^2)$ & $\bigO(n \log n/\epsilon^2)$ & $1+\epsilon$ \\
    Appr. lev. scores sampling (SRHT)~\cite{DMMW12_JMLR} & $\bigO(mn\log m)$ & $\bigO(n \log n/\epsilon^2)$ & $1+\epsilon$ \\
    Appr. lev. scores sampling (CW)~\cite{CW13sparse_STOC, DMMW12_JMLR} & $\bigO(\texttt{nnz}(A)) \log m$ & $\bigO(n \log n/\epsilon^2)$ & $1+\epsilon$ 
   \end{tabular}
  \caption{Summary of data-oblivious and data-aware $\ell_2$ embeddings. 
  Above, $s$ denotes the embedding dimension. 
  By running time, we mean the time needed to compute $\Phi A$.
  For each method, we set the failure rate to be a constant.
  Moreover, ``Exact lev. scores sampling'' means sampling algorithm based on 
  using the exact leverage scores (as importance sampling probabilities); 
  and ``Appr. lev. scores sampling (SRHT)'' 
  and ``Appr. lev. scores sampling (CW)'' are sampling algorithms based on 
  approximate leverage scores estimated by using SRHT and CW (using the 
  algorithm of~\cite{DMMW12_JMLR}) as the underlying random projections, 
  respectively. 
  Note that within the algorithm (of~\cite{DMMW12_JMLR}) for approximating 
  the leverage scores, the target approximation accuracy is set to be a 
  constant. }
    \label{tab:l2_embed}
 \end{table}

\subsubsection{Data-oblivious low-distortion $\ell_1$ subspace embeddings}
\label{sxn:data-oblivious-l1-embed}

General $\ell_p$ subspace embedding and even $\ell_1$ subspace embedding 
is quite different from $\ell_2$ subspace embedding.
Here, we briefly introduce some existing results on $\ell_1$ subspace
embedding; for more general $\ell_p$ subspace embedding, 
Meng and Mahoney\cite{meng2013low} and Clarkson and Woodruff \cite{CW13sparse_STOC}.

 \begin{table}[t]
  \centering
   \begin{tabular}{c|ccc}
     name  &  running time & $s$ & $\kappa_\Phi$ \\
    \hline
    CT~\cite{sohler2011subspace} & $\bigO(mn^2\log n)$ & $\bigO(n\log n)$ & $\bigO(n \log n)$  \\
    FCT~\cite{clarkson2013fast} & $\bigO(mn\log n)$ & $\bigO(n \log n)$ & $\bigO(n^4 \log^4 n)$  \\
    SPCT~\cite{meng2013low} & $\nnz(A)$ & $\bigO(n^5 \log^5 n)$ & $\bigO(n^3 \log^3 n)$ \\
    Reciprocal Exponential~\cite{WZ13}  & $\nnz(A)$ & $\bigO(n \log n)$ & $\bigO(n^2 \log^2 n)$ \\
    Sampling (FCT)~\cite{clarkson2013fast,YMM14_SISC} & $\bigO(mn\log n)$ & $\bigO(n^{13/2} \log^{9/2} n \log(1/\epsilon) / \epsilon^2)$ & $1+\epsilon$ \\
    Sampling (SPCT)~\cite{meng2013low, clarkson2013fast,YMM14_SISC} & $\bigO(\nnz(A) \cdot \log n)$ & $\bigO(n^{15/2} \log^{11/2} n \log(1/\epsilon) / \epsilon^2)$ & $1+\epsilon$ \\
    Sampling (RET)~\cite{WZ13,YMM14_SISC} & $\bigO(\nnz(A) \cdot \log n)$ & $\bigO(n^{9/2} \log^{5/2} n \log(1/\epsilon) / \epsilon^2)$ & $1+\epsilon$
   \end{tabular}
   \caption{Summary of data-oblivious and data-aware $\ell_1$ embeddings. 
   Above, $s$ denotes the embedding dimension.
   By running time, we mean the time needed to compute $\Pi A$.
   For each method, we set the failure rate to be a constant.
   Moreover, ``Sampling (FCT)'', ``Sampling (SPCT)'' and ``Sampling (RET)'' denote the $\ell_1$ 
   sampling algorithms obtained by using FCT, SPCT and RET as the underlying 
   preconditioning methods, respectively.
   }
    \label{tab:l1_embed}
 \end{table}

For $\ell_1$, the first question to ask is whether there exists an J-L transform
equivalent.
This question was answered in the negative by Charikar and Sahai~\cite{charikar2002dimension}.

\begin{lemma}[Charikar and Sahai~\cite{charikar2002dimension}]
  There exists a set of $\bigO(m)$ points in $\ell_1^m$ such that any linear
  embedding into $\ell_1^s$ has distortion at least $\sqrt{m/s}$.
  The trade-off between dimension and distortion for linear embeddings is tight
  up to a logarithmic factor.
  There exists a linear embedding of any set of $N$ points in $\ell_1^m$ to
  $\ell_1^{s'}$ where $s' = \bigO(s \log N)$ and the distortion is
  $\bigO(\sqrt{m/s})$.
\end{lemma}

\noindent
This result shows that linear embeddings are particularly ``bad'' in 
$\ell_1$, compared to the particularly ``good'' results provided by the J-L 
lemma for $\ell_2$.
To obtain a constant distortion, we need $s \geq C m$ for some constant $C$.
So the embedding dimension cannot be independent of $m$.
However, the negative result is obtained by considering arbitrary point 
sets.
In many applications, we are dealing with structured point sets, 
\emph{e.g.}, vectors from a low-dimensional subspace.
In this case, Sohler and Woodruff~\cite{sohler2011subspace} give the first 
linear oblivious embedding of a $n$-dimensional subspace of $\ell_1^m$ into 
$\ell_1^{\bigO(n \log n)}$ with distortion $\bigO(n \log n)$, where both the 
embedding dimension and the distortion are independent of $m$.
In particular, they prove the following quality bounds.

\begin{lemma}[Cauchy transform (CT), Sohler and Woodruff~\cite{sohler2011subspace}]
  \label{lemma:cauchy_subspace}
  Let $\A_1$ be an arbitrary $n$-dimensional linear subspace of $\R^m$.
  Then there is an $s_0 = s_0(n) = \bigO(n \log n)$ and a sufficiently large
  constant $C_0 > 0$, such that for any $s$ with $s_0 \leq s \leq n^{\bigO(1)}$,
  and any constant $C \geq C_0$, if $\Phi \in \R^{s \times m}$ is a random
  matrix whose entries are choose independently from the standard Cauchy
  distribution and are scaled by $C/s$, then with probability at least $0.99$,
  \begin{equation*}
    \|x\|_1 \leq \|\Phi x\|_1 \leq \bigO(n \log n) \cdot \|x\|_1, \quad \forall x \in \A_1.
  \end{equation*}
\end{lemma}

\noindent
The proof is by constructing tail inequalities for the sum of half Cauchy 
random variables~\cite{sohler2011subspace}.
The construction here is quite similar to the construction of the
dense Gaussian embedding for $\ell_2$ in 
Lemma~\ref{lemma:gaussian_jl_subspace}, with several important differences.
The most important differences are the~following:
\begin{itemize}
\item Cauchy random variables replace standard normal random variables;
\item a larger embedding dimension does not always lead to better 
distortion quality; and
\item the failure rate becomes harder to control.
\end{itemize}

As CT is the $\ell_1$ counterpart of the dense Gaussian transform, the Fast
Cauchy Transform (FCT) proposed by Clarkson \emph{et al.}~\cite{clarkson2013fast} 
is the $\ell_1$ counterpart of FJLT.
There are several related constructions.
For example, this FCT construction first preprocesses by a deterministic low-coherence
matrix, then rescales by Cauchy random variables, and finally samples linear
combinations of the rows. 
Then, they construct $\Phi$ as
\begin{equation*}
\Phi = 4B C H,  
\end{equation*}
where:
\begin{itemize}
\item $B\in\R^{s\times 2 m}$ has each column chosen independently and uniformly
  from the $s$ standard basis vectors for $\R^{s}$; for $\alpha$ sufficiently
  large, the parameter is set as $s=\alpha n \log(n/\delta)$, where $\delta
  \in (0, 1)$ controls the probability that the algorithm fails;
\item $C \in \R^{2m \times 2m}$ is a diagonal matrix with diagonal entries
  chosen independently from a Cauchy distribution; and
\item $H \in \R^{2m \times m}$ is a block-diagonal matrix comprised of $m/t$
  blocks along the diagonal.
  Each block is the $2t\times t$ matrix $G_s = \begin{psmallmatrix} H_{t}\\
    I_t\end{psmallmatrix}$, where $I_t$ is the $t\times t$ identity matrix, and
  $H_t$ is the normalized Hadamard matrix.
  (For simplicity, assume $t$ is a power of two and $m/t$ is an~integer.)
  \begin{equation*}
    H = 
  \begin{pmatrix}
    G_s & &\\
    & G_s &&\\
    && \ddots & \\
    &&& G_s \\
  \end{pmatrix}.
\end{equation*}
\end{itemize}

Informally, the effect of $H$ in the above FCT construction is to spread the   
weight of a vector, so that $H y$ has many entries that are not too small.
This means that the vector $CH y$ comprises Cauchy random variables with 
scale factors that are not too small; and finally these variables are 
summed up by $B$, yielding a vector $BCH y$, whose $\ell_1$ norm won't be 
too small relative to $\|y\|_1$.
They prove the following quality bounds.
\begin{lemma}[Fast Cauchy Transform (FCT), Clarkson et
  al.~\cite{clarkson2013fast}]
\label{lemma:fct}
There is a distribution (given by the above construction) over matrices
$\Phi\in\R^{s\times m}$, with $s=\bigO(n \log n + n \log(1/\delta))$, such that
for an arbitrary (but fixed) $A \in \R^{m \times n}$, and for all $x\in\R^{n}$,
the inequalities
\begin{equation*}
\norm{Ax}_1\le\norm{\Phi Ax}_1\le \kappa\norm{Ax}_1
\end{equation*}
hold with probability $1-\delta$, where 
\begin{equation*}
  \kappa=\bigO\left(\frac{n\sqrt{t}}{\delta}\log (s n)\right).
\end{equation*}
Further, for any $y\in\R^m$, the product $\Phi y$ can be 
computed in $\bigO(m\log s)$ time.
\end{lemma}
To make the algorithm work with high probability, one has to set $t$ to be at the order of $s^6$ and 
$s=\bigO(n \log n)$. It follows that $\kappa=\bigO(n^4\log^4 n)$ in the above 
theorem.
That is, while faster in terms of FLOPS than the CT, the FCT leads to worse 
embedding/preconditioning quality.
Importantly, this result is different from how FJLT compares to dense 
Gaussian transform:
FJLT is faster than the dense Gaussian transform, while both provide the 
same order of distortion; but 
FCT becomes faster than the dense Cauchy transform, at the cost of somewhat 
worse distortion quality.

Similar to
\cite{CW13sparse_STOC,meng2013low,nelson2012osnap} for computing an $\ell_2$ subspace embedding, Meng and Mahoney~\cite{meng2013low} developed an algorithm for computing an $\ell_1$ subspace embedding matrix in input-sparsity time, \emph{i.e.}, in $\bigO(\nnz(A))$ time.
They used a CountSketch-like matrix which can be written as the product of two matrices $\Phi = S C \in \R^{s \times
  m}$, where $S \in \R^{s \times m}$ has each column chosen independently and
uniformly from the $s$ standard basis vectors of $\R^s$ and $C \in \R^{m \times
  m}$ is a diagonal matrix with diagonal entries chosen independently from the standard Cauchy distribution.
We summarize the main theoretical results in the following lemma.
\begin{lemma}[Sparse Cauchy Transform (SPCT), Meng and Mahoney~\cite{meng2013low}]
  Given an $n$-dimensional subspace $\A_1 \subset \R^m$ and $\epsilon \in (0, 1)$,
  there is $s = \bigO(n^5 \log^5 n)$ such that with a constant
  probability,
  \begin{equation*}
    1/\bigO(n^2\log^2 n) \|x\|_1 \leq \|\Phi x\|_1 \leq \bigO(n\log n) \|x\|_1, \quad \forall x \in \A_1,
  \end{equation*}
  where $\Phi$ is the sparse Cauchy transform described above.
\end{lemma}

More recently, Woodruff and Zhang~\cite{WZ13} proposed another algorithm that computes an $\ell_1$ subspace embedding matrix in input-sparsity time. Its construction is similar to that of sparse Cauchy transform. That is, $\Phi = SD$ where $D$ is a diagonal matrix with diagonal entries $1/u_1, 1/u_2, \ldots, 1/u_n$ where $u_i$ are exponential variables. Comparing to sparse Cauchy transform, the embedding dimension and embedding quality have been improved. We summarize the main results in the following lemma.
\begin{lemma}[Woodruff and Zhang~\cite{WZ13}]
  Given an $n$-dimensional subspace $\A_1 \subset \R^m$ and $\epsilon \in (0, 1)$,
  there is $s = \bigO(n \log n)$ such that with a constant
  probability,
  \begin{equation*}
    1/\bigO(n\log n) \|x\|_1 \leq \|\Phi x\|_1 \leq \bigO(n\log n) \|x\|_1, \quad \forall x \in \A_1,
  \end{equation*}
  where $\Phi$ is the sparse transform using reciprocal exponential variables described above.
\end{lemma}

To summarize, in Table~\ref{tab:l1_embed}, we provide a summary of the basic 
properties of several data-oblivious $\ell_1$ subspace embeddings discussed 
here (as well as of several data-aware $\ell_1$ subspace-preserving 
embeddings that will be discussed in Section~\ref{sxn:data-aware-l1-embed}).

\subsubsection{Data-aware low-distortion $\ell_2$ subspace embeddings}
\label{sxn:data-aware-l2-embed}

All of the linear subspace embedding algorithms mentioned in previous 
subsections are oblivious, \emph{i.e.}, independent of the input subspace.
That has obvious algorithmic advantages, \emph{e.g.}, one can construct the 
embedding matrix without even looking at the data.
Since using an oblivious embedding is not a hard requirement for the 
downstream task of solving $\ell_p$ regression problems (and since one can 
use random projection embeddings to construct importance sampling 
probabilities~\cite{DMMW12_JMLR} in essentially ``random projection 
time,'' up to small constant factors), a natural question is whether 
non-oblivious or data-aware embeddings could give better conditioning 
performance.
In general, the answer is yes.

As mentioned in Section~\ref{sxn:rla_ram-lev_cond},
Drineas \emph{et al.}~\cite{drineas2006sampling} developed a sampling 
algorithm for solving $\ell_2$ regression by constructing a 
$(1\pm\epsilon)$-distortion $\ell_2$ subspace-preserving sampling matrix.
The underlying sampling distribution is defined based on the statistical 
leverage scores of the design matrix which can be viewed as the 
``influence'' of that row on the LS fit.
That is, the sampling distribution is a distribution $\{p_i\}_{i=1}^m$ 
satisfying
\begin{equation}
\label{eq:samp_distr}
  p_i \geq \beta \cdot \frac{\ell_i}{\sum_j \ell_j}, ~~~ i= 1,\ldots,m.
\end{equation}
Above $\{\ell_i\}_{i=1}^m$ are the leverage scores of $A$ and $\beta \in (0,1]$. When $\beta =1$ and $\beta<1$, \eqref{eq:samp_distr} implies we define $\{p_i\}_{i=1}^m$ according to the exact and estimated leverage scores, respectively. 
 
More importantly, theoretical results indicate that, given a target desired 
accuracy, the required sampling complexity is independent of the higher 
dimension of the matrix.
Similar construction of the sampling matrix appeared in several subsequent 
works, \emph{e.g.}, \cite{drineas2006sampling,drineas2011faster,DMMW12_JMLR}, 
with improved analysis of the sampling complexity.
For completeness, we include the the main theoretical result regarding the 
subspace-preserving quality below, stated here for $\ell_2$.
\begin{theorem}[$\ell_2$ subspace-preserving sampling~\cite{drineas2006sampling,drineas2011faster,DMMW12_JMLR}]
\label{thm:l2_samp}
Given an $n$-dimensional subspace $\A_2 \subset \R^m$ represented by a
    matrix $A \in \R^{m \times n}$ and $\epsilon \in (0,
    1)$, choose
    $s = \bigO(n\log n\log(1/\delta)/\beta \epsilon^2)$,
    and construct a sampling matrix $S \in \R^{m \times m}$ with diagonals
    \begin{equation*}
      s_{ii} =
      \begin{cases}
        1/\sqrt{q_i} & \text{with probability } q_i, \\
        0 & \text{otherwise},
      \end{cases} \quad i = 1,\ldots,m,
    \end{equation*}
    where
    \begin{equation*}
      q_i \geq \min \left\{ 1, s \cdot p_i \right\}, \quad i=1,\ldots,m,
    \end{equation*}
    and $\{p_i\}_{i=1}^m$ satisfies \eqref{eq:samp_distr}.
    Then, with probability at least $0.7$,
    \begin{equation*}
      (1-\epsilon) \|y\|_2 \leq \|S y\|_2 \leq (1+\epsilon) \|y\|_2, \quad \forall y \in \A_2.
    \end{equation*}
\end{theorem}

\noindent
An obvious (but surmountable) challenge to applying this result is that 
computing the leverage scores \emph{exactly} involves forming an 
orthonormal basis for $A$ first.
Normally, this step will take $\bigO(mn^2)$ time which becomes undesirable 
when for large-scale applications.

On the other hand, by using the algorithm of~\cite{DMMW12_JMLR}, computing 
the leverage scores \emph{approximately} can be done in essentially the 
time it takes to perform a random projection:
in particular, Drineas et al.~\cite{DMMW12_JMLR} suggested that one can estimate the 
leverage scores by replacing $A$ with a ``similar'' matrix in the 
computation of the pseudo-inverse (which is the main computational 
bottleneck in the exact computation of the leverage scores).
To be more specific, by noticing that the leverage scores can be expressed 
as the row norms of $AA^\dagger$, we can use $\ell_2$ subspace embeddings to 
estimate them.
The high-level idea is,
$$ \|e_i A A^\dagger\|_2 \approx \|e_i A(\Pi_1 A)^\dagger\|_2 \approx \|e_i A(\Pi_1 A)^\dagger \Pi_2\|_2, $$
where $e_i$ is a vector with zeros but $1$ in the $i$-th coordinate, $\Pi_1 \in \R^{r_1 \times m}$ is a FJLT and 
$\Pi_2 \in \R^{n\times r_2}$ is a JLT which preserve the $\ell_2$ norms of 
certain set of points.
If the estimation of the leverage scores $\tilde \ell_i$ satisfies 
$$(1-\gamma)\ell_i \leq \tilde \ell_i \leq (1+\gamma)\ell_i, ~~~i=1,\ldots,m,$$ 
then it is not hard to show that a sampling distribution 
$\{p_i\}_{i=1}^m$ defined according to 
$p_i = \frac{\tilde \ell_i}{\sum_j \tilde \ell_j}$ satisfies 
\eqref{eq:samp_distr} with $\beta = \frac{1-\gamma}{1+\gamma}$.
When $\gamma$ is constant, say $0.5$, from Theorem~\ref{thm:l2_samp}, the 
required sampling complexity will only need to be increased by a constant 
factor $1/\beta = 3$.
This is less expensive, compared to the gain in the computation cost.

Suppose, now, we use SRHT (Lemma~\ref{lemma:srht}) or CW (Lemma~\ref{thm:sparse_l2}) method as the underlying FJLT, \emph{i.e.}, 
$\Pi_1$, in the approximation of the leverage scores.
Then, combining the theory suggested in~\cite{DMMW12_JMLR} and 
Theorem~\ref{thm:l2_samp}, we have the following lemma.
\begin{lemma}[Fast $\ell_2$ subspace-preserving sampling (SRHT)~\cite{DMMW12_JMLR}]
  Given an $n$-dimensional subspace $\A_2 \subset \R^m$ represented by a
    matrix $A \in \R^{m \times n}$ and $\epsilon \in (0,
    1)$,
  it takes $\bigO(mn \log m)$ time to compute a sampling matrix $S
  \in \R^{s' \times m}$ (with only one nonzero element per row) with $s' =
  \bigO(n \log n / \epsilon^2)$
  such that with constant probability
  \begin{equation*}
    (1-\epsilon) \|y\|_2 \leq \|S y\|_2 \leq (1+\epsilon) \|y\|_2, \quad \forall y \in \A_2.
  \end{equation*}
\end{lemma}

\begin{lemma}[Fast $\ell_2$ subspace-preserving sampling (CW)\cite{DMMW12_JMLR, CW13sparse_STOC}]
  Given an $n$-dimensional subspace $\A_2 \subset \R^m$ represented by a
    matrix $A \in \R^{m \times n}$ and $\epsilon \in (0,
    1)$,
  it takes $\bigO(\nnz(A) \cdot \log m)$ time to compute a sampling matrix $S
  \in \R^{s' \times m}$ (with only one nonzero element per row) with $s' =
  \bigO(n \log n / \epsilon^2)$
  such that with constant probability
  \begin{equation*}
    (1-\epsilon) \|y\|_2 \leq \|S y\|_2 \leq (1+\epsilon) \|y\|_2, \quad \forall y \in \A_2.
  \end{equation*}
\end{lemma}

\begin{remark}
Although using CW runs asymptotically faster than using SRHT, due to the 
poorer embedding quality of CW, in order to achieve the same embedding 
quality, and relatedly the same quality results in applications to 
$\ell_2$ regression, it may need a higher embedding dimension, \emph{i.e.}, 
$r_1$.
This results in a substantially longer QR factorization time for CW-based 
methods.
\end{remark}

Finally, recall that a summary of both data-oblivious and data-aware subspace 
embedding for $\ell_2$ norm can be found in Table~\ref{tab:l2_embed}.

\subsubsection{Data-aware low-distortion $\ell_1$ subspace embeddings}
\label{sxn:data-aware-l1-embed}

In the same way as we can use data-aware embeddings for $\ell_2$ regression, 
we can also use data-aware embeddings for $\ell_1$ regression.
Indeed, the idea of using data-aware sampling to obtain 
$(1\pm\epsilon)$-distortion subspace embeddings for $\ell_1$ regression was 
first used in \cite{clarkson2005subgradient}, where it was shown that an 
$\ell_1$ subspace embedding can be done by weighted sampling after 
preprocessing the matrix, including preconditioning, using ellipsoidal 
rounding.
Sampling probabilities depend on the $\ell_1$ norms of the rows of the 
preconditioned matrix.
Moreover, the resulting sample has each coordinate weighted by the 
reciprocal of
its sampling probability.
Different from oblivious $\ell_1$ subspace embeddings, the sampling approach 
can achieve a much better distortion.
\begin{lemma}[Clarkson~\cite{clarkson2005subgradient}]
  Given an $n$-dimensional subspace $\A_1 \subset \R^m$ represented by a
    matrix $A \in \R^{m \times n}$ and $\epsilon, \delta
  \in (0, 1)$, 
  it takes $\bigO(m n^5 \log m)$ time to compute a sampling matrix $S
  \in \R^{s' \times m}$ (with only one nonzero element per row) with $s' =
  \bigO(n^{3.5} \log(n / (\delta \epsilon))/\epsilon^2)$
  such that, with probability at least $1 - \delta$,
  \begin{equation*}
    (1-\epsilon) \|y\|_1 \leq \|S y\|_1 \leq (1+\epsilon) \|y\|_1, \quad \forall y \in \A_1.
  \end{equation*}
\end{lemma}
Therefore, to estimate the $\ell_1$ norms of any vector from a $n$-dimensional
subspace of $\R^m$, we only need to compute the weighted sum of the absolute
values of a few coordinates of this vector.

Recall that the $\ell_2$ leverage scores used in the $\ell_2$ sampling 
algorithm described in Theorem~\ref{thm:l2_samp} are the squared row norms 
of a orthonormal basis of $\A_2$ which can be a viewed as a ``nice'' basis 
for the subspace of interest. 
Dasgupta \emph{et al.}~\cite{dasgupta2009sampling} generalized this method 
to the general $\ell_p$ case; in particular, they proposed to sample rows 
according to the $\ell_p$ row norms of $AR^{-1}$, where $AR^{-1}$ is a 
well-conditioned (in the $\ell_p$ sense of well-conditioning) basis for 
$\A_p$. 
Different from $\ell_1$ sampling algorithm~\cite{clarkson2005subgradient} 
described above, computing such matrix $R$ is usually sufficient, meaning 
it is not needed to preprocess $A$ and form the basis $AR^{-1}$ explicitly.

\begin{theorem}[$\ell_p$ subspace-preserving sampling, Dasgupta \emph{et
  al.}~\cite{dasgupta2009sampling}]
    \label{thm:subspace_preserve}
    Given an $n$-dimensional subspace $\A_p \subset \R^m$ represented by a
    matrix $A \in \R^{m \times n}$ and a matrix $R \in \R^{n \times n}$ such that $A R^{-1}$ is well-conditioned, $p \in [1, \infty)$, $\epsilon \in (0,
    1/7)$, and $\delta \in (0, 1)$, choose
    \begin{equation*}
      s \geq 16 (2^p + 2) \bar{\kappa}_p^p(AR^{-1}) (n \log(12/\epsilon) + \log(2/\delta)) / (p^2 \epsilon^2),
    \end{equation*}
    and construct a sampling matrix $S \in \R^{m \times m}$ with diagonals
    \begin{equation*}
      s_{ii} =
      \begin{cases}
        1/p_i^{1/p} & \text{with probability } p_i, \\
        0 & \text{otherwise},
      \end{cases} \quad i = 1,\ldots,m,
    \end{equation*}
    where
    \begin{equation*}
      p_i \geq \min \left\{ 1, s \cdot \|a_iR^{-1}\|_p^p/|AR^{-1}|_p^p \right\}, \quad i=1,\ldots,m.
    \end{equation*}
    Then, with probability at least $1 - \delta$,
    \begin{equation*}
      (1-\epsilon) \|y\|_p \leq \|S y\|_p \leq (1+\epsilon) \|y\|_p, \quad \forall y \in \A_p.
    \end{equation*}
\end{theorem}

In fact, Theorem~\ref{thm:subspace_preserve} holds for any choice of $R$.
When $R = I$, it implies sampling according to the $\ell_p$ row norms of $A$ and the sampling complexity replies on $\bar \kappa_p^p(A)$. 
However, it is worth mentioning that a large condition number for $A$ will 
leads to a large sampling size, which in turn affects the running time of the 
subsequent operations. 
Therefore, preconditioning is typically necessary.
That is, one must find a matrix $R \in \R^{n \times n}$ such that $\bar{\kappa}_p(AR^{-1}) = \bigO(\poly(n))$, which could be done by the preconditioning algorithms introduced in the previous sections.

Given $R$ such that $AR^{-1}$ is well-conditioned, computing the row norms of $AR^{-1}$ takes $\bigO(\nnz(A) \cdot n)$ time.
Clarkson \emph{et al.}~\cite{clarkson2013fast} improve this running time by 
estimating the row norms of $A R^{-1}$ instead of computing them exactly.
The central idea is to post-multiply a random projection matrix $\Pi_2 \in \R^{n \times r}$ with $r = \bigO(\log m)$ which takes only $\bigO(\nnz(A) \cdot \log m)$ time.

If one uses FCT or SPCT in Table~\ref{tab:cond} to compute a matrix $R$ such 
that $AR^{-1}$ is well-conditioned and then uses the above idea to 
estimate quickly the $\ell_1$ row norms of $AR^{-1}$ to define the sampling 
distribution, then by combining with Theorem~\ref{thm:subspace_preserve}, we 
have the following two results.

\begin{lemma}[Fast $\ell_1$ subspace-preserving sampling (FCT)~\cite{clarkson2013fast,YMM14_SISC}]
  \label{lemma:fast_sampling_1}
  Given an $n$-dimensional subspace $\A_1 \subset \R^m$ represented by a matrix $A \in \R^{m \times n}$ and $\epsilon \in (0,
    1)$,
  it takes $\bigO(mn\log m)$ time to compute a sampling matrix $S
  \in \R^{s' \times m}$ (with only one nonzero element per row) with $s' =
  \bigO( n^{\frac{13}{2}} \log^\frac{9}{2} n \log(1/\epsilon) / \epsilon^2)$
  such that with a constant probability,
  \begin{equation*}
    (1-\epsilon) \|x\|_1 \leq \|Sx\|_1\leq (1+\epsilon) \|x\|_1, \quad \forall x \in \A_1.
  \end{equation*}
\end{lemma}

\begin{lemma}[Fast $\ell_1$ subspace-preserving sampling (SPCT)~\cite{meng2013low, clarkson2013fast,YMM14_SISC}]
  \label{lemma:fast_sampling_2}
  Given an $n$-dimensional subspace $\A_1 \subset \R^m$ represented by a matrix $A \in \R^{m \times n}$ and $\epsilon \in (0,
    1)$,
  it takes $\bigO(\nnz(A) \cdot \log m)$ time to compute a sampling matrix $S
  \in \R^{s' \times m}$ (with only one nonzero element per row) with $s' =
  \bigO( n^{\frac{15}{2}} \log^\frac{11}{2} n \log(1/\epsilon) / \epsilon^2)$
  such that with a constant probability,
  \begin{equation*}
    (1-\epsilon) \|x\|_1 \leq \|Sx\|_1\leq (1+\epsilon) \|x\|_1, \quad \forall x \in \A_1.
  \end{equation*}
\end{lemma}

\begin{remark}
Fast sampling algorithm also exists for $\ell_p$ regression.
That is, after computing a matrix $R$ such that $AR^{-1}$ is well-conditioned, one can use a similar idea to approximate the $\ell_2$ row norms of $AR^{-1}$, \emph{e.g.}, post-multiplying a random matrix with independent Gaussian variables (JLT), which lead to estimation of the $\ell_p$ row norms of $AR^{-1}$ up to small factors; see~\cite{clarkson2013fast} for more details.
\end{remark}

\begin{remark}
We note that the speed-up comes at the cost of increased sampling 
complexity, which does not substantially affect most theoretical 
formulations, since the sampling complexity is still
$\bigO(\poly(n) \log(1/\epsilon)/\epsilon^2)$.
In practice, however, it might be worth computing $U = A R^{-1}$ and its row 
norms explicitly to obtain a smaller sample size.
One should be aware of this trade-off when implementing a 
subspace-preserving sampling algorithm.
\end{remark}

Finally, recall that a summary of both data-oblivious and data-aware subspace 
embeddings for $\ell_1$ norm can be found in Table~\ref{tab:l1_embed}.

\subsection{Application of rounding/embedding methods to $\ell_1$ and $\ell_2$ regression}
\label{sxn:app_to_lp}

In this subsection, we will describe how the ellipsoidal rounding and 
subspace embedding methods described in the previous subsections can be 
applied to solve $\ell_2$ and $\ell_1$ regression problems.
In particular, by combining the tools we have introduced in the previous two 
subsections, \emph{e.g.}, solving subproblems and constructing 
preconditioners with ellipsoid rounding and subspace-embedding methods, we 
are able to describe several approaches to compute very fine 
$(1+\epsilon)$ relative-error solutions to $\ell_p$ regression problems.

Depending on the downstream task of interest, \emph{e.g.}, how the solution
to the regression problem will be used, one might be interested in obtaining
low-precision solutions, \emph{e.g.}, $\epsilon = 10^{-1}$, medium-precision 
solutions, \emph{e.g.}, $\epsilon = 10^{-4}$, or high-precision solutions, 
\emph{e.g.}, $\epsilon = 10^{-10}$.
As described in Section~\ref{sxn:rla_ram}, the design principles for these 
cases are somewhat different.
In particular, the use of $\ell_2$ and $\ell_1$ well-conditioned bases is 
somewhat different, depending on whether or not one is interested in 
low precision.
Here, we elaborate on how we can use the methods described previously 
construct low-precision solvers and high-precision solvers for solving 
$\ell_p$ regression problems.
As a reference, see Table~\ref{tab:l2_solve} and Table~\ref{tab:l1_solve} 
for a summary of several representative RandNLA algorithms for solving 
$\ell_2$ and $\ell_1$ regression problems, respectively.
(Most of these have been previously introduced for smaller-scale 
computations in RAM; and in Section~\ref{sxn:implementations} we will 
describe several variants that extend to larger-scale parallel and 
distributed environments.)

 \begin{table}[t]
  \centering
   \begin{tabular}{c|c|cc}
     type & precision & example & reference \\
    \hline
   \multirow{2}{*}{embedding + solving subproblem} & \multirow{2}{*}{low} & CW + (FJLT+SVD) & \cite{CW13sparse_STOC} \\
     &  & appr. lev. samp. (SRHT) + SVD & \cite{DMMW12_JMLR} \\
    direct solver & high &  SVD or QR & \cite{golub1996matrix} \\
    PC + direct solver & high & PC (Gaussian) + normal equation & \cite{coakley2011fast} \\
    PC + iterative alg. & high & PC (FJLT) + LSQR & \cite{avron2010blendenpik,
    rokhlin2008fast} 
   \end{tabular}
   \caption{Summary of RandNLA-based $\ell_2$ regression solvers; PC stands for preconditioning.
   }
    \label{tab:l2_solve}
 \end{table}

 \begin{table}[t]
  \centering
   \begin{tabular}{c|c|cc}
     type & precision & example & reference \\
    \hline
      \multirow{2}{*}{(PC + sampling) + solving subproblem} & \multirow{2}{*}{low} & (ER/fast ER + sampling) + IPM & 
     \cite{clarkson2005subgradient, dasgupta2009sampling} \\
    & & (SCT/FCT + sampling) + IPM & \cite{sohler2011subspace, clarkson2013fast} \\
    second-order & high & IPM & \cite{nesterov1994interior} \\
    PC + first-order & high & ER + accelerated gradient descent & \cite{nesterov2008rounding}
   \end{tabular}
   \caption{Summary of RandNLA-based $\ell_1$ regression solvers; PC stands for preconditioning.
   }
    \label{tab:l1_solve}
 \end{table}

\subsubsection{Low-precision solvers}
\label{sxn:low-precision-solvers}

The most straightforward use of these methods (and the one to which most of 
the theory has been developed) is to construct a subspace-preserving 
embedding matrix and then solve the resulting reduced-sized problem exactly, 
thereby obtaining an approximate solution to the original problem.
In somewhat more detail, this algorithmic approach performs the following two 
steps.


\begin{enumerate}
\item 
Construct a subspace-preserving embedding matrix $\Pi$ with distortion $1\pm \frac{\epsilon}{4}$.
\item
Using a black-box solver, solve the reduced-sized problem exactly, 
\emph{i.e.}, exactly solve
           $$ \hat x = \min_{x\in \mathbb{R}^{n}} \|\Pi Ax - \Pi b\|_p. $$
\end{enumerate}

\noindent
(We refer to this approach as \emph{low-precision} since the running time 
complexity with respect to the error parameter $\epsilon$ is 
$\mbox{poly}(1/\epsilon)$.
Thus, while this approach can be analyzed for a fixed $\epsilon$, this 
dependence means that as a practical matter this approach cannot 
achieve high-precision~solutions.)

To see why this approach gives us a $(1 + \epsilon)$-approximate solution to 
the original problem, recall that a subspace-preserving embedding matrix 
$\Pi$ with distortion factor $(1\pm \frac{\epsilon}{4})$ satisfies 
\begin{equation*}
    (1-\epsilon/4) \cdot \|Ax\|_p \leq \|\Pi Ax\|_p \leq (1+\epsilon/4) \cdot \|Ax\|_p, \quad \forall x \in \R^n.
  \end{equation*}
Therefore, the following simple reasoning shows that $\hat x$ is indeed a $(1+\epsilon)$-approximation solution.
\begin{equation*}
  \|A \hat x\|_p \leq \frac{1}{1-\epsilon/4} \|\Pi A\hat x\|_p 
  \leq \frac{1}{1-\epsilon/4} \|\Pi A x^\ast\|_p \leq \frac{1+\epsilon/4}{1-\epsilon/4} \|Ax^\ast\|_p < (1+\epsilon)\|A x^\ast\|_p.
\end{equation*}
For completeness, we include the following lemma stating this result more 
precisely.
\begin{lemma}
  \label{lemma:fast_reg}
  Given an $\ell_p$ regression problem specified by $A \in \R^{m \times n}$ and
  $p \in [1, \infty)$ using the constrained formulation~\eqref{eq:lp_reg_homo},
  let $\Phi$ be a $(1\pm\epsilon/4)$-distortion embedding of $\A_p$, and $\hat{x}$
  be an optimal solution to the reduced-sized problem $\min_{c^T x = 1} \|\Phi A
  x\|_p$.
  Then $\hat{x}$ is a $(1+\epsilon)$-approximate solution to
  the original problem.
\end{lemma}

A great deal of work has followed this general approach.
In particular, the meta-algorithm for $\ell_2$ regression from 
Section~\ref{sxn:rla_ram} is of this general form. 
Many other authors have proposed related algorithms that require solving the 
subproblem by first computing a subspace-preserving sampling matrix.
See, \emph{e.g.}, \cite{mahoney2011randomized} and references therein.
Here, we simply cite several of the most immediately-relevant for our 
subsequent discussion.
\begin{itemize}
\item
Sampling for $\ell_2$ regression. 
One could use the original algorithm 
of~\cite{drineas2006sampling,DMM08_CURtheory_JRNL}, which performs a 
data-aware random sampling and solves the subproblem in 
$\bigO(mn^2)$ time to obtain an approximate solution.
Using the algorithm of~\cite{DMMW12_JMLR}, the running time of this method 
was improved to roughly $\bigO(m n \log(n))$ time, and by combining the algorithm 
of~\cite{DMMW12_JMLR} with the algorithm of~\cite{CW13sparse_STOC}, the 
running time was still further improved to input-sparsity~time.
\item
Projections for $\ell_2$ regression. 
Alternatively, one could use the algorithm 
of~\cite{sarlos2006improved,drineas2011faster}, which performs a 
data-oblivious Hadamard-based random projection and solves the subproblem in 
roughly $\bigO(m n \log(n))$ time, or one could use the algorithm 
of~\cite{CW13sparse_STOC}, which runs in input-sparsity~time.
\item
Sampling and projections for $\ell_1$ and $\ell_p$ regression.
See~\cite{clarkson2005subgradient,sohler2011subspace,clarkson2013fast} and 
see~\cite{dasgupta2009sampling,meng2013low,CW13sparse_STOC} and 
references therein for both data-oblivious and data-aware methods.
\end{itemize}
To summarize these and other results, depending on whether the idealization 
that $m \gg n$ holds, either the Hadamard-based projections for $\ell_2$ 
regression (\emph{e.g.}, the projection algorithm of~\cite{drineas2011faster} 
or the sampling algorithm of~\cite{drineas2006sampling} combined with the 
algorithm of~\cite{DMMW12_JMLR}) and $\ell_1$ regression (\emph{e.g.}, the 
algorithm of~\cite{clarkson2013fast}) or the input-sparsity time algorithms 
for $\ell_2$ and $\ell_1$ regression (\emph{e.g.}, the algorithms 
of~\cite{CW13sparse_STOC} and~\cite{meng2013low}) lead to the best 
worst-case asymptotic performance.
There are, however, practical tradeoffs, both in RAM and in 
parallel-distributed environments, and the most appropriate method to use 
in any particular situation is still a matter of ongoing research.


\subsubsection{High-precision solvers}
\label{sxn:high-precision-solvers}

A more refined use of these methods (and the one that has been used most in
implementations) is to construct a subspace-preserving embedding matrix 
and then use that to construct a preconditioner for the original $\ell_p$ 
regression problem, thereby obtaining an approximate solution to the 
original problem.
In somewhat more detail, this algorithmic approach performs the following 
two~steps.


\begin{enumerate}
\item
Construct a randomized preconditioner for $A$, called $N$.
\item
Invoke an iterative algorithm whose convergence rate depends on the 
condition number of the problem being solved (a linear system for $\ell_2$ 
regression, and a linear or convex program for $\ell_1$ regression) on the 
preconditioned system $AN$.
\end{enumerate}

\noindent
(We refer to this approach as \emph{high-precision} since the running time 
complexity with respect to the error parameter $\epsilon$ is 
$\log(1/\epsilon)$.
Among other things, this means that, given a moderately good 
solution---\emph{e.g.}, the one obtained from the embedding that could be 
used in a low-precision solver---one can very easily obtain a very high 
precision solution.)

Most of the work for high-precision RandNLA solvers for $\ell_p$ regression
has been for $\ell_2$ regression (although we mention a few solvers for 
$\ell_1$ regression for completeness and comparison).
\begin{itemize}
\item
For $\ell_2$ regression.
Recall that theoretical (and empirical) results suggest that the required 
number of iterations in many iterative solvers such as 
LSQR~\cite{paige1982lsqr} depends strongly on the condition number of the 
system.
Thus, a natural idea is first to compute a randomized preconditioner and 
then to apply one of these iterative solvers on the preconditioned system.
For example, if we use SRHT (Lemma~\ref{lemma:srht}) to create a
preconditioned system with condition number bounded by a small constant and 
then use LSQR to solve the preconditioned problem iteratively, the total 
running time would be $\bigO(m n \log(m/\epsilon) + n^3 \log n)$, where 
$\bigO(m n \log(m))$ comes from SRHT, $\bigO(n^3 \log n)$ from computing the 
preconditioner matrix, and $\bigO(m n \log(1/\epsilon))$ from LSQR iterations.
Authors in \cite{avron2010blendenpik,rokhlin2008fast} developed algorithms 
that use FJLT for preconditioning and LSQR as an iterative solver.
In~\cite{meng2011lsrn}, the authors developed a randomized solver for 
$\ell_2$ regression using Gaussian transform and LSQR or the Chebyshev 
semi-iterative method; see Section~\ref{sxn:implementations-l2parallel} for 
more details.

As with the low-precision solvers, note that if we use the input-sparsity 
time algorithm of~\cite{CW13sparse_STOC} for embedding and then use an 
(SRHT + LSQR) approach above to solve the reduced-sized problem, then 
under the assumption that $m \geq \poly(n)$ and $\epsilon$ is fixed, this 
particular combination would become the best approach proposed.
However, there are various trade-offs among those approaches.
For instance, there are trade-offs between running time and conditioning 
quality in preconditioning for computing the subspace-preserving sampling 
matrix, and there are trade-offs between embedding dimension/sample size 
and failure rate in embedding/sampling.
Some of the practical trade-offs on different problem types and computing 
platforms will be discussed in Section \ref{sxn:implementations-l2spark}
below.
\item
For $\ell_1$ regression.
While most of the work in RandNLA for high-precision solvers has been for 
$\ell_2$ regression, we should point out related work for $\ell_1$ 
regression.
In particular, Nesterov~\cite{nesterov2008rounding} proposed an algorithm 
that employs a combination of ellipsoid rounding and accelerated 
gradient descent; and second-order methods from \cite{nesterov1994interior} 
use interior point techniques more generally.
See also the related solvers of Portnoy 
\emph{et al.}~\cite{portnoy1997gaussian,Por97}.
For $\ell_1$ regression, Meng and Mahoney~\cite{meng2013robust} coupled 
these ideas with RandNLA ideas to develop an iterative medium-precision 
algorithm for $\ell_1$ regression; see 
Section~\ref{sxn:implementations-l1distributed} for more details.
\end{itemize}


%% file: implementation.tex
\section{Implementations and empirical results}
\label{sxn:implementations}

In this section, we describe several implementations in large-scale 
computational environments of the theory described in 
Section~\ref{sxn:round_embed}.
In particular, in Section~\ref{sxn:implementations-l2parallel}, we will 
describe \texttt{LSRN}, an $\ell_2$ regression solver appropriate for 
parallel environments using multi-threads and MPI; 
and then in Section~\ref{sxn:implementations-l1distributed}, we will 
describe the results of both a low-precision algorithm as well as a
related medium-precision iterative algorithm for the $\ell_1$ regression 
problem.
Both of these subsections summarize recent previous work, and they both 
illustrate how implementing RandNLA algorithms in parallel and 
distributed environments requires paying careful attention to 
computation-communication tradeoffs.
These prior results do not, however, provide a comprehensive evaluation of 
any particular RandNLA method.
Thus, for completeness, we also describe in 
Section~\ref{sxn:implementations-l2spark} several new results: 
\emph{a comprehensive empirical evaluation of low-precision, 
medium-precision, and high-precision random sampling and random projection 
algorithms for the very overdetermined $\ell_2$ regression problem.}
Hereby, by ``medium-precision'', typically we mean calling a high-precision solver but executing less iterations in the underlying iterative solver.
These implementations were done in Apache 
Spark\footnote{\tiny \url{http://spark.apache.org}}; they have been applied
to matrices of up to terabyte size; and they illustrate several points that
will be important to understand as other RandNLA algorithms are implemented 
in very large-scale computational environments.

\subsection{Solving $\ell_2$ regression in parallel environments}
\label{sxn:implementations-l2parallel}

In this subsection, we describe implementation details for a high-precision
$\ell_2$ regression solver designed for large-scale parallel environments.
\texttt{LSRN}~\cite{meng2011lsrn} is designed to solve the minimum-length 
least squares problem \eqref{eq:ls_min_length} to high precision; and it 
works for linear systems that are either strongly over-determined, 
\emph{i.e.}, $m \gg n$ or strongly under-determined, \emph{i.e.}, $m \ll n$, 
and possibly rank-deficient.
\texttt{LSRN} uses random normal projections to compute a preconditioner 
matrix such that the preconditioned system is provably extremely 
well-conditioned.  
In particular, either LSQR~\cite{paige1982lsqr} (a conjugate gradient based 
method) or the Chebyshev semi-iterative (CS) method~\cite{golub1961chebyshev} 
can be used at the iterative step to compute the min-length solution within 
just a few iterations.
As we will describe, the latter method is preferred on clusters with high 
communication cost.
Here, we only present the formal description of the Algorithm \texttt{LSRN} 
for strongly over-determined systems in Algorithm~\ref{alg:lsrn}.

\begin{algorithm}
  \begin{algorithmic}[1]
    \STATE Choose an oversampling factor $\gamma > 1$, \emph{e.g.}, $\gamma =
    2$. Set $s = \lceil \gamma n \rceil$.

    \STATE Generate $G = \text{randn}(s,m)$, a Gaussian matrix.

    \STATE Compute $\tilde{A} = G A$.
    \STATE Compute $\tilde{A}$'s economy-sized SVD: $\tilde{U} \tilde{\Sigma}
    \tilde{V}^T$.
    
    \STATE Let $N = \tilde{V} \tilde{\Sigma}^{-1}$.
           (Note: that this is basically $R^{-1}$ from QR on the embedding, but it is written here ito the SVD.)
    
    \STATE Iteratively compute the min-length solution $\hat{y}$ to 
    \begin{equation*}
      \text{minimize}_{y \in \mathbb{R}^r} \quad \| A N y - b \|_2.
    \end{equation*}

    \STATE Return $\hat{x} = N \hat{y}$.
  \end{algorithmic}
  \caption{\texttt{LSRN} for strongly over-determined systems}
  \label{alg:lsrn}
\end{algorithm}

Two important aspects of \texttt{LSRN} are the use of the Gaussian 
transform and the CS method, and they are coupled in a nontrivial way.
In the remainder of this subsection, we discuss these~issues.

To start, note that, among the available choices for the random projection 
matrix, the Gaussian transform has particularly-good conditioning properties.
In particular, the distribution of the spectrum of the preconditioned 
system depends only on that of a certain Gaussian matrix, not the original 
linear system.
In addition, one can show that
\begin{equation*}
    \mathcal{P} \left( \kappa(AN) \leq \frac{1+\alpha+\sqrt{r/s}}{1-\alpha-\sqrt{r/s}} \right) \geq 1 - 2 e^{- \alpha^2 s / 2},
  \end{equation*}
where $\kappa(AN)$ is the condition number of the preconditioned system, 
$r$ is the rank of $A$, and $\alpha$ is a parameter~\cite{meng2011lsrn}.
For example, if we choose the oversampling factor $\gamma$ in 
Algorithm~\ref{alg:lsrn} to be $2$, then the condition number of the new 
linear system is less than $6$ with high probability. 
In addition, a result on bounds on the singular values provided 
in \cite{meng2011lsrn} enable CS to work more efficiently.


Moreover, while slower in terms of FLOPS than FFT-based fast transforms, the 
Gaussian transform comes with several other advantages for large-scale 
environments.
First, it automatically speeds up with sparse input matrices and fast linear 
operators (in which case FFT-based fast transforms are no longer ``fast'').
Second, the preconditioning process is embarrassingly parallel and thus 
scales well in parallel environments.
Relatedly, it is easy to implement using multi-threads or MPI.
Third, it still works (with an extra ``allreduce'' operation) when $A$ is 
partitioned along its bigger dimension.
Lastly, when implemented properly, Gaussian random variables can be 
generated very fast~\cite{marsaglia2000ziggurat} (which is nontrivial, given 
that the dominant cost in na\"{\i}vely-implemented Gaussian-based 
projections can be generating the random variables).
For example, it takes less than 2 seconds to generate $10^9$ random Gaussian 
numbers using 12 CPU cores~\cite{meng2011lsrn}.

To understand why CS is preferable as a choice of iterative solver compared 
to other methods such as the conjugate gradient based LSRN, one has to take 
the convergence rate and computation/communication costs into account.
In general, if (a bound for) the condition number of the linear system is 
large or not known precisely, then the CS method will fail ungracefully 
(while LSQR will just converge very slowly).
However, with the \emph{very} strong preconditioning guarantee of the 
Gaussian transform, we have very strong control on the condition number of 
the embedding, and thus the CS method can be expected to converge within a 
very few iterations. 
In addition, since CS doesn't have vector inner products that require 
synchronization between nodes (while the conjugate gradient based LSQR 
does), CS has one less synchronization point per iteration, \emph{i.e.},
it has improved communication properties.
See Figure~\ref{fig:code-snippets} for the Python code snippets of LSQR and 
CS, respectively.
On each iteration, both methods have to do two matrix-vector multiplications,
while CS only needs one cluster-wide synchronization compared to two in LSQR.
Thus, the more communication-efficient CS method is enabled by the very 
strong control on conditioning that is provided by the more expensive 
Gaussian projection.
It is this advantage that makes CS favorable in the distributed environments, 
where communication costs are considered more expensive.  

\begin{figure}

\begin{lstlisting}[frame=single,language=Python,caption=One iteration in LSQR]
u    = A.matvec(v) - alpha*u
beta = sqrt(comm.allreduce(np.dot(u,u)))
...
v    = comm.allreduce(A.rmatvec(u)) - beta*v
\end{lstlisting}

\begin{lstlisting}[frame=single,language=Python,caption=One iteration in CS]
v  = comm.allreduce(A.rmatvec(r)) - beta*v
x += alpha*v
r -= alpha*A.matvec(v)
\end{lstlisting}

\caption{Python code snippets for LSQR-based and CS-based iterations, 
respectively, illustrating that the latter has one synchronization 
point periteration, while the former has two.}
\label{fig:code-snippets}
\end{figure}

\subsection{Solving $\ell_1$ regression in distributed environments}
\label{sxn:implementations-l1distributed}

In this subsection, we describe implementation details for both 
low-precision and high-precision solvers for the $\ell_1$ regression 
problem in large-scale distributed environments.
These algorithms were implemented using MapReduce 
framework~\cite{dean2004mapreduce} which (at least until the relatively 
recent development of the Apache Spark framework) was the {\it de facto} 
standard parallel environment for analyzing massive datasets.

\paragraph{Low-precision solver}
Recall that one can use the sampling algorithm described in 
Section~\ref{sxn:app_to_lp} to obtain a low-precision approximate solution 
for $\ell_1$ regression.
This can be summarized in the following three steps.


\begin{enumerate}
\item
Compute an $\ell_1$-well-conditioned basis $U = AR^{-1}$ for $A$.
\item
Construct an importance sampling distribution $\{p_i\}_{i=1}^{m}$ 
based on the $\ell_1$ row norms of $U$.
Randomly sample a small number of constraints according to 
$\{p_i\}_{i=1}^{m}$ to construct a subproblem.
\item
Solve the $\ell_1$-regression problem on the subproblem.
\end{enumerate}

Next, we will discuss some of the implementation details of the above three 
steps in the MapReduce framework.  
The key thing to note is that, for the problems we are considering, the dominant cost 
is the cost of input/output, \emph{i.e.}, communicating the data, and hence 
we want to extract as much information as possible for each pass over the 
data.

The first step, as described in Section~\ref{sxn:app_to_lp}, is to 
construct an $\ell_1$ well-conditioned basis for $A$; and for this one can 
use one of the following three methods---ellipsoid rounding (ER), a QR 
factorization of $\Pi A$, where $\Pi A$ is a low-distortion subspace 
embedding matrix in terms of $\ell_1$ norm (QR), or a combination of these 
two (QR+ER method).
See Table~\ref{tab:cond} for summary of these approaches to conditioning.
Note that many conditioning methods are embarrassingly parallel, in which 
case it is straightforward to implement them in MapReduce.
For example, the Cauchy transform (CT) with embedding dimension $r$ can 
be implemented in the following manner.

\vspace{3mm}
\noindent
\textbf{Mapper:}
  \begin{algorithmic}[1]
    \STATE For each row $a_i$ of $A$, generate a vector $c_i \in \R^{r\times 1}$ consisting $r$ standard Cauchy random variables.
    \STATE For $j=1,\ldots r$, emit $(j, c_{i,j} a_i)$ where
     $c_{i,j}$ denotes the $j$-th element of $c_i$.
  \end{algorithmic}
  
\noindent  \textbf{Reducer:}
  \begin{algorithmic}[1]
    \STATE Reduce vectors associated with key $k$ to $v_k$ with addition operation.
    \STATE Return $v_k$.
        \end{algorithmic}
\vspace{3mm}

\noindent
After collecting all the vectors $v_k$, for $k = 1,\ldots,r$, one only has 
to assemble these vectors and perform QR decomposition on the resulting 
matrix, which completes the preconditioning process.


With the matrix $R^{-1}$ such that $AR^{-1}$ is well-conditioned, a second 
pass over the dataset is sufficient to construct a subproblem and obtain 
several approximate solutions to the original problem, \emph{i.e.}, the 
second and three steps of the sampling algorithm above.
Note that since computation is a less precious resource than communication 
here, one can exploit this to compute multiple subsampled solutions in this 
single pass.
(E.g., performing, say, $100$ matrix-vector products is only marginally more 
expensive than performing $1$, and thus one we can solve multiple subsampled 
solutions in a single ``pass'' with almost no extra effort.
To provide an example, on a $10$-node Hadoop cluster, with a matrix 
of size ca. $10^8 \times 50$, a single query took $282$ seconds, while $100$ 
queries took only $383$ seconds, meaning that the extra $99$ queries come 
almost ``for free.'')
We summarize the basic steps as follows.
Assume that $A \in \R^{m \times n}$ has condition number $\kappa_1$, $s$ is 
the sampling size and $n_x$ is the number of approximate solutions desired.
Then the following algorithm returns $n_x$ approximate solutions to the 
original problem.

\vspace{3mm}
\noindent
\textbf{Mapper:}
  \begin{algorithmic}[1]
    \STATE For each row $a_i$ of $A$, let $p_i = \min \{ s \|a_i\|_1 / (\kappa_1 n^{1/2}), 1 \}$.

    \STATE For $k = 1, \ldots, n_x$, emit $(k, a_i/p_i)$ with probability $p_i$.
  \end{algorithmic}

\noindent
  \textbf{Reducer:}
  \begin{algorithmic}[1]
    \STATE Collect row vectors associated with key $k$ and assemble $A_k$.
    \STATE Compute $\hat{x}_k = \arg \min_{c^T x = 1} \|A_k x\|_1$ using
      interior-point methods.
    \STATE Return $\hat{x}_k$.
    \end{algorithmic}
\vspace{3mm}
\noindent
Note here, in the second step of the reducer above, since the size of the subsampled matrix $A_k$ typically only depends on the low dimension $n$, the subproblem can be fit into the memory of a single machine and can be solved locally.

As an aside, note that such an algorithm can be used to compute approximate 
solutions for other problems such as the quantile regression problem by 
only increasing the sampling size by a constant factor. 
In \cite{YMM14_SISC}, the authors evaluate the empirical performance of 
this algorithm by using several different underlying preconditioners, 
\emph{e.g.}, CT, FCT, etc., on a terabyte-size dataset in Hadoop to solve 
$\ell_1$ regression and other quantile regression problems.

\paragraph{High-precision solver}
To obtain a high-precision solution for the $\ell_1$ regression problem, we 
have to resort to iterative algorithms.
See Table~\ref{table:iter_l1}, where we summarize several iterative 
algorithms in terms of their convergence rates and complexity per iteration.
Note that, among these methods, although IPCPM (interior point cutting plane 
methods) needs additional work at each iteration, the needed of number of 
passes is linear in the low dimension $n$ and it only has a dependence on 
$\log(1/\epsilon)$.
Again, since communication is a much more precious resource than computation
in the distributed application where this was implemented, this can be an 
acceptable tradeoff when, \emph{e.g.}, a medium-precision solution is needed. 

\begin{table}
  \begin{center}
    \begin{tabular}{c|c|c}
      & passes & extra work per pass \\
      \hline
      subgradient~\cite{clarkson2005subgradient} & $\mathcal{O}(n^4 / \epsilon^2)$ &  \\
      gradient~\cite{nesterov2009unconstrained} & $\mathcal{O}(m^{1/2} / \epsilon)$ & \\
      ellipsoid~\cite{grotschel1981ellipsoid} & $\mathcal{O}(n^2 \log (\kappa_1 /\epsilon))$ \\
      IPCPM~\cite{tarasov1988method}
      & $\mathcal{O}(n \log (\kappa_1 /\epsilon))$ & $\mathcal{O}(n^{7/2} \log n)$ 
    \end{tabular}
  \end{center}
  \caption{Iterative algorithms for solving $\ell_1$ regression.}
  \label{table:iter_l1}
\end{table}

Meng and Mahoney~\cite{meng2013robust} proposed a randomized IPCPM algorithm 
to solve the $\ell_1$ regression problem to medium precision in large-scale
distributed environments.
It includes several features specially-designed for MapReduce and distributed 
computation.
(To describe the method, recall that IPCPM is similar to a bisection method,
except that it works in a high dimensional space.
It starts with a search region $\mathcal{S}_0 = \{ x \,|\, S x \leq t\}$,
which contains a ball of desired solutions described by a separation oracle. 
At step $k$, we first compute the maximum-volume ellipsoid $\mathcal{E}_k$ 
inscribing $\mathcal{S}_k$. 
Let $y_k$ be the center of $\mathcal{E}_k$, and send $y_k$ to the oracle.
If $y_k$ is not a desired solution, the oracle returns a linear cut that 
refines the search region $\mathcal{S}_k \to \mathcal{S}_{k+1}$.)
The algorithm of~\cite{meng2013robust} is different from the standard IPCPM, 
mainly for the following two reasons.
\begin{itemize}
\item 
\textbf{Initialization using all the solutions returned by sampling algorithms.}
To construct a search region $\mathcal{S}_0$, one can use the multiple 
solutions returned by calling the sampling algorithm, \emph{e.g.}, 
low-precision solutions, to obtain a much better initial condition.
If we denote by $\hat x_1, \ldots \hat x_N$ the $N$ approximation solution, 
then given each $\hat{x}$, let $\hat{f} = \|A \hat{x}\|_1$ and 
$\hat{g} = A^T \text{sign}(A \hat{x})$.
Note that given $\hat x_1, \ldots, \hat x_N$, computing $\hat f_i, \hat g_i$ 
for $i=1,\ldots, N$ can be done in a single pass.
Then we have
  \begin{equation*}
    \|x^* - \hat{x}\|_2 \leq \|A (x^* - \hat{x})\|_1 \leq \|A x^* \|_1 + \|A \hat{x}\|_1 \leq 2 \hat{f}.
  \end{equation*}
Hence, for each subsampled solution $\hat x_i$, we have a hemisphere that 
contains the optimal solution. 
We use all these hemispheres to construct a better initial search region 
$\mathcal{S}_0$, which may potentially reduce the number of iterations 
needed for convergence. 
\item 
\textbf{Performing multiple queries per iteration.}
Instead of sending one query point at each iteration, one can exploit the 
fact that it is inexpensive to compute multiple query points per iteration, 
and one can send multiple query points at a time.
Let us still use $\hat x_i$ to denote the multiple query points.
Notice that by convexity, 
  \begin{equation*}
    \|A x^*\|_1 \geq \|A \hat x\|_1 + \hat g^T ( x^* - \hat x ).
  \end{equation*}
This implies $g^T x^* \leq g^T \hat x$. 
That is, given any query point $\hat x$, the subgradient serves as a 
separation oracle which returns a half-space that contains the desired 
ball.
This means that, for each query point $\hat x_i$, a half-space containing 
the ball of desired solutions will be~returned.
\end{itemize}
Note that both of these differences take advantage of performing extra 
computation while minimizing the number of iterations (which is strongly 
correlated with communication for MapReduce computations).


\subsection{Detailed empirical evaluations of $\ell_2$ regression solvers in parallel/distributed environments}
\label{sxn:implementations-l2spark}

In this subsection, we provide a detailed empirical evaluation of the 
performance of RandNLA algorithms for solving very over-determined very
large-scale $\ell_2$ regression problems.
Recall that the subspace embedding that is a crucial part of RandNLA 
algorithms can be data-aware (\emph{i.e.}, a sampling algorithm) or 
data-oblivious (\emph{i.e.}, a projection algorithm).
Recall also that, after obtaining a subspace embedding matrix, one can 
obtain a low-precision solution by solving the resulting subproblem, or one 
can obtain a high-precision solution by invoking a iterative solver, 
\emph{e.g.}, LSQR~\cite{paige1982lsqr}, 
for $\ell_2$ regression, with a preconditioner constructed from by the 
embedding.  
Thus, in this empirical evaluation, we consider both random sampling and 
random projection algorithms, and we consider solving the problem to 
low-precision, medium-precision, and high-precision on a suite or data sets 
chosen to be challenging for different classes of algorithms.
We consider a range of matrices designed to ``stress test'' all of the 
variants of the basic meta-algorithm of Section~\ref{sxn:rla_ram} that we 
have been describing, and we consider matrices of size ranging up to just 
over the terabyte size scale.

\subsubsection{Experimental setup}

In order to illustrate a range of uniformity and nonuniformity properties 
for both the leverage scores and the condition number, we considered the 
following four types of datasets.
\begin{itemize}
\item UG (matrices with {\it uniform} leverage scores and {\it good} condition number);
\item UB (matrices with {\it uniform} leverage scores and {\it bad} condition number);
\item NG (matrices with {\it nonuniform} leverage scores and {\it good} condition number);
\item NB (matrices with {\it nonuniform} leverage scores and {\it bad} condition number).
\end{itemize}

\noindent
These matrices are generated in the following manner.
For matrices with uniform leverage scores, we generated the matrices 
by using the commands that are listed in Table~\ref{table:matlab-cmds-unif}.
For matrices with nonuniform leverage scores, we considered matrices with 
the following structure:
$$ A = \begin{pmatrix} \alpha B & R \\ \bf{0} & I \end{pmatrix}, $$ 
where $B \in \R^{(m-d/2)\times(d/2)}$ is a random matrix with each element 
sampled from $\N(0,1)$, $I\in \R^{(d/2)\times(d/2)}$ is the identity matrix,
and $R\in \R^{(m-d/2) \times (d/2)}$ is a random matrix generated using 
\texttt{ 1e-8 * rand(m-d/2,d/2). }
In this case, the condition number of $A$ is controlled by $\alpha$.
It is worth mentioning that the last $d/2$ rows of the above matrix have 
leverage scores exactly $1$ and the rest ones are approximately $d/2/(n-d/2)$.
Also, for matrices with bad condition number, the condition number is
approximately $1e6$ (meaning $10^6$); while for matrices with good condition 
number, the condition number is approximately $5$.

\begin{table}[t]
\begin{lstlisting}[frame=single]
U = orth(randn(m,n)); 
S = diag(linspace(1,1/kappa,n));
V = orth(randn(n,n)); 
A = U*S*V';
x = randn(n,1); 
b = A*x;
err = randn(m,1); 
b = b+0.25*norm(b)/norm(err)*err;
\end{lstlisting}
\caption{Commands (presented in MATLAB format) used to generate matrices 
with uniform leverage scores, \emph{i.e.}, the UG and UB matrices.  Here, 
\texttt{kappa} is a parameter used to determine the condition number of the 
generated matrix.}
\label{table:matlab-cmds-unif}
\end{table}

To generate a large-scale matrix that is beyond the capacity of RAM, and to 
evaluate the quality of the solution for these larger inputs, we used two 
methods.
First, we replicate the matrix (and the right hand side vector, when it is
needed to solve regression problems) \texttt{REPNUM} times, and we ``stack'' 
them together vertically.
We call this na\"{\i}ve way of stacking matrices as \texttt{STACK1}.
Alternatively, for NB or NG matrices, we can stack them in the following 
manner:
$$ \tilde A = \begin{pmatrix} \alpha B & R \\
 \cdots \\
 \alpha B & R \\  \bf{0} & I \end{pmatrix}. $$
We call this stacking method \texttt{STACK2}.
The two different stacking methods lead to different properties for the 
linear system being solved---we summarize these in 
Table~\ref{table:stack}---and, while they yielded results that were usually 
similar, as we mention below, the results were different in certain extreme 
cases.  
With either method of stacking matrices, the optimal solution remains the 
same, so that we can evaluate the approximate solutions of the new large 
least-squares problems.
We considered these and other possibilities, but in the results reported 
below, unless otherwise specified we choose the following: for large-scale 
UG and UB matrices, we use \texttt{STACK1} to generate the data; and, for 
large-scale NG and NB matrices, we use \texttt{STACK2} to generate the data.

\begin{table}[t]
\begin{center}
\begin{sc}
\begin{tabular}{c|ccc}
 name & condition number & leverage scores & coherence \\
 \hline
 STACK1 & unchanged & divided by \texttt{REPNUM} & divided by \texttt{REPNUM} \\
 \bigcell{c}{STACK2 \\(for NB and NG only)} & increased & unknown & always $1$
\end{tabular}
\end{sc}
\end{center}
\caption{Summary of methods for stacking matrices, to generate matrices too 
large to fit into RAM; here, \texttt{REPNUM} denotes the number of 
replications and coherence is defined as the largest leverage score of the matrix.}
\label{table:stack}
\end{table}

Recall that Table~\ref{tab:l2_embed} provides several methods for 
computing an $\ell_2$ subspace embedding matrix.
Since a certain type of random projection either can be used to obtain an 
embedding directly or can be used (with the algorithm of~\cite{DMMW12_JMLR}) 
to approximate the leverage scores for use in sampling, we consider 
both data-aware and data-oblivious methods.
Throughout our evaluation, we use the following notations to denote various 
ways of computing the subspace embedding.
\begin{itemize}
\item 
  \texttt{PROJ CW} --- Random projection with the input-sparsity time CW method
\item
  \texttt{PROJ GAUSSIAN} --- Random projection with Gaussian transform
\item
  \texttt{PROJ RADEMACHER} --- Random projection with Rademacher transform
\item
  \texttt{PROJ SRDHT} --- Random projection with Subsampled randomized discrete Hartley transform~\cite{DHT}
\item
 \texttt{SAMP APPR} ---  Random sampling based on approximate leverage scores
\item
  \texttt{SAMP UNIF} --- Random sampling with uniform distribution
\end{itemize}
Note that, instead of using a vanilla SRHT, we perform our evaluation with a
SRDHT (\emph{i.e.}, a subsampled randomized discrete Hartley transform).
(An SRDHT is a related FFT-based transform which has similar properties to a 
SRHT in terms of speed and accuracy but doesn't have the restriction on the dimension to be a power of $2$.) Also note that, instead of using a distributed FFT-based transform to implement SRDHT, we treat the transform as a dense matrix-matrix multiplication, hence we should not expect SRDHT to have computational advantage over other transforms.

Throughout this section, by embedding dimension, we mean the projection 
size for projection based methods and the sampling size for sampling based 
methods.
Also, it is worth mentioning that for sampling algorithm with approximate 
leverage scores, we fix the underlying embedding method to be 
\texttt{PROJ CW} and the projection size $c$ to be $d^2/4$.
In our experiments, we found that---when they were approximated sufficiently 
well---the precise quality of the approximate leverage scores do not have a 
strong influence on the quality of the solution obtained by the sampling 
algorithm.
We will elaborate this more in Section~\ref{sec:appr_lev}.

The computations for Table~\ref{table:lev}, Figure~\ref{fig:lev}, and 
Table~\ref{table:cond} below (\emph{i.e.}, for the smaller-sized problems)
were performed on a shared-memory machine with 12 Intel Xeon CPU cores at 
clock rate 2GHz with 128GB RAM.
In these cases, the algorithms are implemented in MATLAB.
All of the other computations (\emph{i.e.}, for the larger-sized problems) 
were performed on a cluster with 16 nodes (1 master and 15 slaves), each of 
which has 8 CPU cores at clock rate 2.5GHz with 25GB RAM.
For all these cases, the algorithms are implemented in Spark via a Python API.

\subsubsection{Overall performance of low-precision solvers}

\begin{figure}[h!tbp]
\begin{centering}
\begin{tabular}{ccc}
\subfigure[$\|x - x^\ast\|_2/\|x^\ast\|_2$]{
\includegraphics[width=0.3\textwidth]{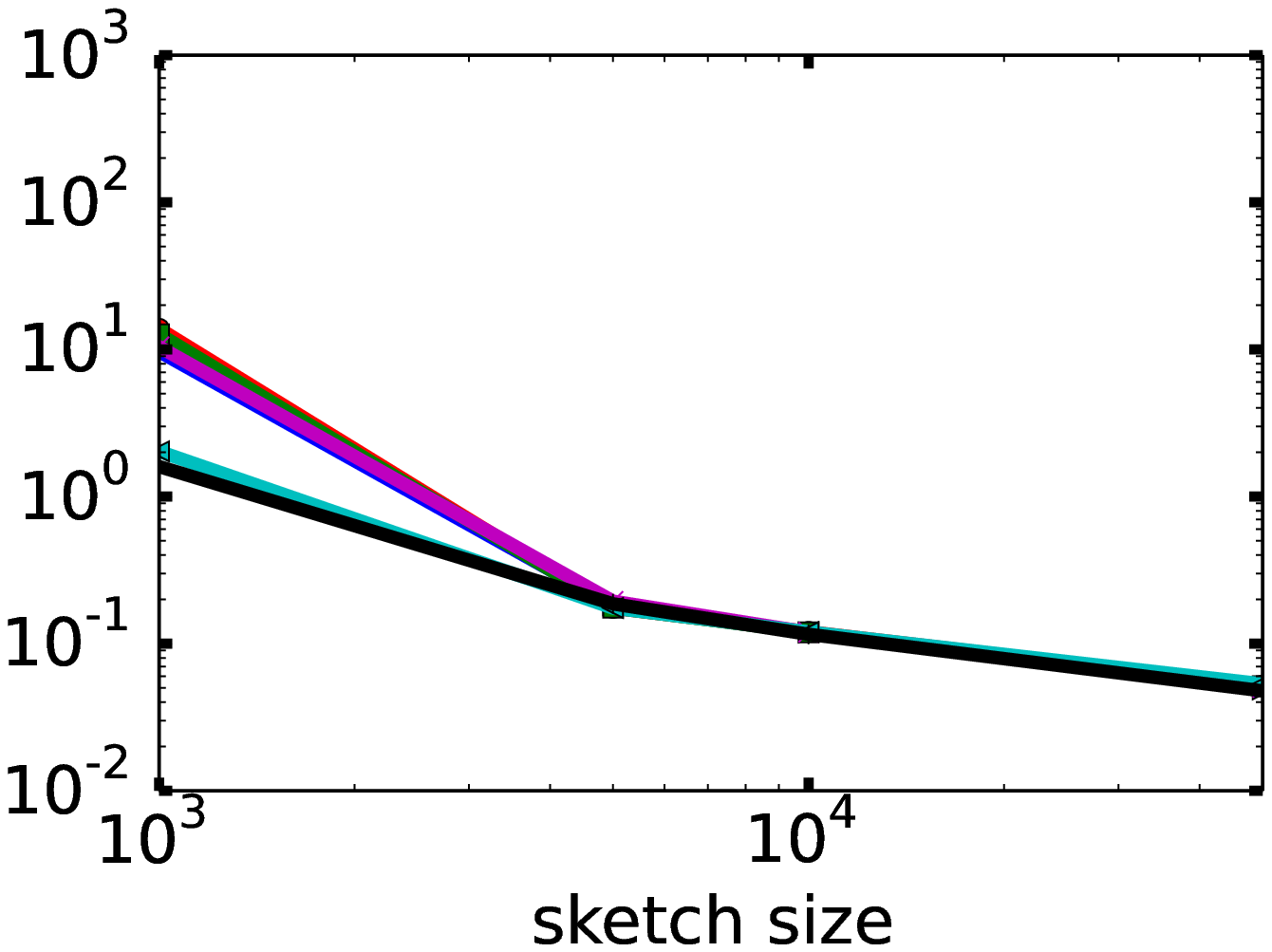}
}
&
\subfigure[$|f-f^\ast|/|f^\ast|$]{
\includegraphics[width=0.3\textwidth]{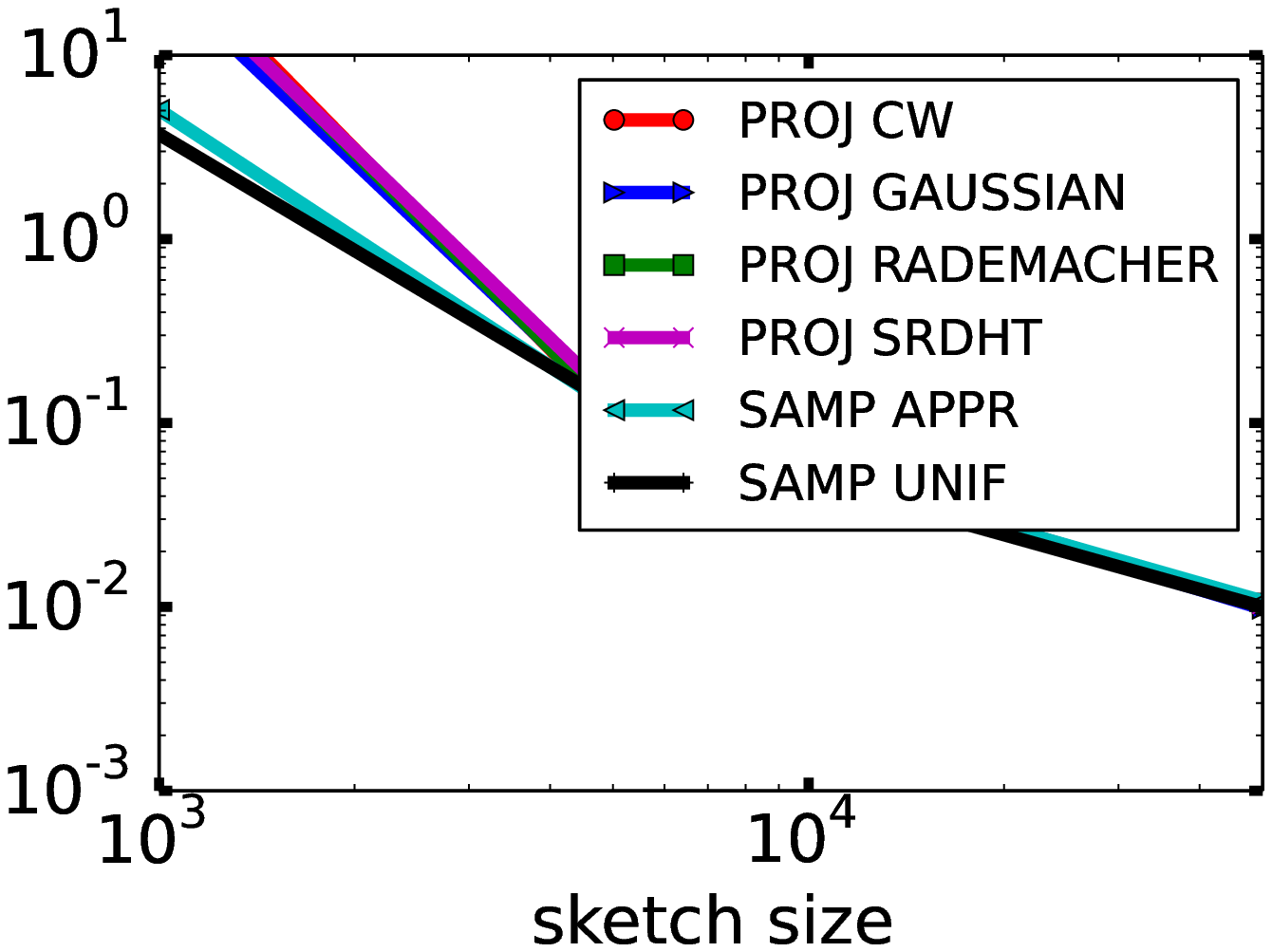}
}
&
\subfigure[Running time(sec)]{
\includegraphics[width=0.3\textwidth]{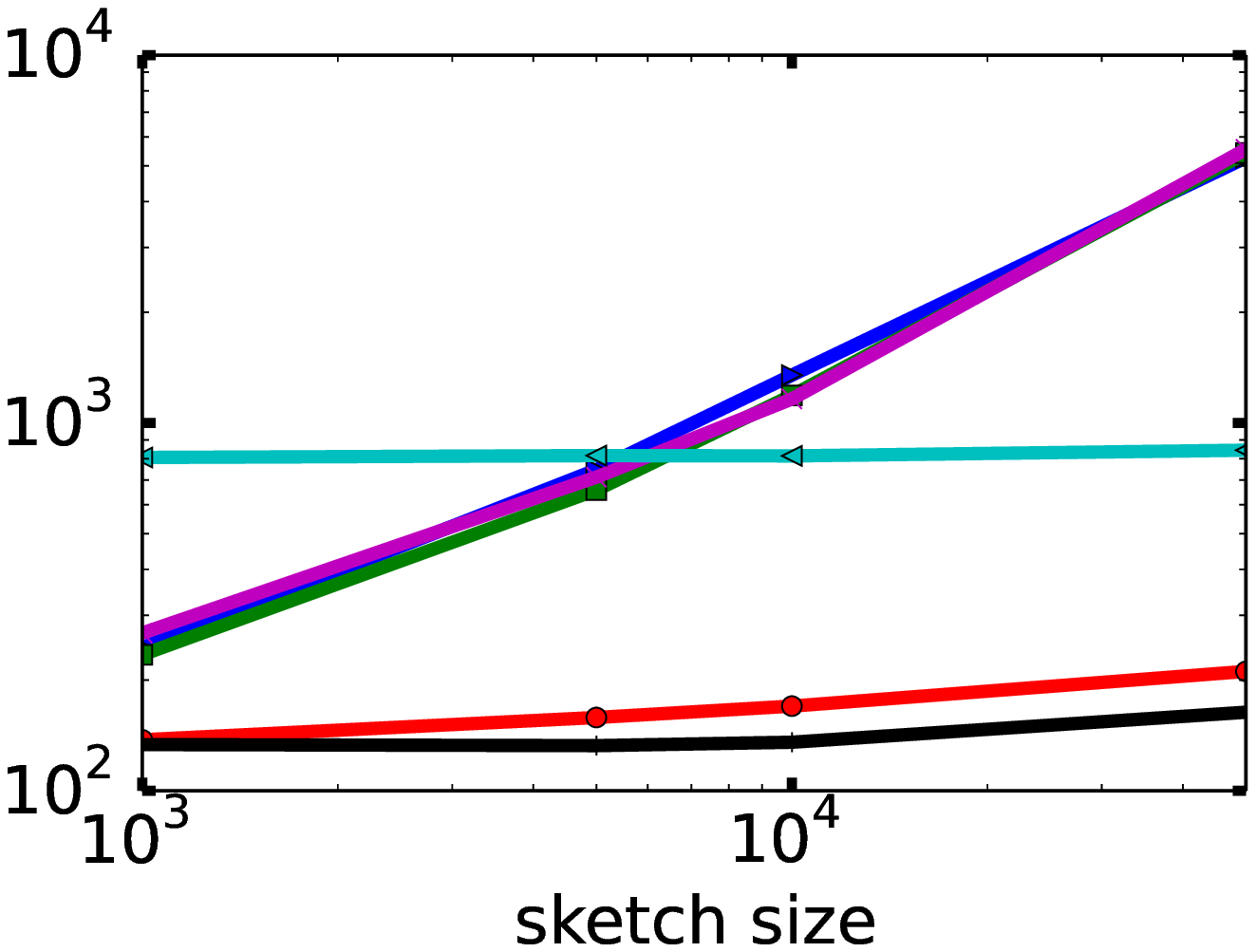}
}
\\
 \multicolumn{3}{c}{\bf $1e7 \times 1000$ UG matrix}
 \\
\subfigure[$\|x - x^\ast\|_2/\|x^\ast\|_2$]{
\includegraphics[width=0.3\textwidth]{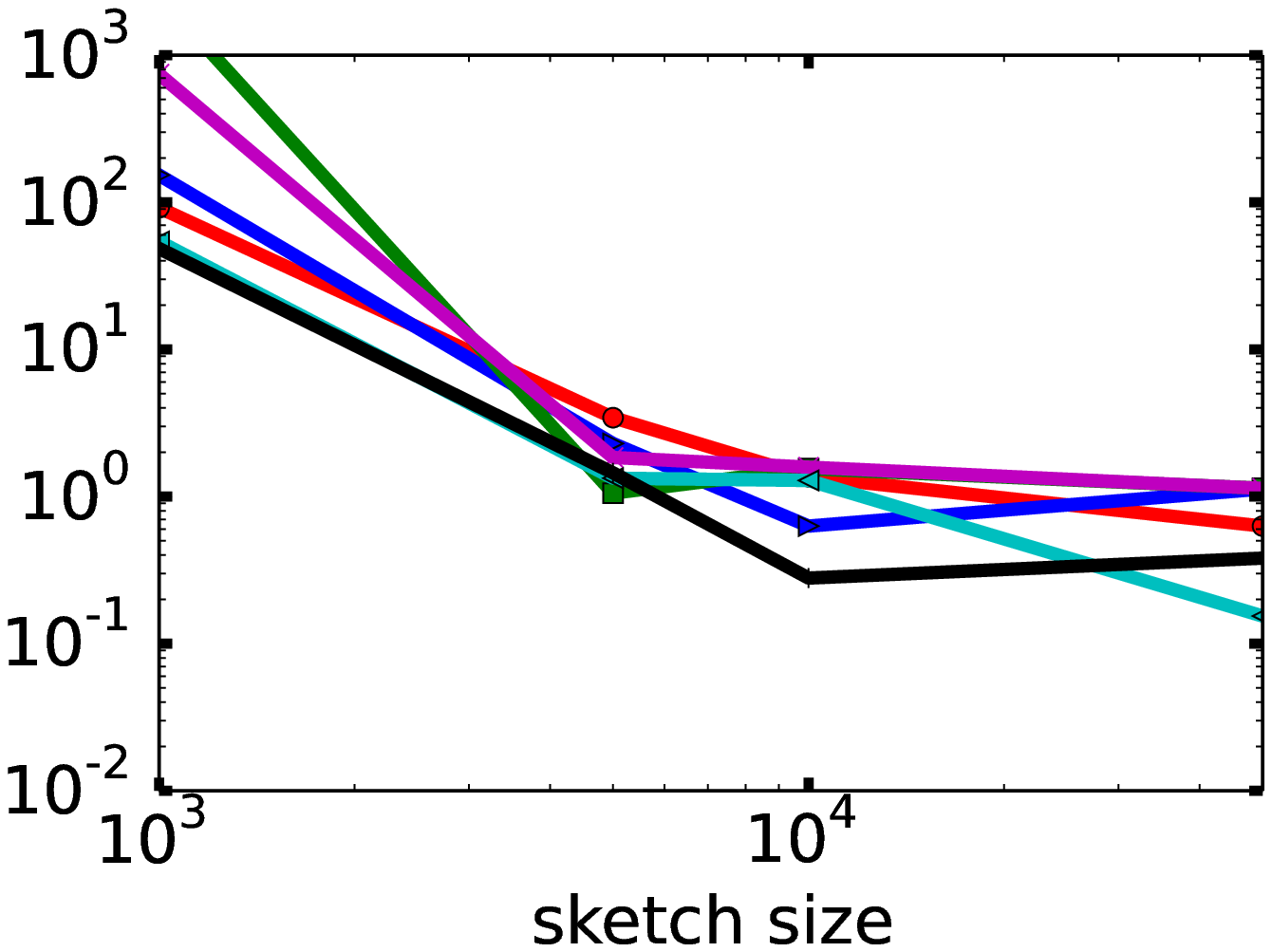}
}
&
\subfigure[$|f-f^\ast|/|f^\ast|$]{
\includegraphics[width=0.3\textwidth]{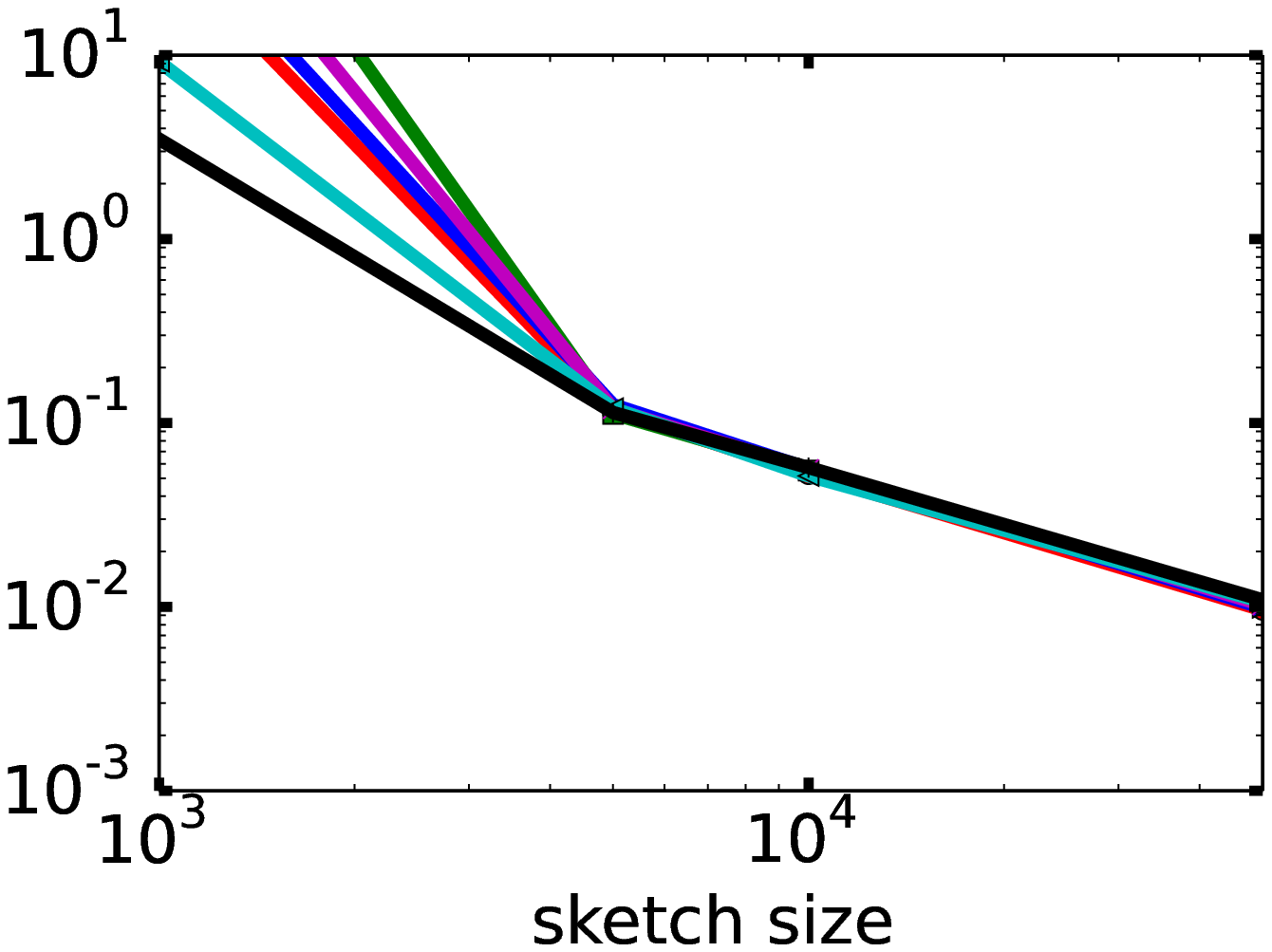}
}
&
\subfigure[Running time(sec)]{
\includegraphics[width=0.3\textwidth]{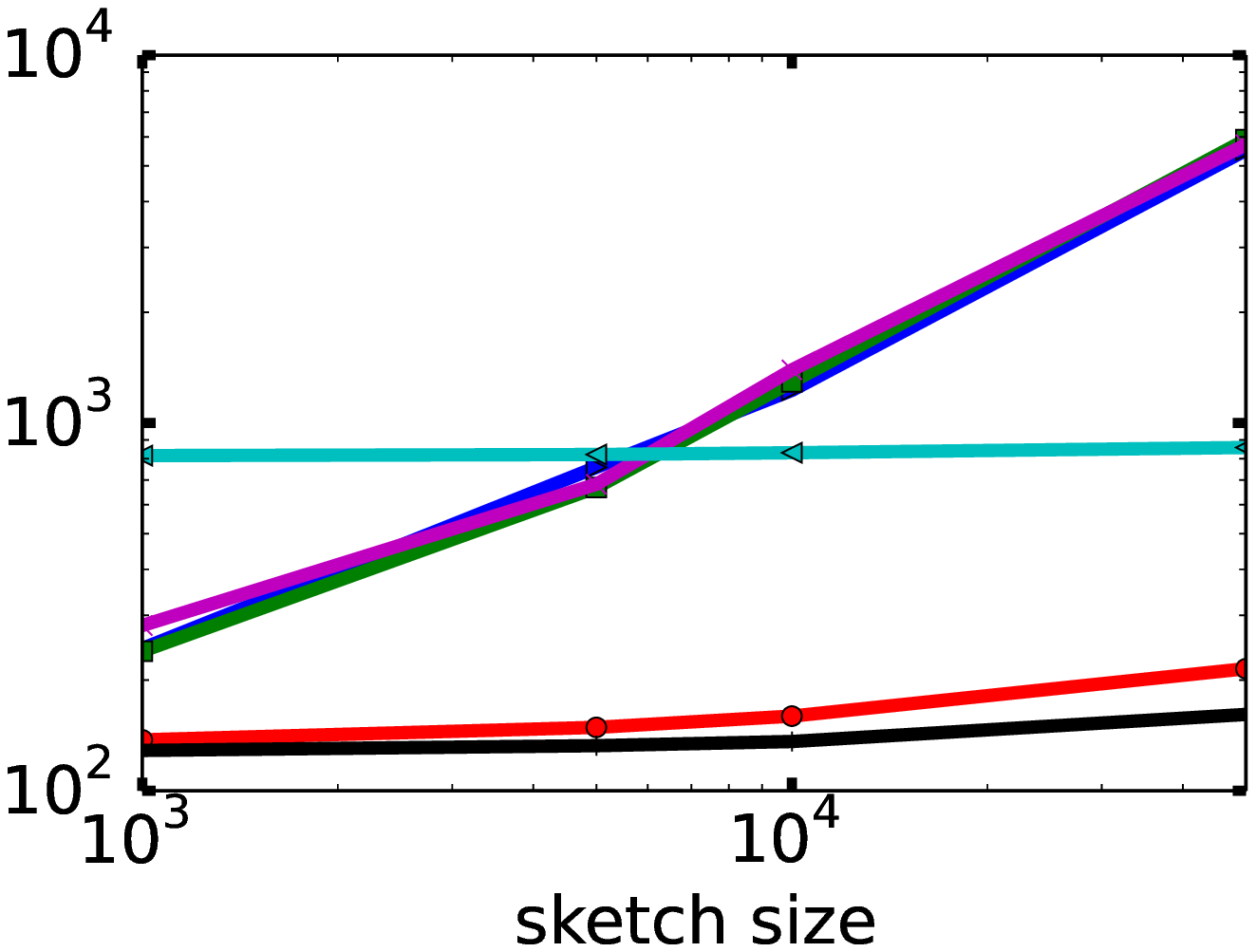}
}
\\
 \multicolumn{3}{c}{\bf $1e7 \times 1000$ UB matrix}
  \\
\subfigure[$\|x - x^\ast\|_2/\|x^\ast\|_2$]{
\includegraphics[width=0.3\textwidth]{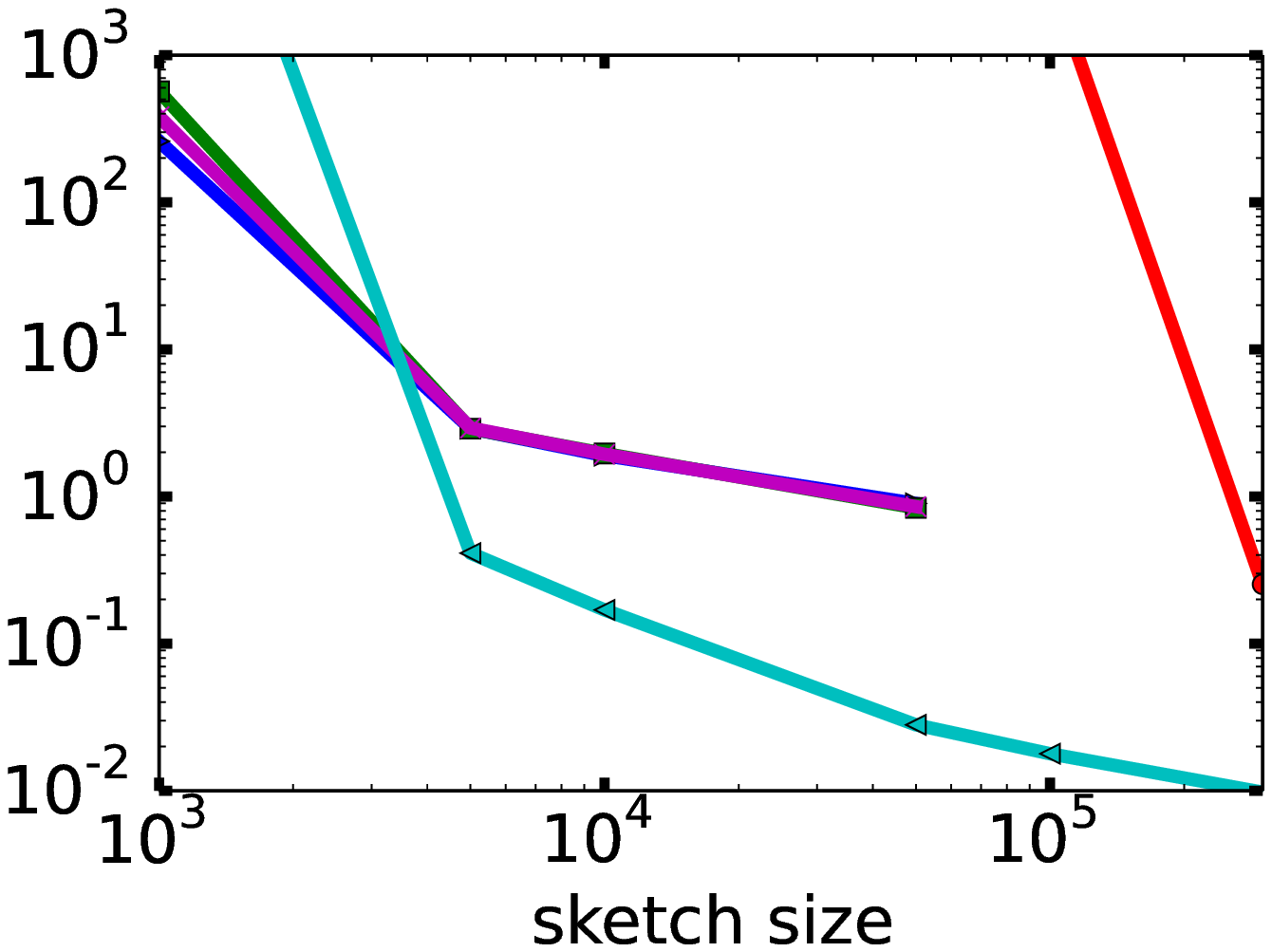}
}
&
\subfigure[$|f-f^\ast|/|f^\ast|$]{
\includegraphics[width=0.3\textwidth]{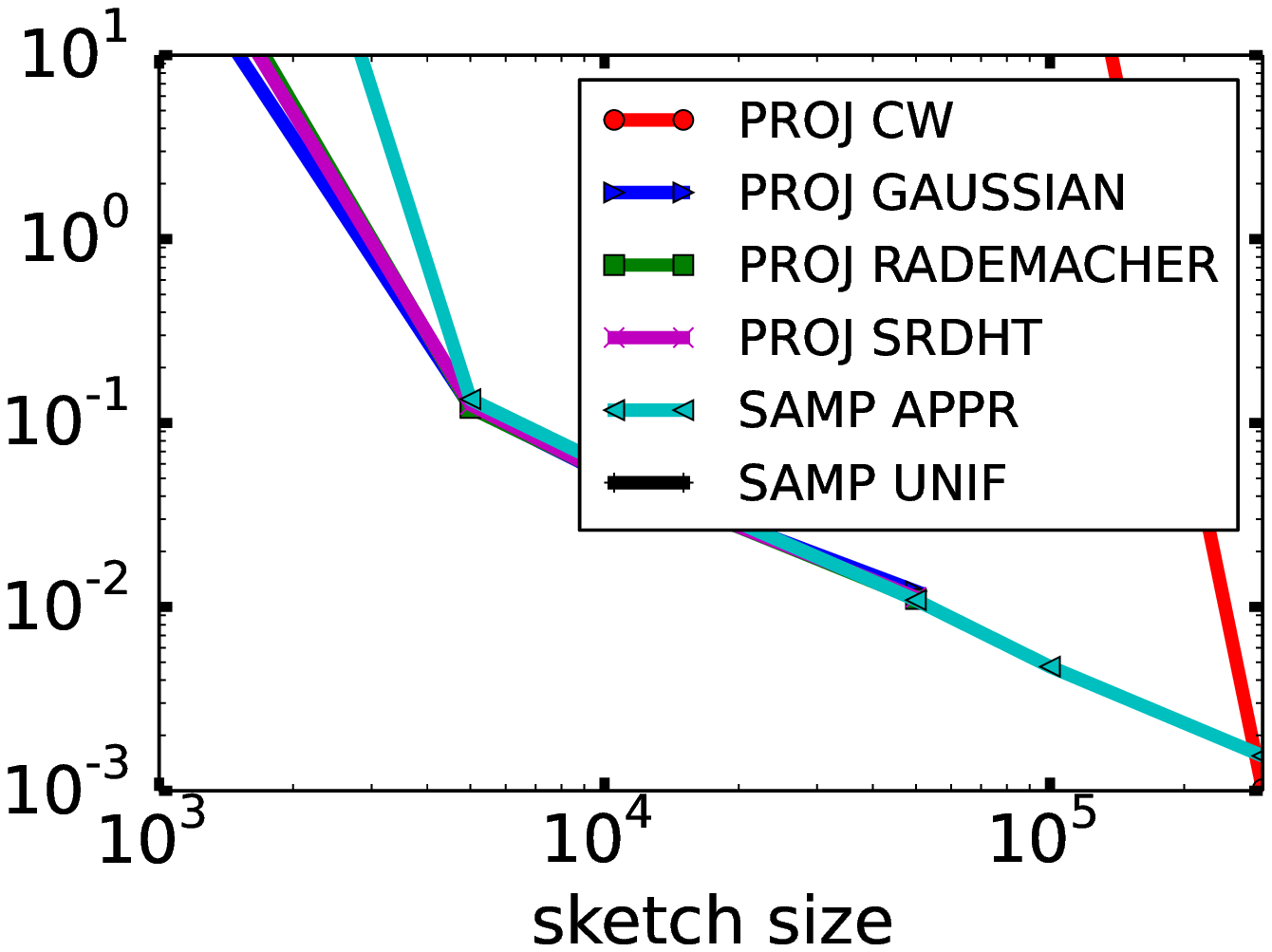}
}
&
\subfigure[Running time(sec)]{
\includegraphics[width=0.3\textwidth]{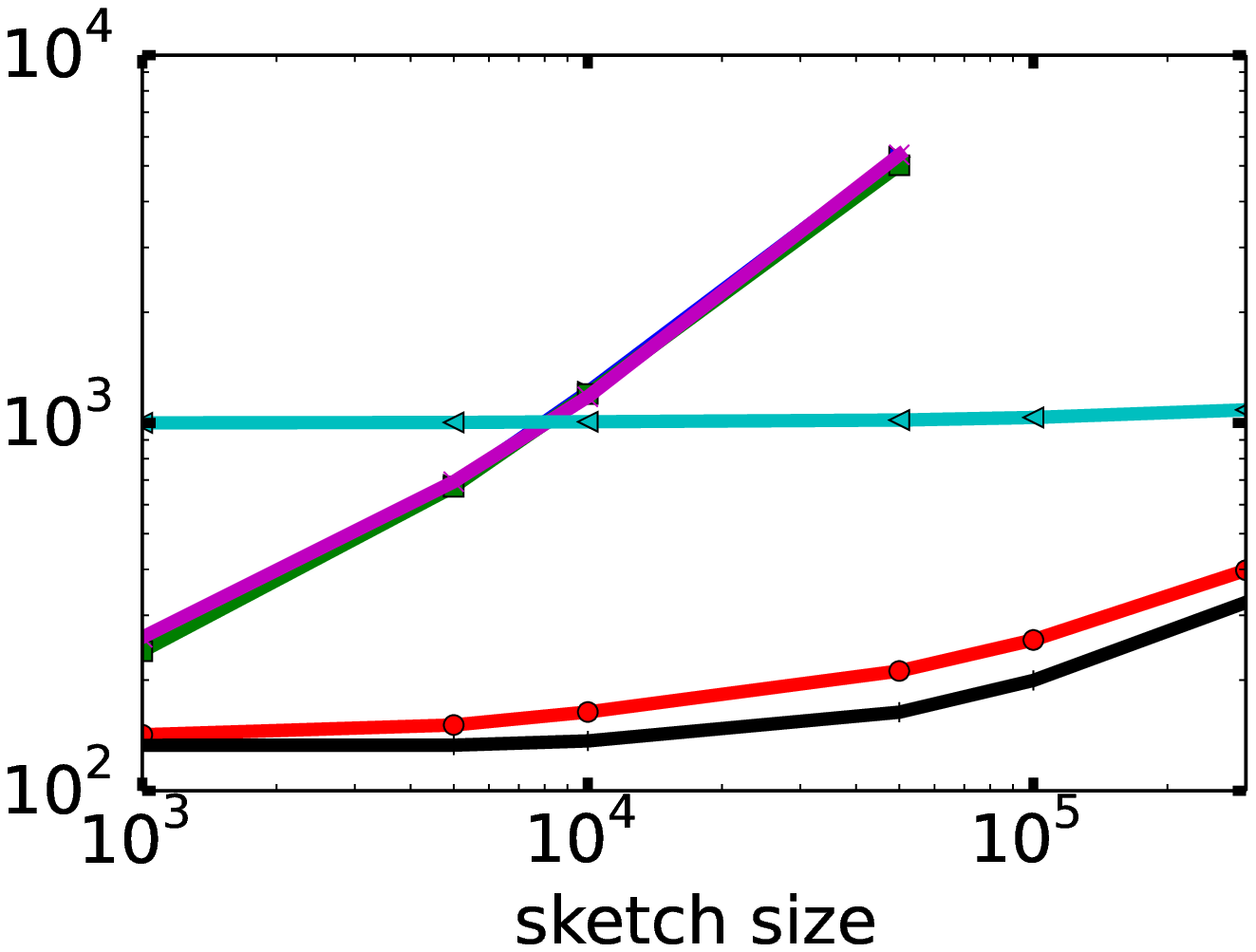}
}
\\
 \multicolumn{3}{c}{\bf $1e7 \times 1000$ NG matrix}
  \\
\subfigure[$\|x - x^\ast\|_2/\|x^\ast\|_2$]{
\includegraphics[width=0.3\textwidth]{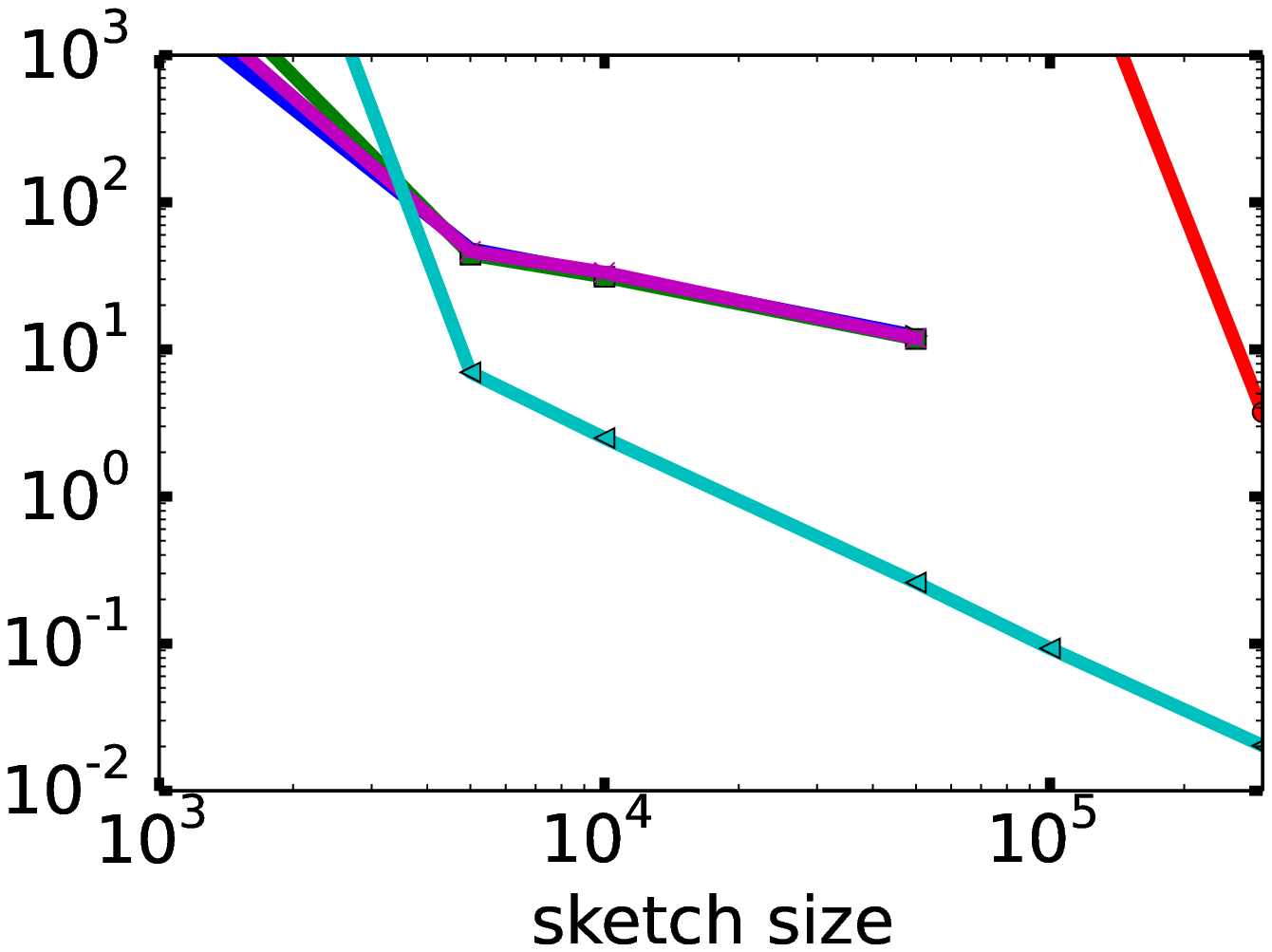}
}
&
\subfigure[$|f-f^\ast|/|f^\ast|$]{
\includegraphics[width=0.3\textwidth]{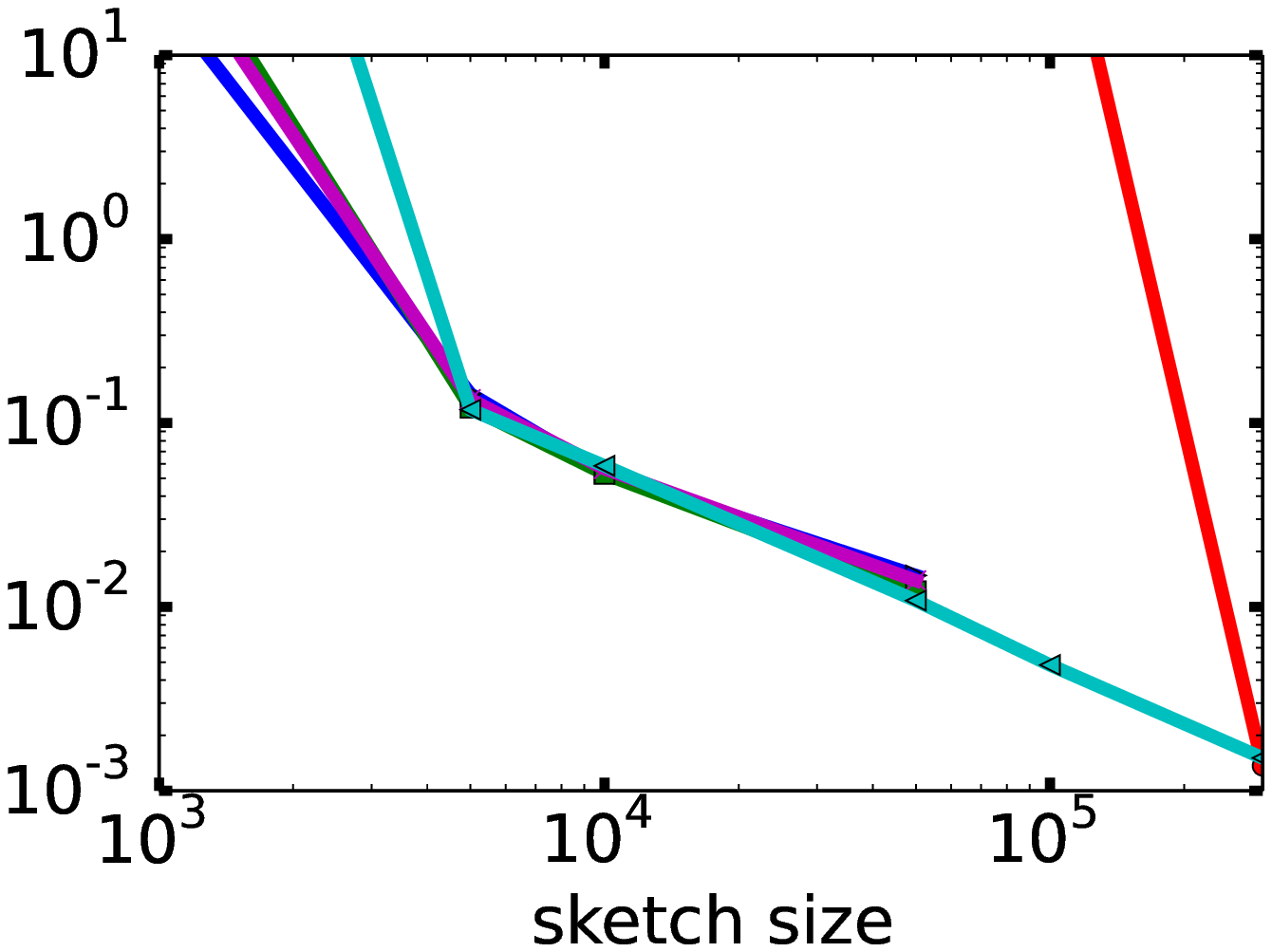}
}
&
\subfigure[Running time(sec)]{
\includegraphics[width=0.3\textwidth]{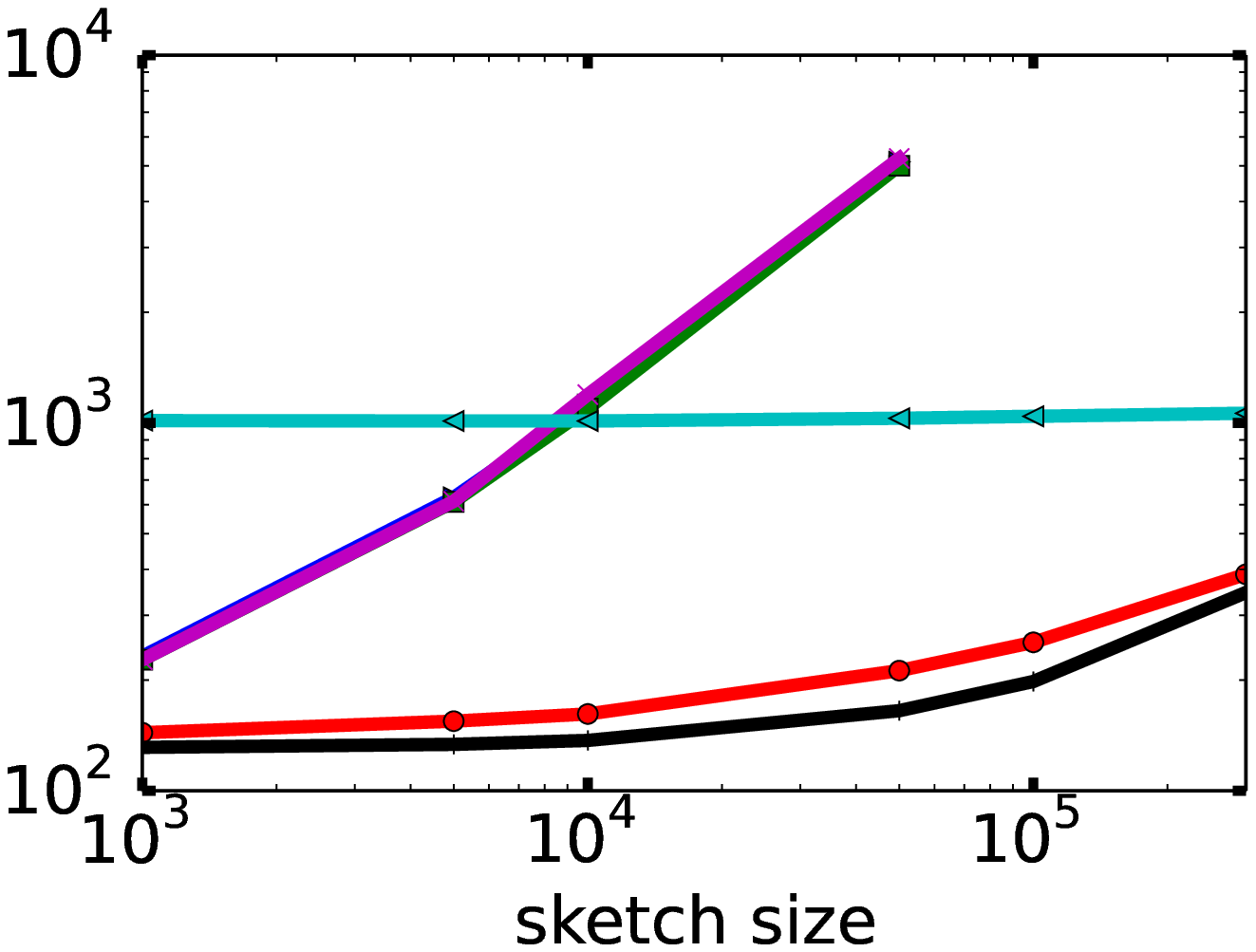}
}
\\
 \multicolumn{3}{c}{\bf $1e7 \times 1000$ NB matrix}
\end{tabular}
\end{centering}
\caption{Evaluation of all 6 of the algorithms on the 4 different types of 
matrices of size $1e7$ by $1000$.
For each method, the following three quantities are computed: 
relative error of the objective $|f-f^\ast|/f^\ast$;
relative error of the certificate $\|x-x^\ast\|_2/\|x^\ast\|_2$;
and the running time to compute the approximate solution.
Each subplot shows one of the above quantities versus the embedding dimension, respectively.
For each setting, 3 independent trials are performed and the median is reported.}
\label{fig:main}
\end{figure}

Here, we evaluate the performance of the 6 kinds of embedding methods 
described above (with different embedding dimension) on the 4 different 
types of dataset described above (with size $1e7$ by $1000$).
For dense transforms, \emph{e.g.}, \texttt{PROJ GAUSSIAN}, due to the memory 
capacity, the largest embedding dimension we can handle is $5e4$.
For each dataset and each kind of the embedding, we compute the following 
three quantities:
relative error of the objective $|f-f^\ast|/f^\ast$;
relative error of the solution certificate $\|x-x^\ast\|_2/\|x^\ast\|_2$;
and the total running time to compute the approximate solution.
The results are presented in Figure~\ref{fig:main}.

As we can see, when the matrices have uniform leverage scores,
all the methods including \texttt{SAMP UNIF} behave similarly.
As expected, \texttt{SAMP UNIF} runs fastest, followed by \texttt{PROJ CW}.
On the other hand, when the leverages scores are nonuniform,
\texttt{SAMP UNIF} breaks down even with large sampling size.
Among the projection based methods, the dense transforms, \emph{i.e.}, 
\texttt{PROJ GAUSSIAN}, \texttt{PROJ RADEMACHER} and \texttt{PROJ SRDHT}, 
behave similarly.
Although \texttt{PROJ CW} runs much faster, it yields very poor results 
until the embedding dimension is large enough, \emph{i.e.}, $c=3e5$.
Meanwhile, sampling algorithm with approximate leverage scores, \emph{i.e.}, 
\texttt{SAMP APPR}, tends to give very reliable solutions. 
(This breaks down if the embedding dimension in the approximate leverage 
score algorithm is chosen to be too small.)
In particular, the relative error is much lower throughout all choices of 
the embedding dimension.
This can be understood in terms of the theory; 
see~\cite{drineas2006sampling,DMM08_CURtheory_JRNL} 
and~\cite{drineas2011faster} for details.
In addition, its running time becomes more favorable when the embedding 
dimension is~larger.

As a more minor point, theoretical results also indicate that the upper 
bound of the relative error of the solution vector depends on the condition 
number of the system as well as the amount of mass of $b$ lies in the range 
space of $A$, denote by $\gamma$~\cite{DMMW12_JMLR}.
Across the four datasets, $\gamma$ is roughly the same.
This is why we see the relative error of the certificate, \emph{i.e.}, the
vector achieving the minimum solution, tends to be larger when the condition 
number of the matrix becomes higher. 

\subsubsection{Quality of the approximate leverage scores}
\label{sec:appr_lev}

Here, we evaluate the quality of the fast approximate leverage score 
algorithm of \cite{DMMW12_JMLR}, and we investigate the quality of the 
approximate leverage scores with several underlying embeddings.
(The algorithm of \cite{DMMW12_JMLR} considered only Hadamard-based 
projections, but other projection methods could be used, leading to 
similar approximation quality but different running times.)
We consider only an NB matrix since leverage scores with nonuniform 
distributions are harder to approximate.
In addition, the size of the matrix we considered is only rather small, 
$1e6$ by $500$, due to the need to compute the exact leverage scores for 
comparison.
Our implementation follows closely the main algorithm 
of~\cite{DMMW12_JMLR}, except that we consider other random projection 
matrices.
In particular, we used the following four ways to compute the underlying 
embedding: namely, \texttt{PROJ CW}, \texttt{PROJ GAUSSIAN}, 
\texttt{PROJ RADEMACHER}, and \texttt{PROJ SRDHT}.
For each kind of embedding and embedding dimension, we compute a series of 
quantities which characterize the statistical properties of the approximate 
leverage scores.
The results are summarized in Table~\ref{table:lev}.

\begin{table}[H]
\begin{center}
\begin{sc}
\small
\begin{tabular}{c|cccc}
  $c$ & PROJ CW & PROJ GAUSSIAN & PROJ RADEMACHER & PROJ SRDH \\
\hline
 & \multicolumn{4}{c}{$ \| \hat p - p^\ast \|_2/\|p^\ast\|_2$} \\
\hline
  5e2 & 0.9205 & 0.7738 & 0.7510 & 0.5008\\
  1e3 & 0.9082 & 0.0617 & 0.0447 & 0.0716\\
  5e3 & 0.9825 & 0.0204 & 0.0072 & 0.0117\\
  1e4 & 0.9883 & 0.0143 & 0.0031 & 0.0075\\
  5e4 & 0.9962 & 0.0061 & 0.0006 & 0.0030 \\
  1e5 & 0.0016 & 0.0046 & 0.0003 & 0.0023\\
\hline
 & \multicolumn{4}{c}{ $D_{KL}(p^\ast \vert \vert \hat p)$ } \\
\hline
  5e2 & 18.5241 & 0.0710 & 0.6372 & 0.1852\\
  1e3 & 19.7773 & 0.0020 & 0.0015 & 0.0029\\
  5e3 & 20.3450 & 0.0002 & 0.0001 & 0.0001\\
  1e4 & 20.0017 & 0.0001 & 0.0001 & 0.0001\\
  5e4 & 19.2417 & 1.9e-5 & 1.0e-5 & 1.0e-5\\
  1e5 & 0.0001 & 1.0e-5 & 5e-6 & 5e-6\\
\hline
 & \multicolumn{4}{c}{$\alpha_L = \max_i\{\hat p^L_i/p_i^{\ast,L}\}$} \\
\hline
  5e2 & 28.6930 & 7.0267 & 7.3124 & 4.0005\\
  1e3 & 11.4425 & 1.1596 & 1.1468 & 1.2201\\
  5e3 & 50.3311 & 1.0584 & 1.0189 & 1.0379\\
  1e4 & 82.6574 & 1.0449 & 1.0099 & 1.0199\\
  5e4 & 218.9658 & 1.0192 & 1.0018 & 1.0094 \\ 
  1e5 & 1.0016 & 1.0108 & 1.0009 & 1.0060\\
\hline
 & \multicolumn{4}{c}{$\alpha_S = \max_i\{\hat p^S_i/p_i^{\ast,S}\}$ } \\
\hline
  5e2 & 0 & 24.4511 & 16.8698 & 4.5227\\
  1e3 & 0 & 1.3923 & 1.3718 & 1.3006\\
  5e3 & 0 & 1.1078 & 1.1040 & 1.1077\\
  1e4 & 0 & 1.0743 & 1.0691 & 1.0698\\
  5e4 & 0 & 1.0332 & 1.0317 & 1.0310\\
  1e5 & 1.0236 & 1.0220 & 1.0218 & 1.0198\\
\hline
 & \multicolumn{4}{c}{$\beta_L = \min_i\{\hat p^L_i/p_i^{\ast,L}\}$} \\
\hline
  5e2 & 0 & 0.0216 & 0.0448 & 0.4094\\
  1e3 & 0 & 0.8473 & 0.8827 & 0.8906\\
  5e3 & 0 & 0.9456 & 0.9825 & 0.9702\\
  1e4 & 0 & 0.9539 & 0.9916 & 0.9827\\
  5e4 & 0 & 0.9851 & 0.9982 & 0.9922 \\
  1e5 & 0.9969 & 0.9878 & 0.9993 & 0.9934\\
\hline
 & \multicolumn{4}{c}{$\beta_S = \min_i\{\hat p^S_i/p_i^{\ast,S}\}$ } \\
\hline
  5e2 & 0 & 0.0077 & 0.0141 & 0.1884\\
  1e3 & 0 & 0.7503 & 0.7551 & 0.7172\\
  5e3 & 0 & 0.9037 & 0.9065 & 0.9065\\
  1e4 & 0 & 0.9328 & 0.9306 & 0.9356\\
  5e4 & 0 & 0.9704 & 0.9691 & 0.9710\\
  1e5 & 0.9800 & 0.9787 & 0.9789 & 0.9803
\end{tabular}
\end{sc}
\end{center}
\caption{Quality of the approximate leverage scores.
The test was performed on an NB matrix with size $1e6$ by $500$.
In above, $\hat p$ denotes the distribution by normalizing the approximate 
leverage scores and $p^\ast$ denotes the exact leverage score distribution.
$D_{KL}(p \vert \vert q)$ is the KL divergence~\cite{KL51} of $q$ from $p$ defined as 
$\sum_i p_i \ln \frac{p_i}{q_i}$.
Let $L = \{ i \vert p^\ast_i = 1 \}$ and $S = \{ i \vert p^\ast_i < 1 \}$.
In this case, $\hat p^L$ denotes the corresponding slice of $\hat p$, and
the quantities $\hat p^S, p^{\ast,L}, p^{\ast.S}$ are defined similarly. }
\label{table:lev}
\end{table}

As we can see, when the projection size is large enough, all the 
projection-based methods to compute approximations to the leverage scores 
produce highly accurate leverage scores.
Among these projection methods, \texttt{PROJ CW} is typically faster but 
also requires a much larger projection size in order to yield reliable 
approximate leverage scores.
The other three random projections perform similarly.
In general, the algorithms approximate the large leverage scores (those that
equal or are close to $1$) better than the small leverage scores, since 
$\alpha_L$ and $\beta_L$ are closer to $1$.
This is crucial when calling \texttt{SAMP APPR} since the important rows 
shall not be missed, and it is a sufficient condition for the theory 
underlying the algorithm of~\cite{DMMW12_JMLR} to apply.

\begin{figure}[H]
\begin{centering}
\begin{tabular}{ccc}
\subfigure[$\|x - x^\ast\|_2/\|x^\ast\|_2$]{
\includegraphics[width=0.3\textwidth]{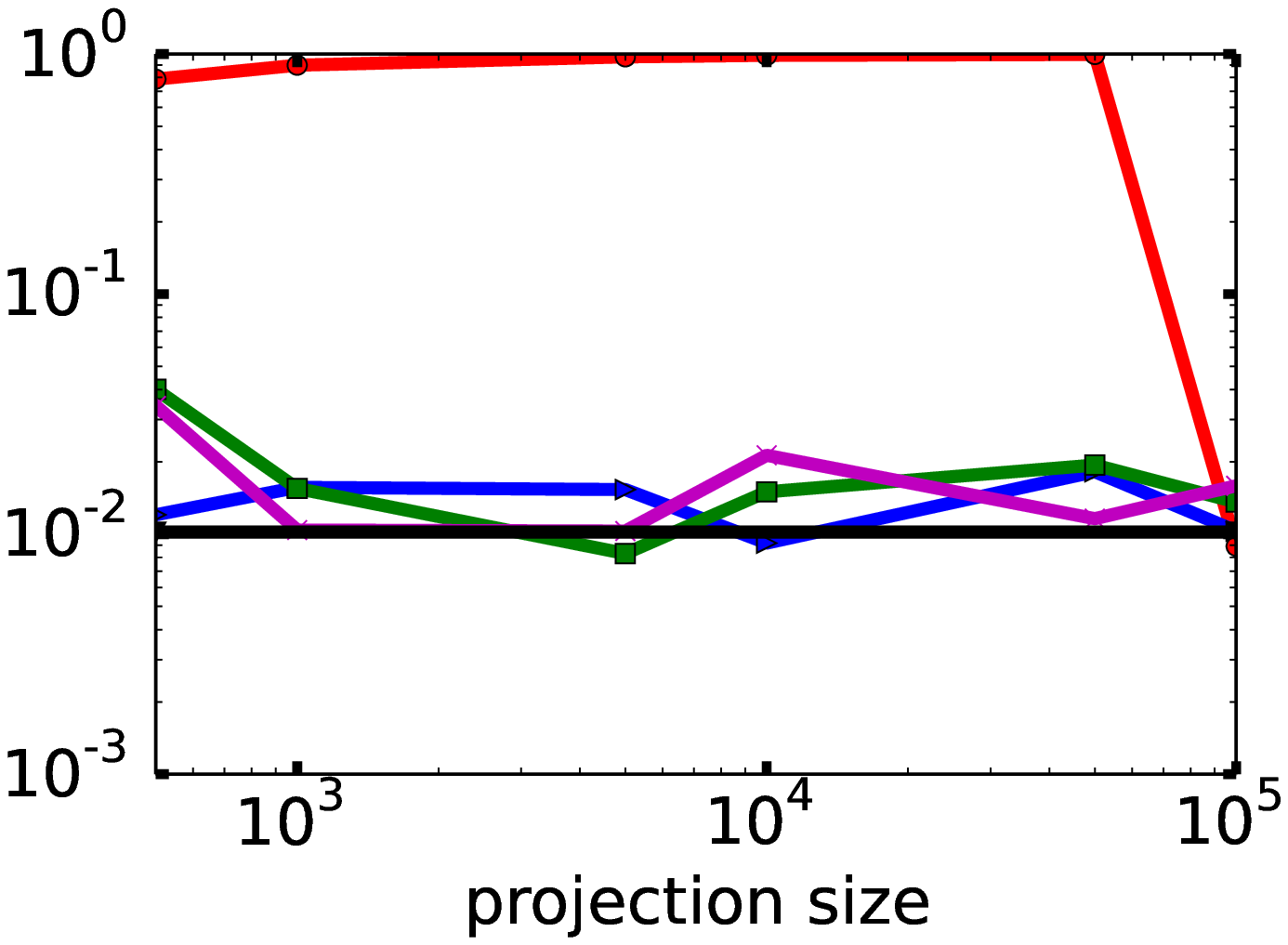}
}
&
\subfigure[$|f-f^\ast|/|f^\ast|$]{
\includegraphics[width=0.3\textwidth]{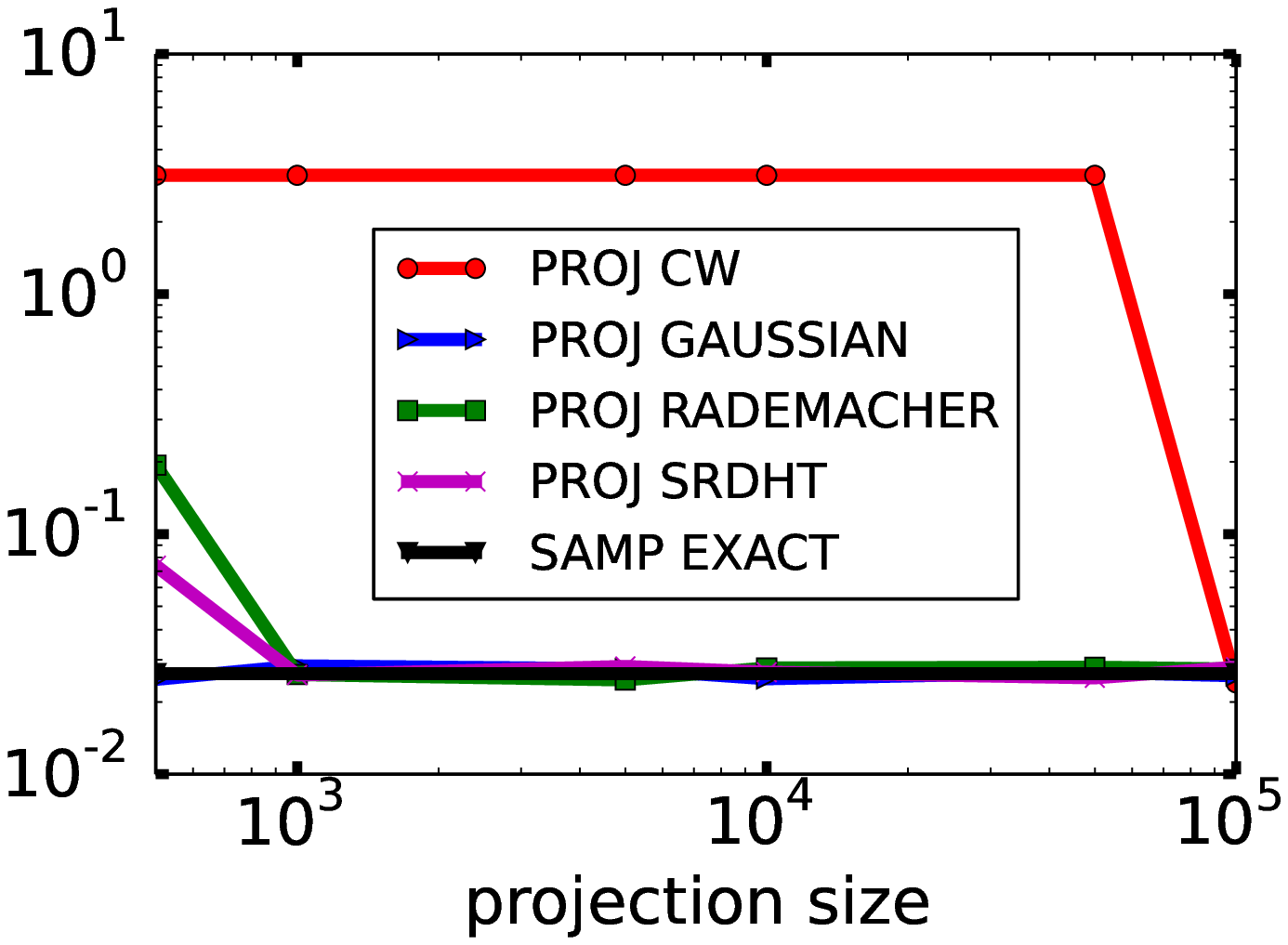}
}
&
\subfigure[Running time(sec)]{
\includegraphics[width=0.3\textwidth]{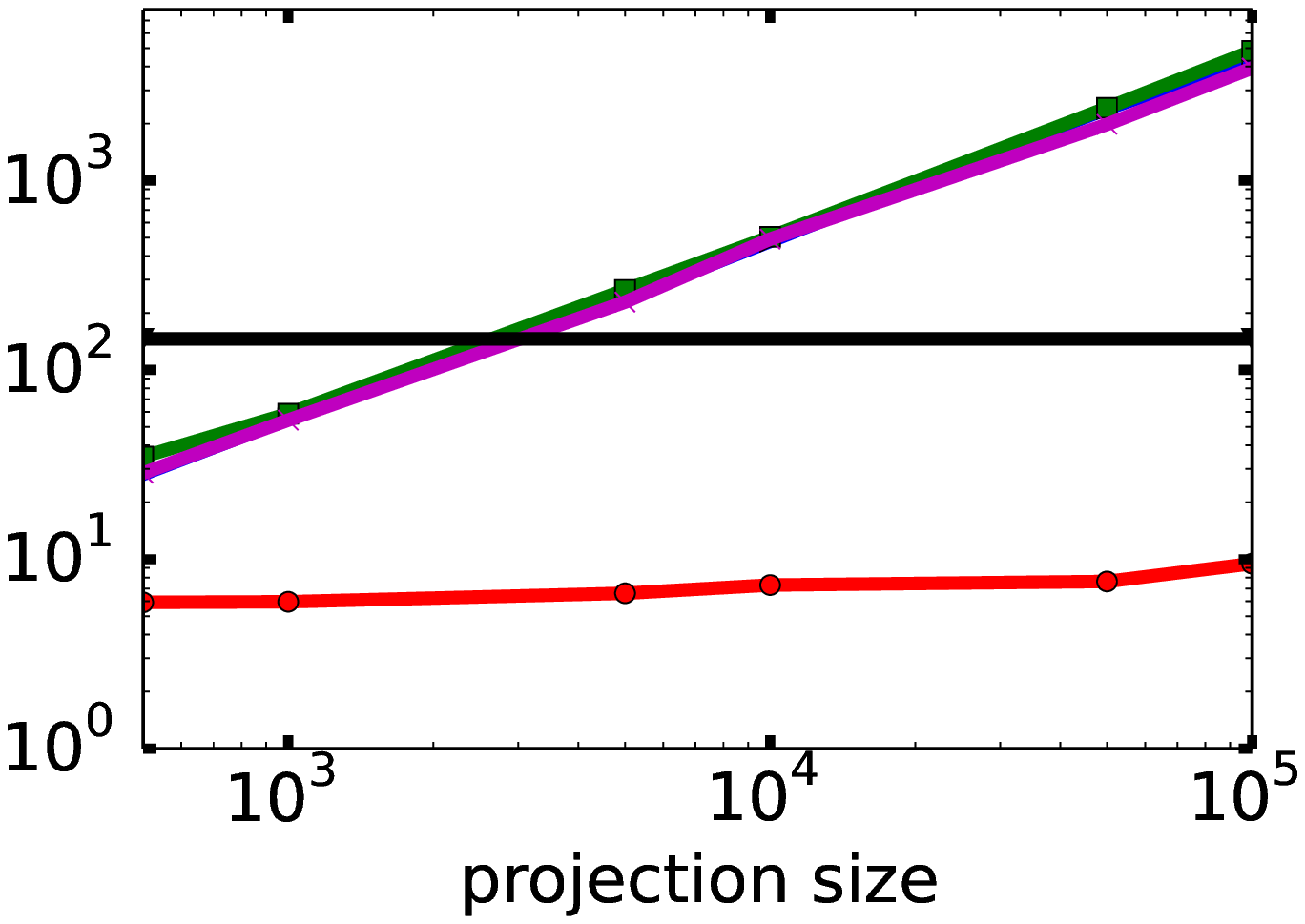}
}
\end{tabular}
\end{centering}
\caption{ 
Performance of sampling algorithms with approximate leverage scores, as 
computed by several different underlying projections. 
The test was performed on an NB matrix of size $1e6$ by $500$ and the 
sampling size was $1e4$.
Each subplot shows one of the following three quantities versus the 
projection size used in the underlying random projection phase:
relative error of the objective $|f-f^\ast|/f^\ast$;
relative error of the certificate $\|x-x^\ast\|_2/\|x^\ast\|_2$;
and the running time.
For each setting, 5 independent trials are performed and the median is 
reported.}
\label{fig:lev}
\end{figure}

Next, we invoke the sampling algorithm for the $\ell_2$ regression problem, 
with sampling size $s=1e4$ by using these approximate leverage scores.
We evaluate the relative error on both the solution vector and objective and 
the total running time.
For completeness and in order to evaluate the quality of the approximate 
leverage score algorithm, we also include the results by using the exact 
leverage scores.
The results are presented in Figure~\ref{fig:lev}.

These results suggest that the precise quality of the approximate leverage 
scores does not substantially affect the downstream error, \emph{i.e.}, 
sampling-based algorithms are robust to imperfectly-approximated leverage 
scores, as long as the largest scores are not too poorly approximated.
(Clearly, however, we could have chosen parameters such that some of the 
larger scores were very poorly approximated, \emph{e.g.}, by choosing the 
embedding dimension to be too small, in which case the quality would matter.
In our experience, the quality matters less since these approximate 
leverage scores are sufficient to solve $\ell_2$ regression problems.)
Finally, and importantly, note that the solution quality obtained by using 
approximate leverage scores is as good as that of using exact leverage 
scores, while the running time can be much less.



\subsubsection{Performance of low-precision solvers when $n$ changes}

\begin{figure}[h]
\begin{centering}
\begin{tabular}{ccc}
\subfigure[$\|x - x^\ast\|_2/\|x^\ast\|_2$]{
\includegraphics[width=0.3\textwidth]{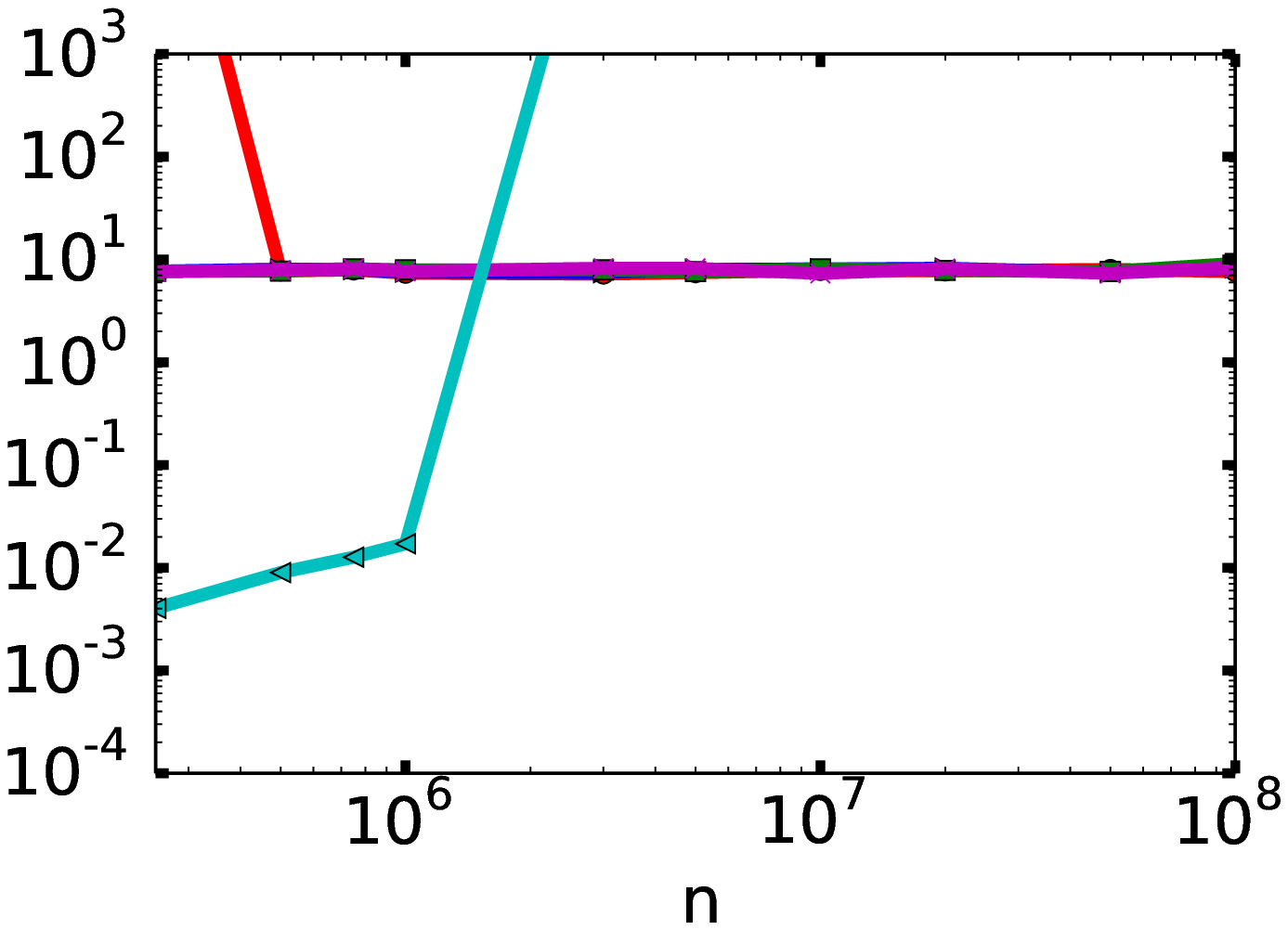}
}
&
\subfigure[$|f-f^\ast|/|f^\ast|$]{
\includegraphics[width=0.3\textwidth]{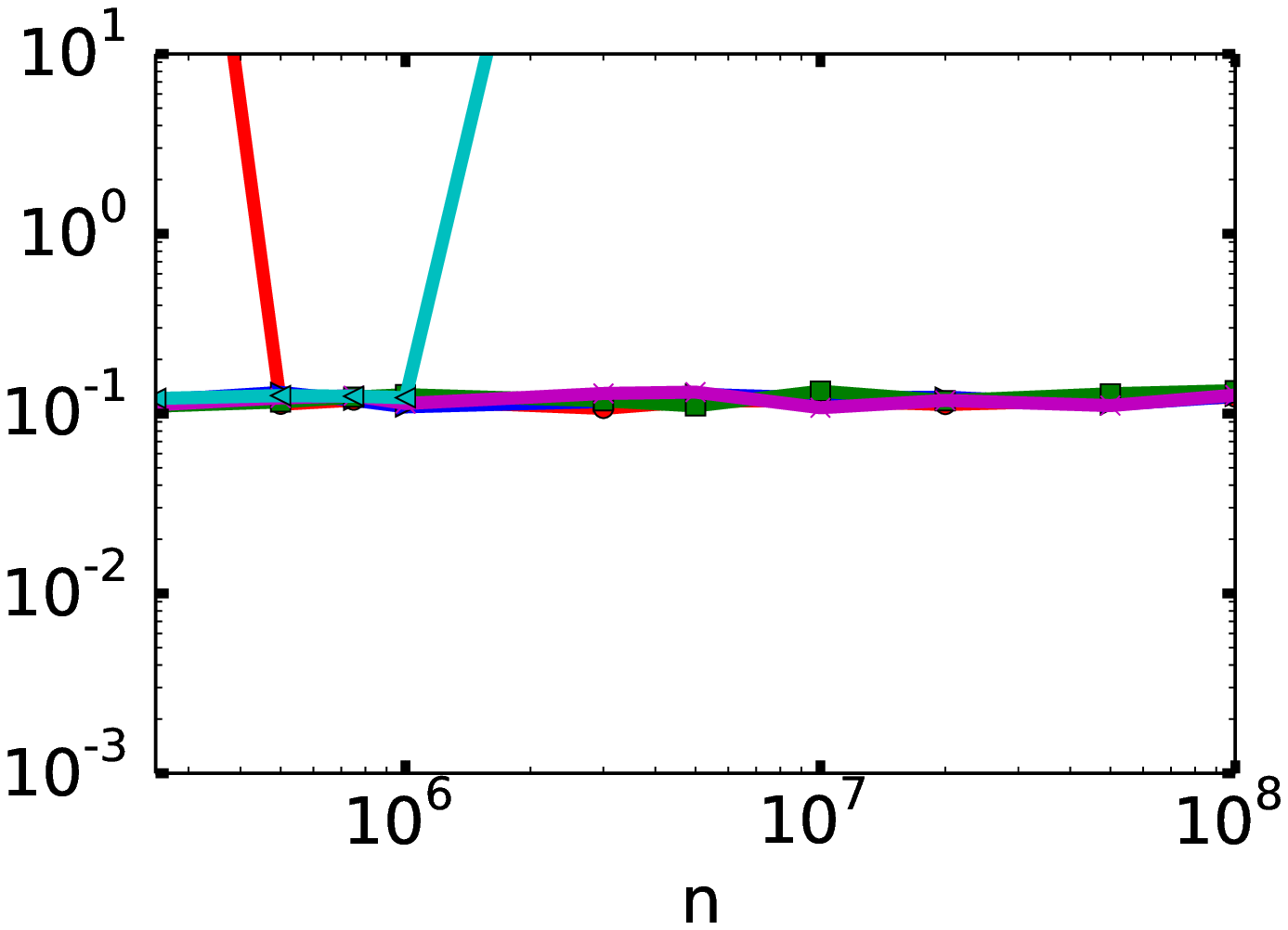}
}
&
\subfigure[Running time(sec)]{
\includegraphics[width=0.3\textwidth]{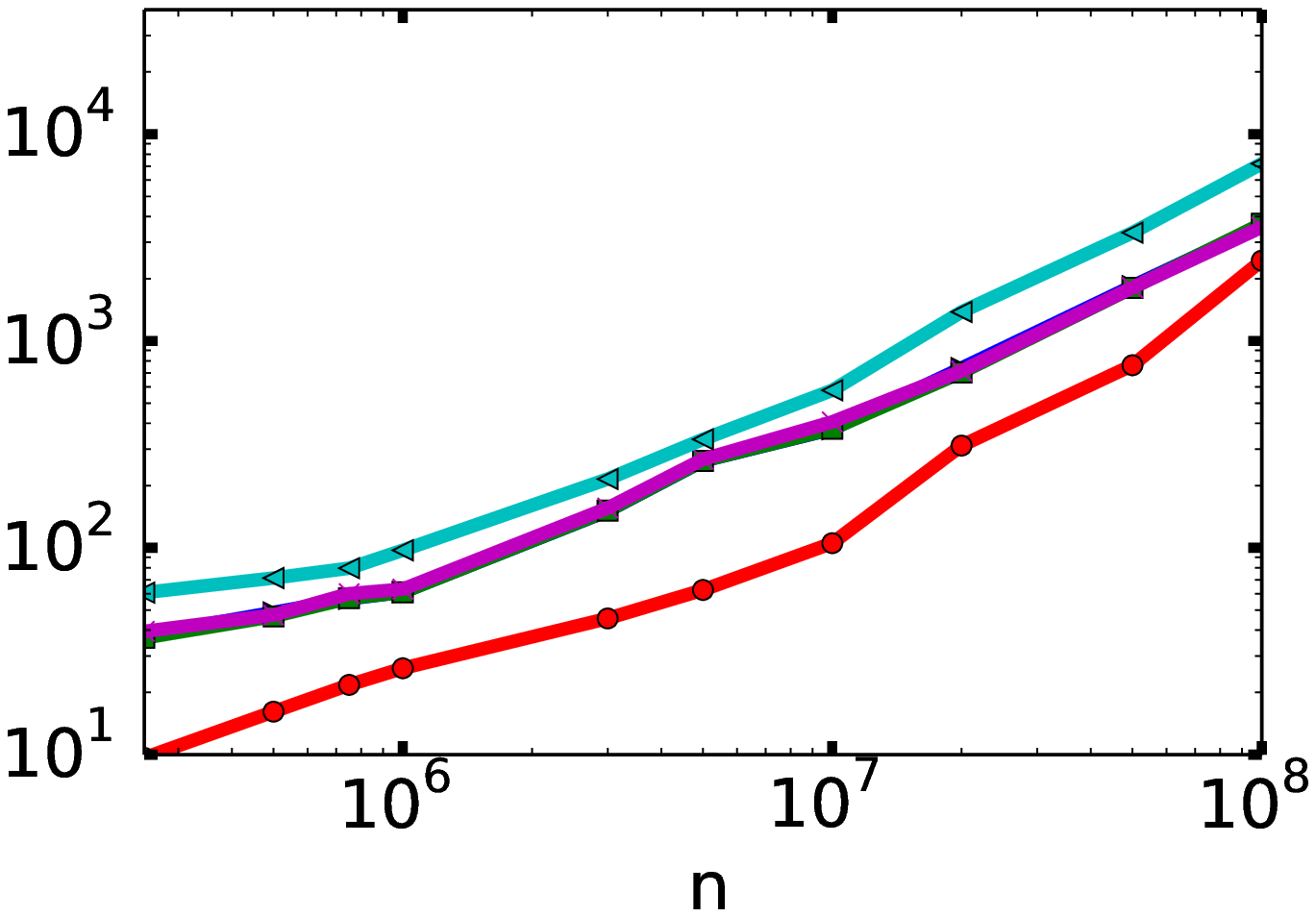}
}
\\
 \multicolumn{3}{c}{\bf $s=5e3$}
 \\
\subfigure[$\|x - x^\ast\|_2/\|x^\ast\|_2$]{
\includegraphics[width=0.3\textwidth]{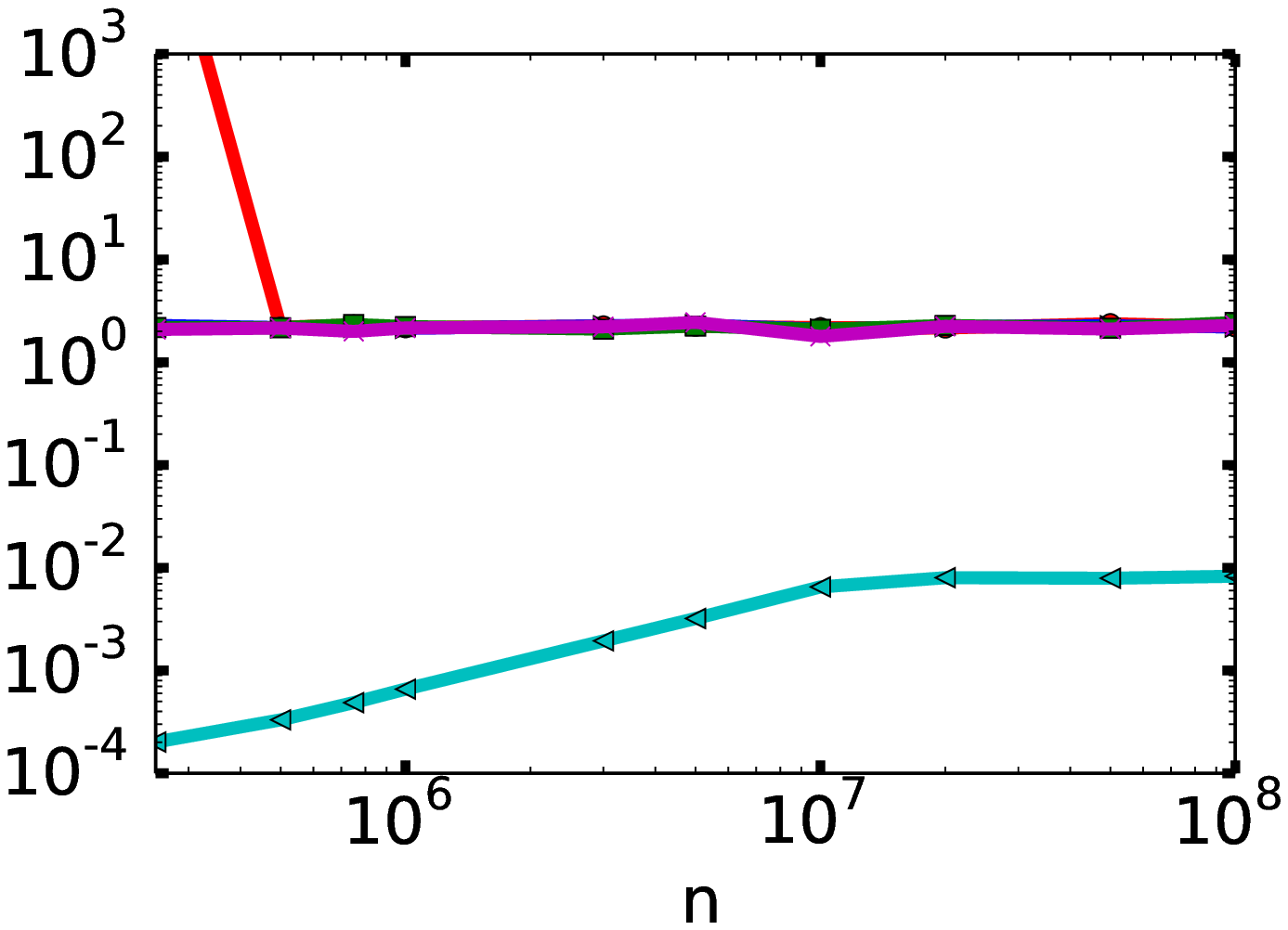}
}
&
\subfigure[$|f-f^\ast|/|f^\ast|$]{
\includegraphics[width=0.3\textwidth]{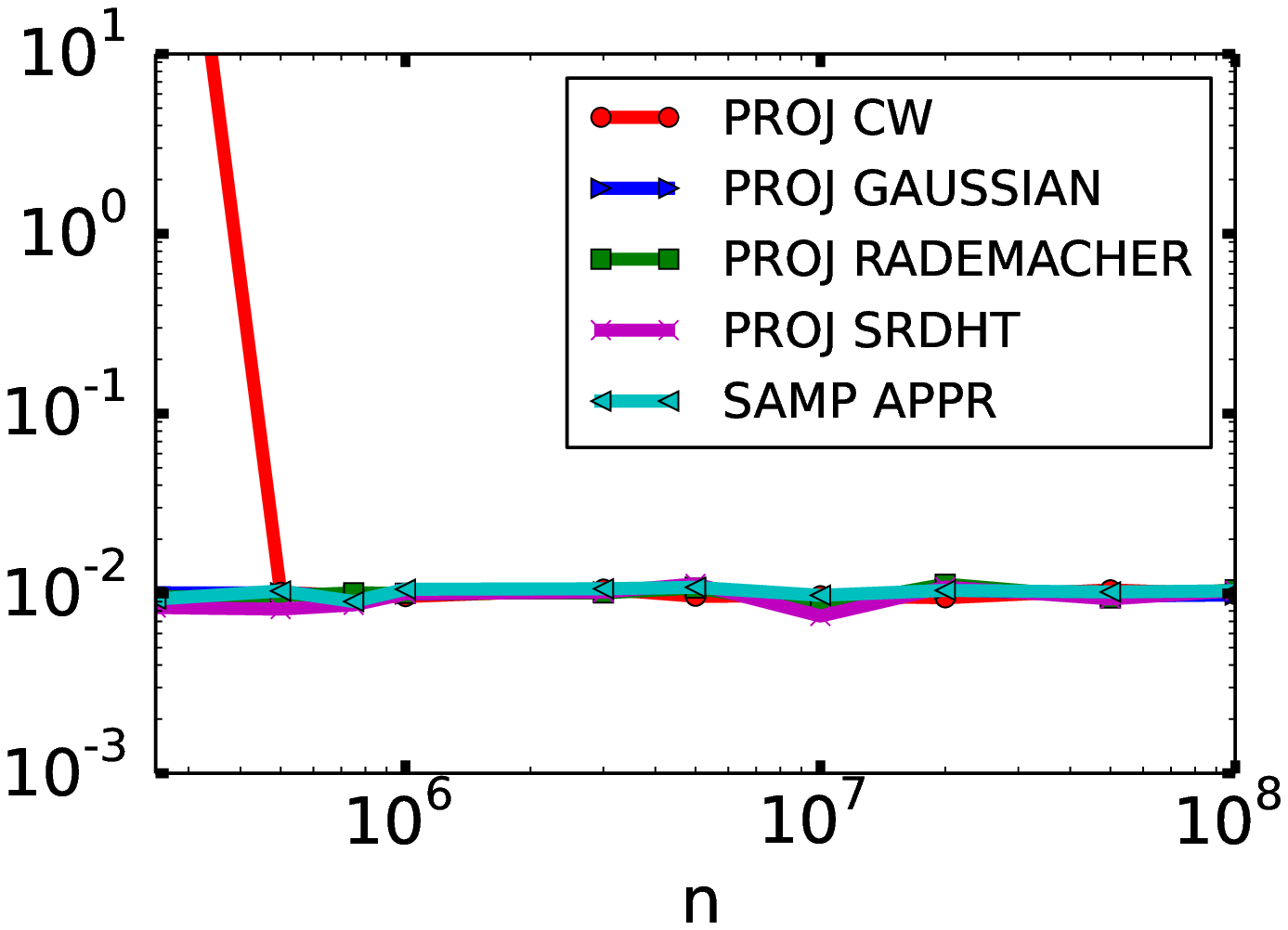}
}
&
\subfigure[Running time(sec)]{
\includegraphics[width=0.3\textwidth]{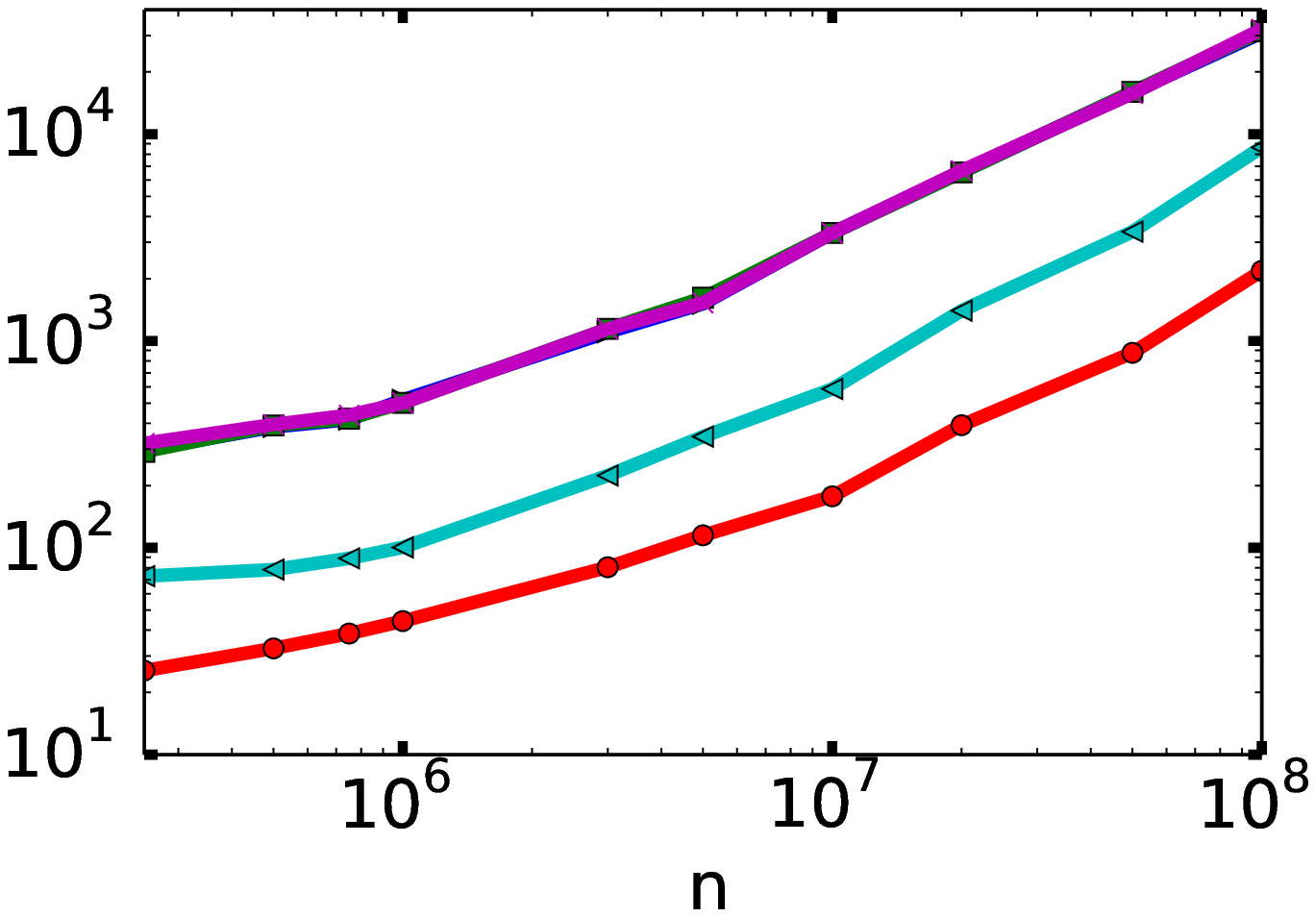}
}
\\
 \multicolumn{3}{c}{\bf $s=5e4$}
 \\
\end{tabular}
\end{centering}
\caption{ Performance of all the algorithms on NB matrices with varying $n$ from $2.5e5$ to $1e8$ and fixed $d = 1000$.
The matrix is generated using \texttt{STACK1}.
For each method, the embedding dimension is fixed to be $5e3$ or $5e4$.
The following three quantities are computed: 
relative error of the objective $|f-f^\ast|/f^\ast$;
relative error of the certificate $\|x-x^\ast\|_2/\|x^\ast\|_2$;
and the running time to compute the approximate solution.
For each setting, 3 independent trials are performed and the median is reported.}
\label{fig:n}
\end{figure}

Here, we explore the scalability of the low-precision solvers by evaluating
the performance of all the embeddings on NB matrices with varying $n$.
We fix $d=1000$ and let $n$ take values from $2.5e5$ to $1e8$.
These matrices are generated by stacking an NB matrix with size $2.5e5$ by 
$1000$ \texttt{REPNUM} times, with \texttt{REPNUM} varying from $1$ to $400$ 
using \texttt{STACK1}.
For conciseness, we fix the embedding dimension of each method to be either 
$5e3$ or $5e4$.
The relative error on certificate and objective and running time are 
evaluated.
The results are presented in Figure~\ref{fig:n}.

Especially worthy mentioning is that when using \texttt{STACK1}, by increasing \texttt{REPNUM}, as we 
pointed out, the coherence of the matrix, \emph{i.e.}, the maximum leverage 
score, is decreasing, as the size is increased.
We can clearly see that, when $n=2.5e5$, \emph{i.e.}, the coherence is $1$, 
\texttt{PROJ CW} fails.
Once the coherence gets smaller, \emph{i.e.}, $n$ gets larger, the 
projection-based methods behave similarly and the relative error remains 
roughly the same as we increased $n$.
This is because \texttt{STACK1} doesn't alter the condition number and
the amount of mass of the right hand side vector that lies in the range 
space of the design matrix and the lower dimension $d$ remains the same.
However, \texttt{SAMP APPR} tends to yield larger error on approximating the 
certificate as we increase \texttt{REPNUM}, \emph{i.e.}, the coherence gets 
smaller.
Moreover, it breaks down when the embedding dimension is very small.

\subsubsection{Performance of low-precision solvers when $d$ changes}

\begin{figure}[h]
\begin{centering}
\begin{tabular}{ccc}
\subfigure[$\|x - x^\ast\|_2/\|x^\ast\|_2$]{
\includegraphics[width=0.3\textwidth]{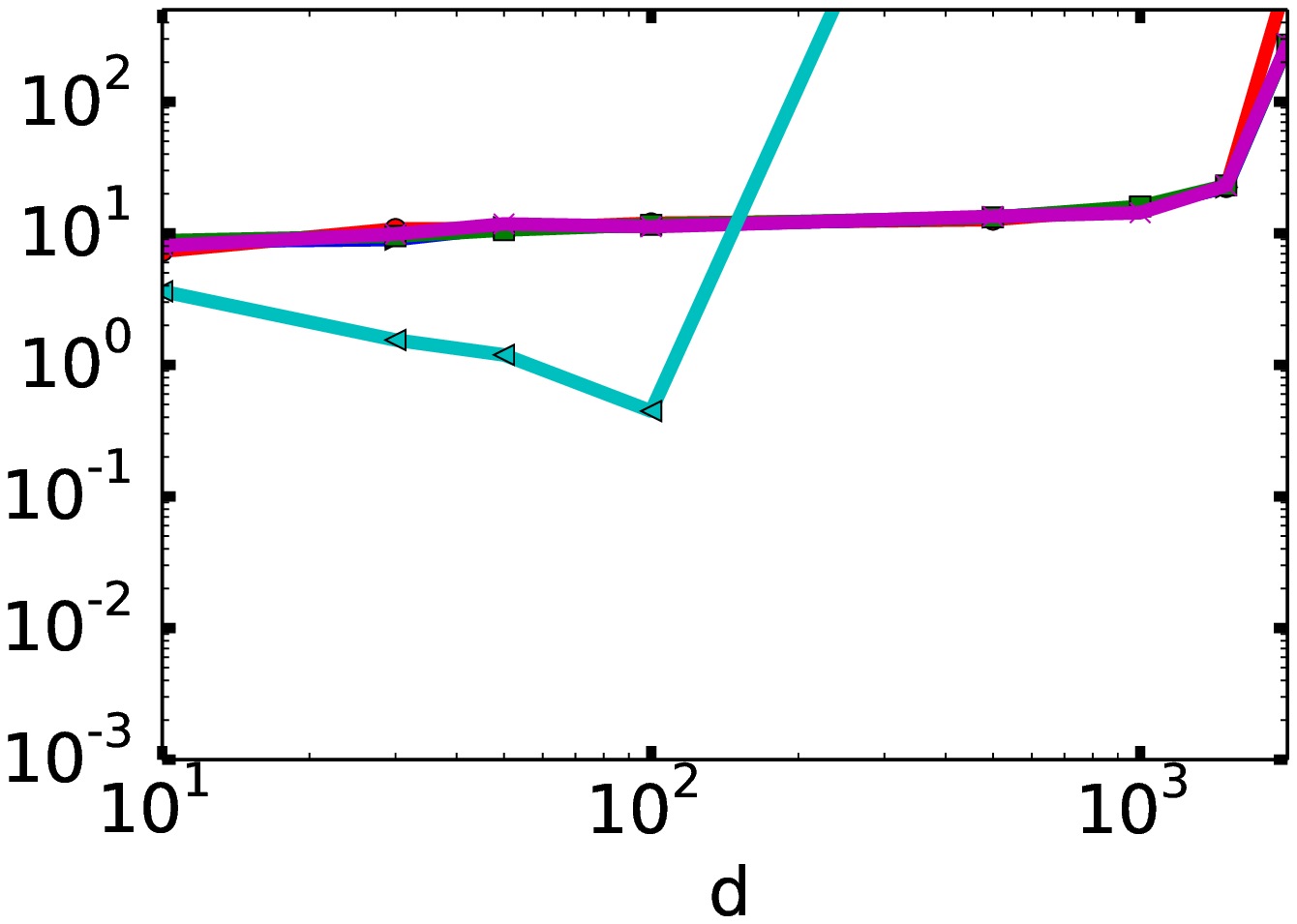}
}
&
\subfigure[$|f-f^\ast|/|f^\ast|$]{
\includegraphics[width=0.3\textwidth]{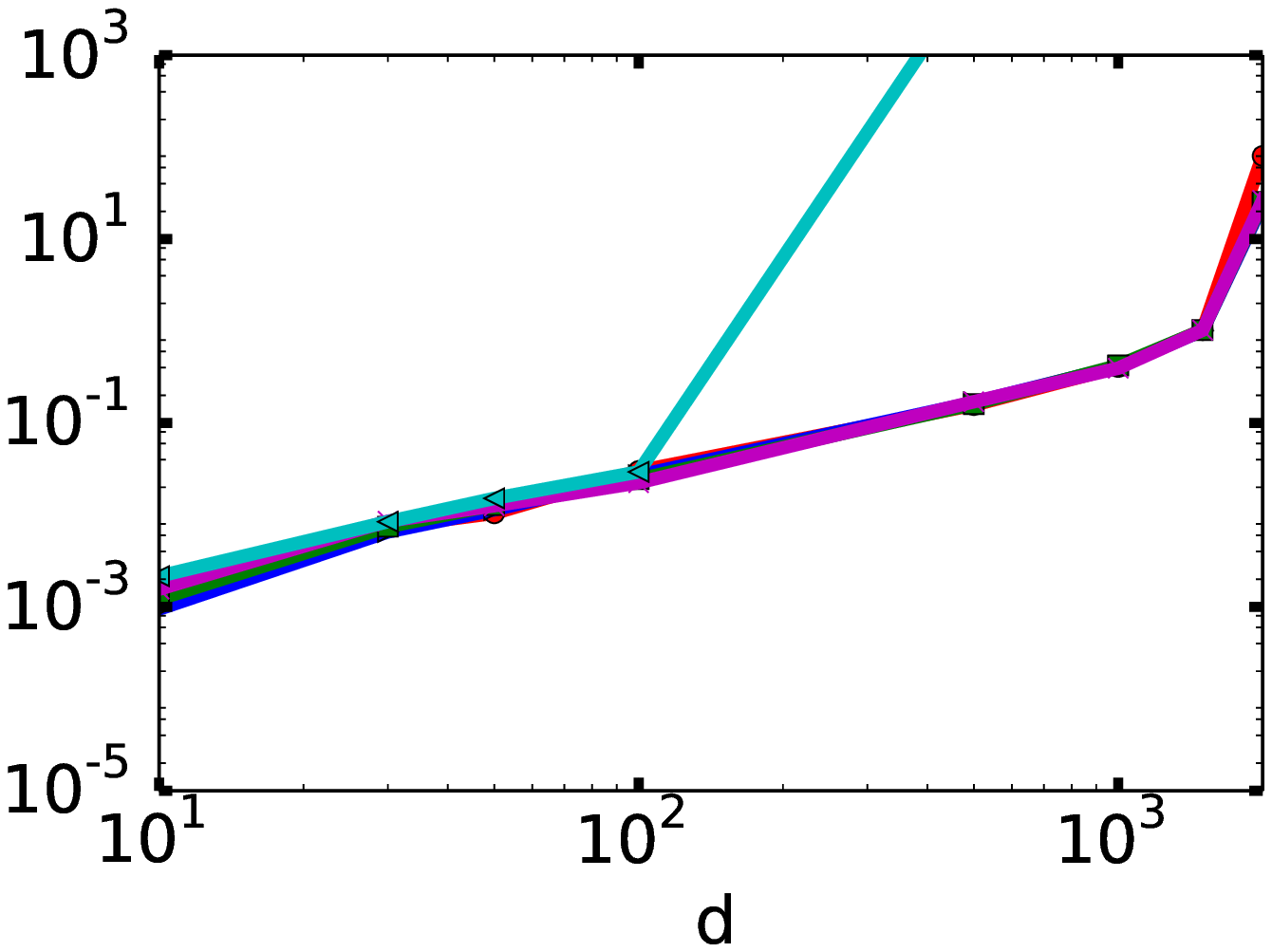}
}
&
\subfigure[Running time(sec)]{
\includegraphics[width=0.3\textwidth]{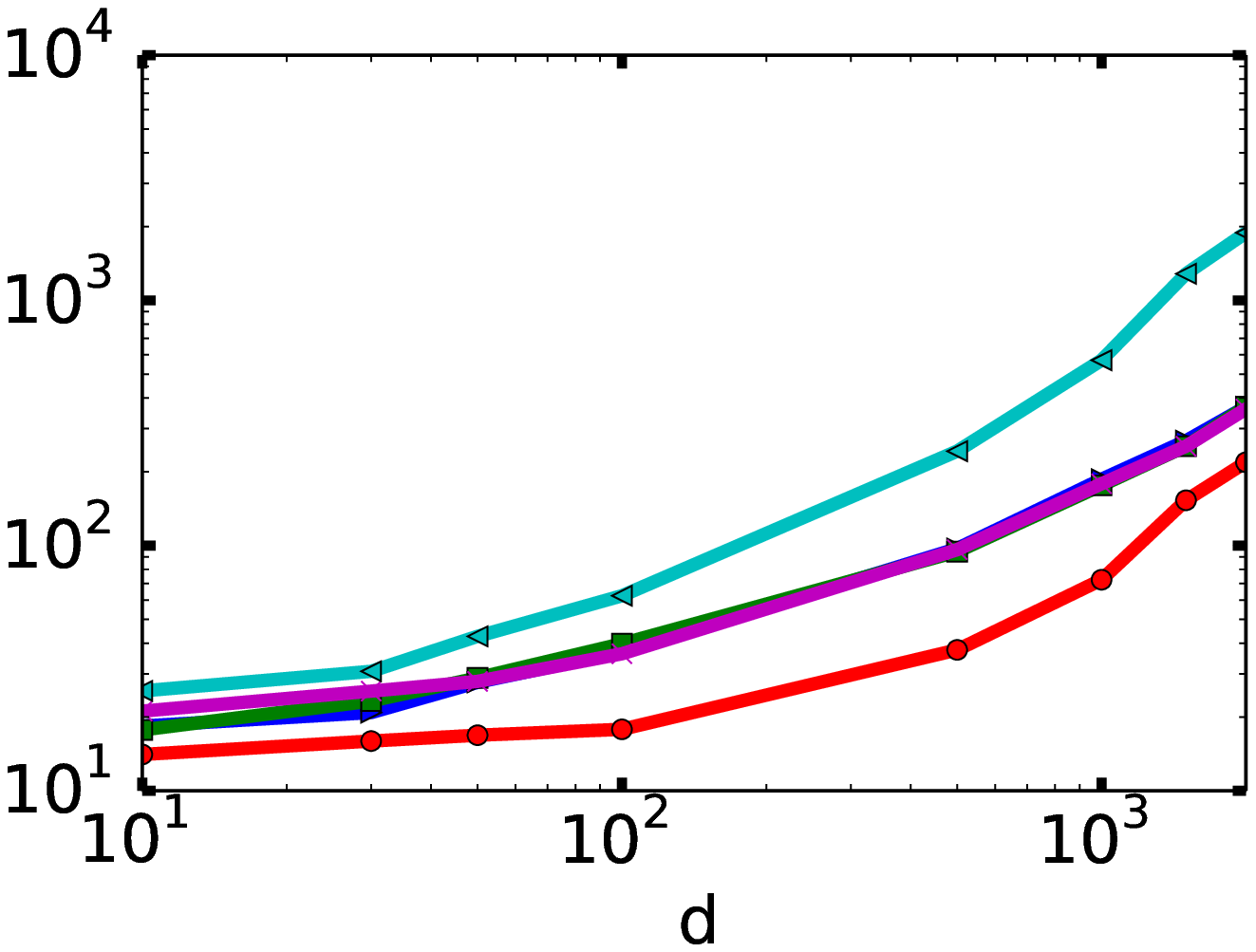}
}
\\
 \multicolumn{3}{c}{\bf $s=2e3$}
 \\
\subfigure[$\|x - x^\ast\|_2/\|x^\ast\|_2$]{
\includegraphics[width=0.3\textwidth]{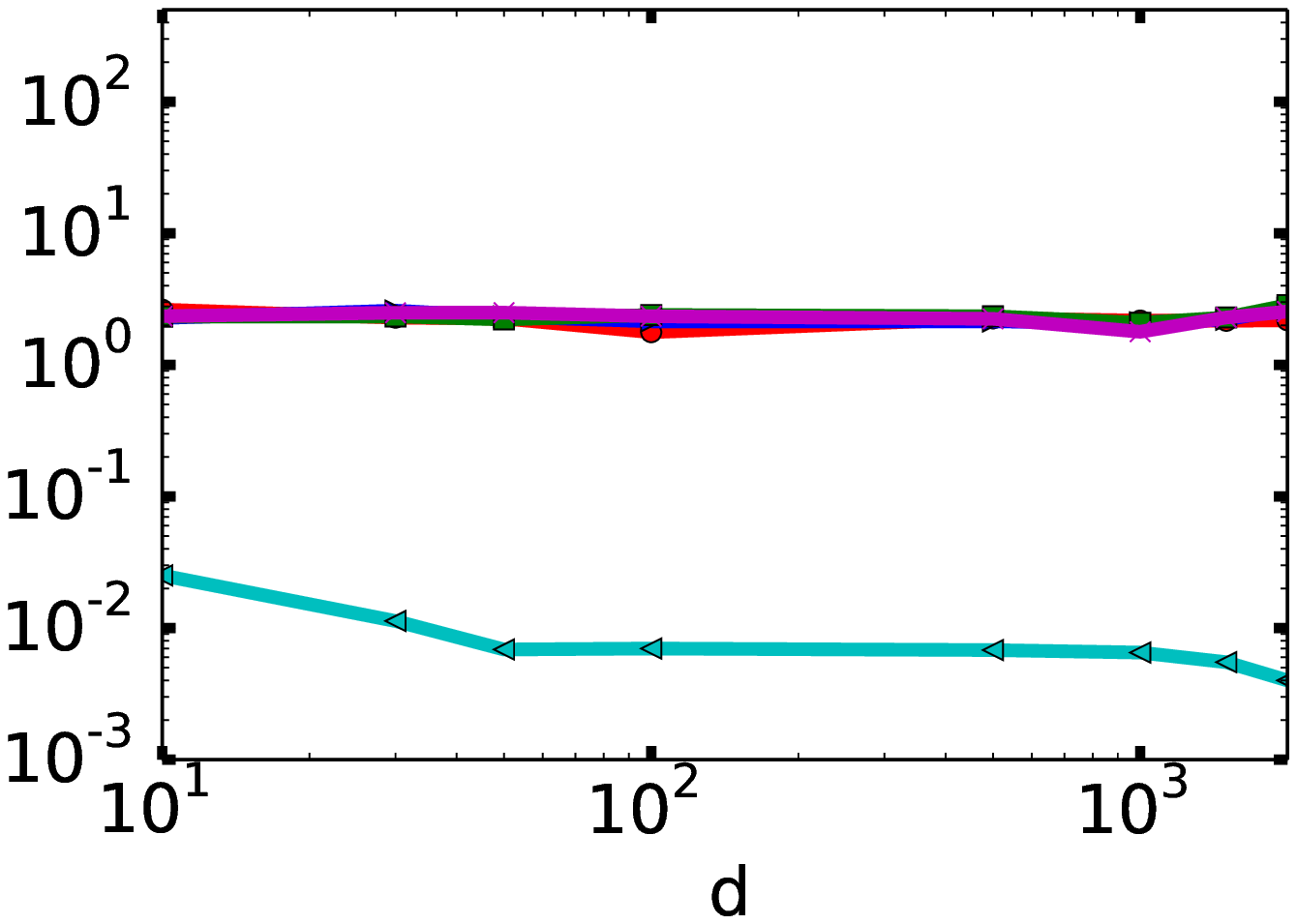}
}
&
\subfigure[$|f-f^\ast|/|f^\ast|$]{
\includegraphics[width=0.3\textwidth]{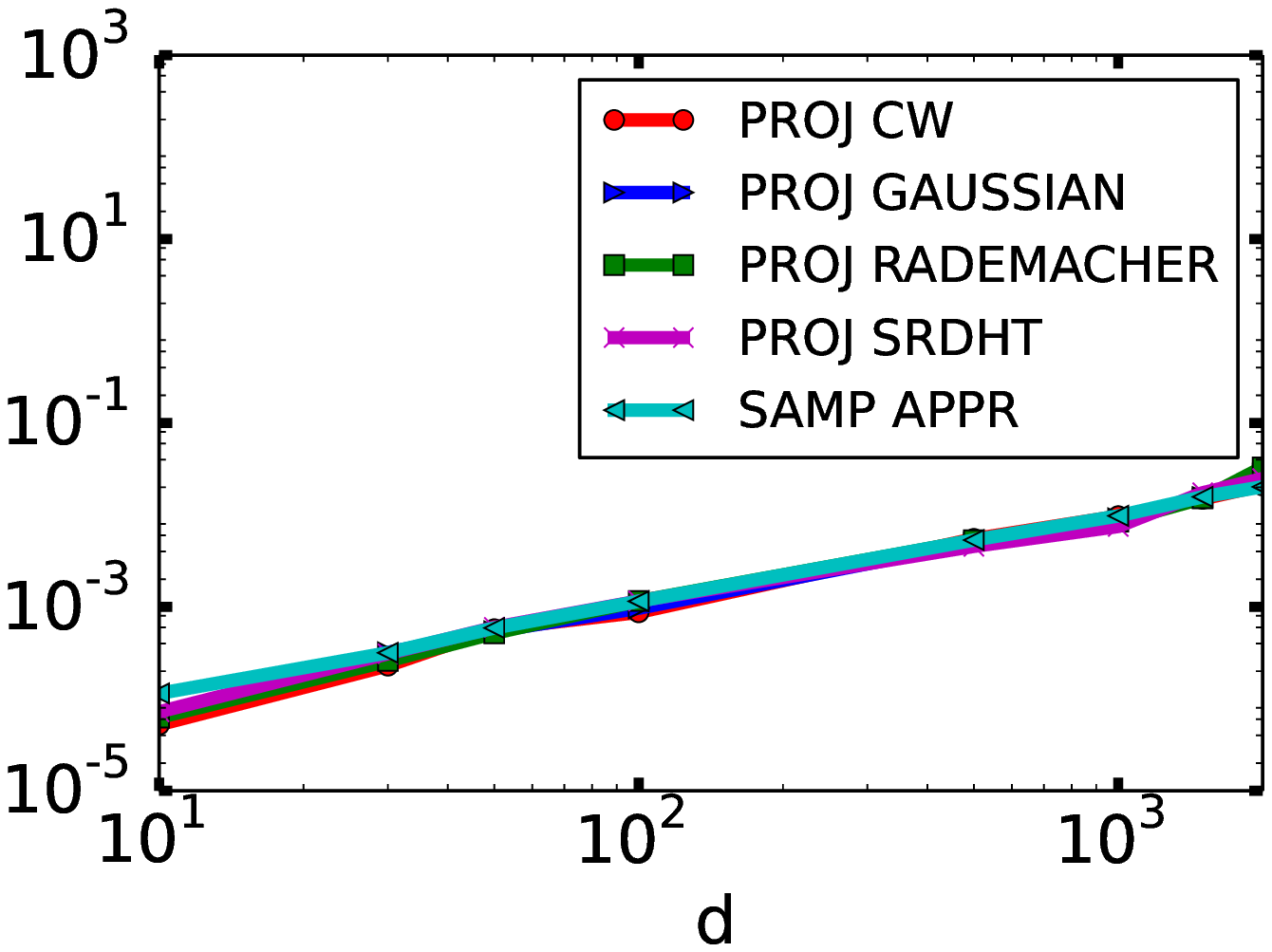}
}
&
\subfigure[Running time(sec)]{
\includegraphics[width=0.3\textwidth]{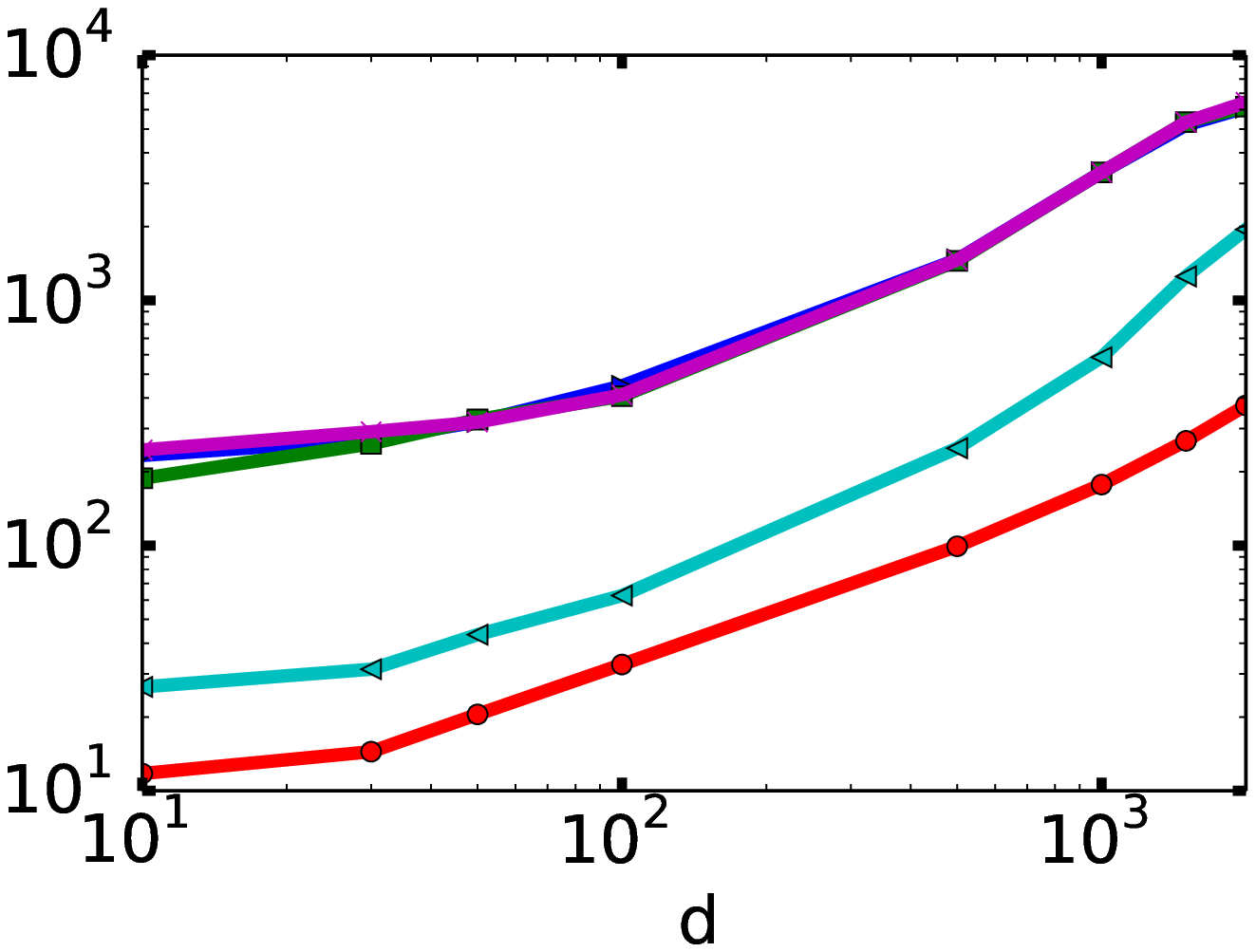}
}
\\
 \multicolumn{3}{c}{\bf $s=5e4$}
 \\
\end{tabular}
\end{centering}
\caption{ Performance of all the algorithms on NB matrices with varying $d$ from $10$ to $2000$ and fixed $n = 1e7$.
The matrix is generated using \texttt{STACK1}.
For each method, the embedding dimension is fixed to be $2e3$ or $5e4$.
The following three quantities are computed: 
relative error of the objective $|f-f^\ast|/f^\ast$;
relative error of the certificate $\|x-x^\ast\|_2/\|x^\ast\|_2$;
and the running time to compute the approximate solution.
For each setting, 3 independent trials are performed and the median is reported.}
\label{fig:d}
\end{figure}

Here, we evaluate the performance of the low-precision solvers by evaluating 
the performance of all the embeddings on NB matrices with changing $d$.
We fix $n = 1e7$ and let $d$ take values from $10$ to $2000$.
For each $d$, the matrix is generated by stacking an NB matrix with size 
$2.5e5$ by $d$ 40 times using \texttt{STACK1}, so that the coherence of the 
matrix is $1/40$.
For conciseness, we fix the embedding of each method to be $2e3$ or $5e4$.
The relative error on certificate and objective and running time are 
evaluated.
The results are shown in Figure~\ref{fig:d}.

As can be seen, overall, all the projection-based methods behave similarly.
As expected, the relative error goes up as $d$ gets larger.
Meanwhile, \texttt{SAMP APPR} yields lower error as $d$ increases.
However, it seems to have a stronger dependence on the lower dimension of 
the matrix, as it breaks down when $d$ is $100$ for small sampling size, 
\emph{i.e.}, $s=2e3$.

\subsubsection{Performance of high precision solvers}

\begin{figure}[h]
\begin{centering}
\begin{tabular}{ccc}
\subfigure[$\|x - x^\ast\|_2/\|x^\ast\|_2$]{
\includegraphics[width=0.3\textwidth]{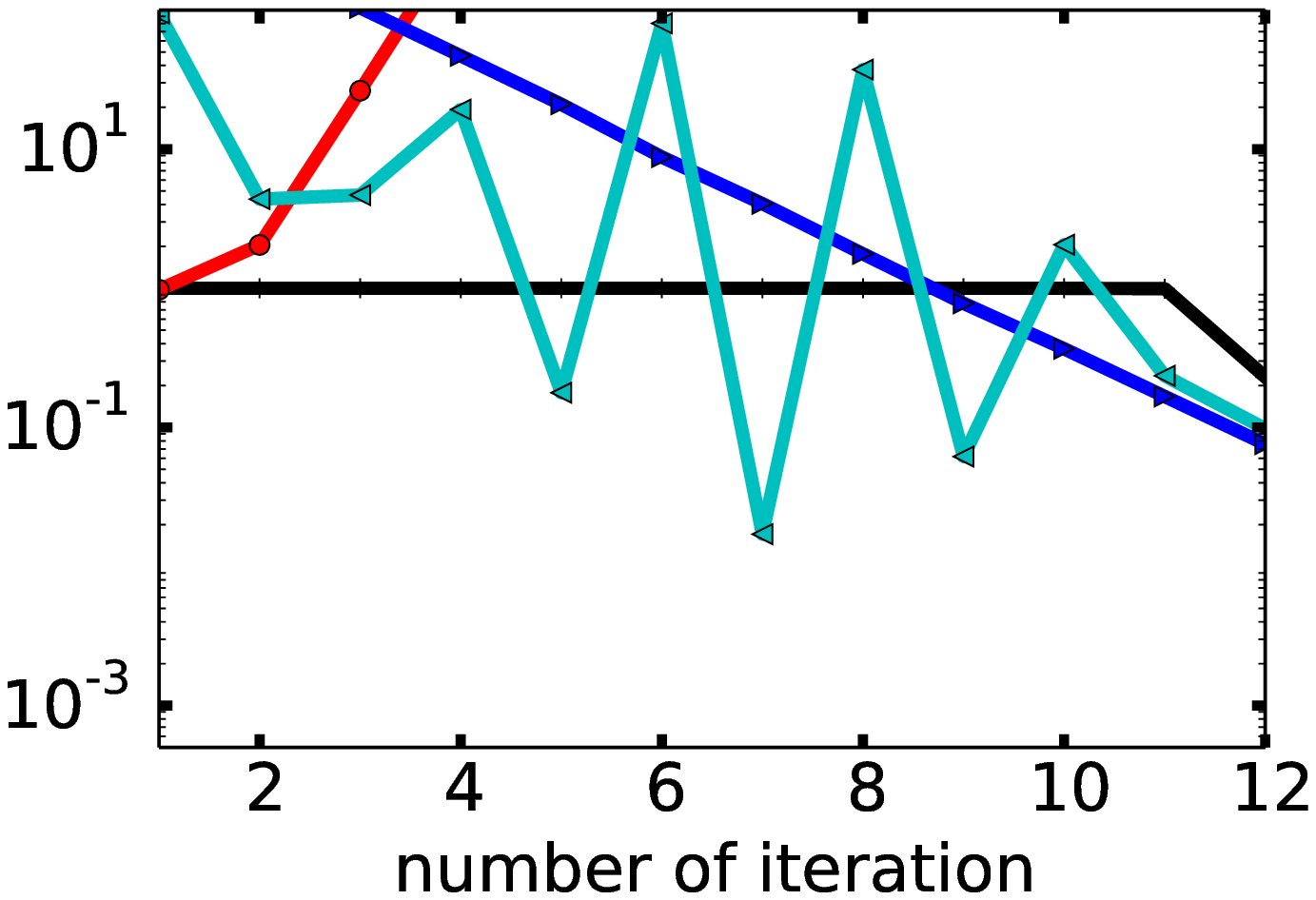}
}
&
\subfigure[$|f-f^\ast|/|f^\ast|$]{
\includegraphics[width=0.3\textwidth]{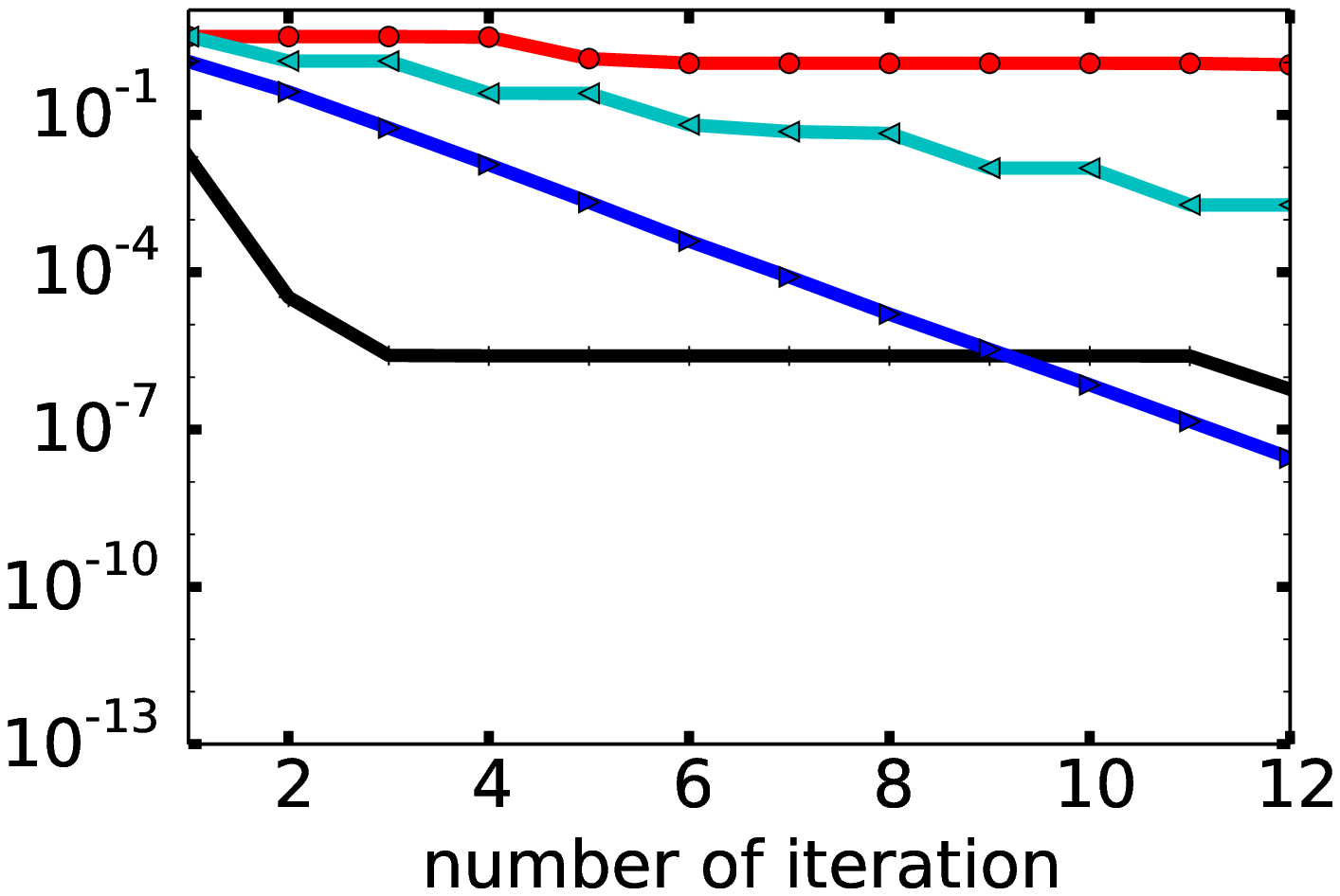}
}
&
\subfigure[Running time(sec)]{
\includegraphics[width=0.3\textwidth]{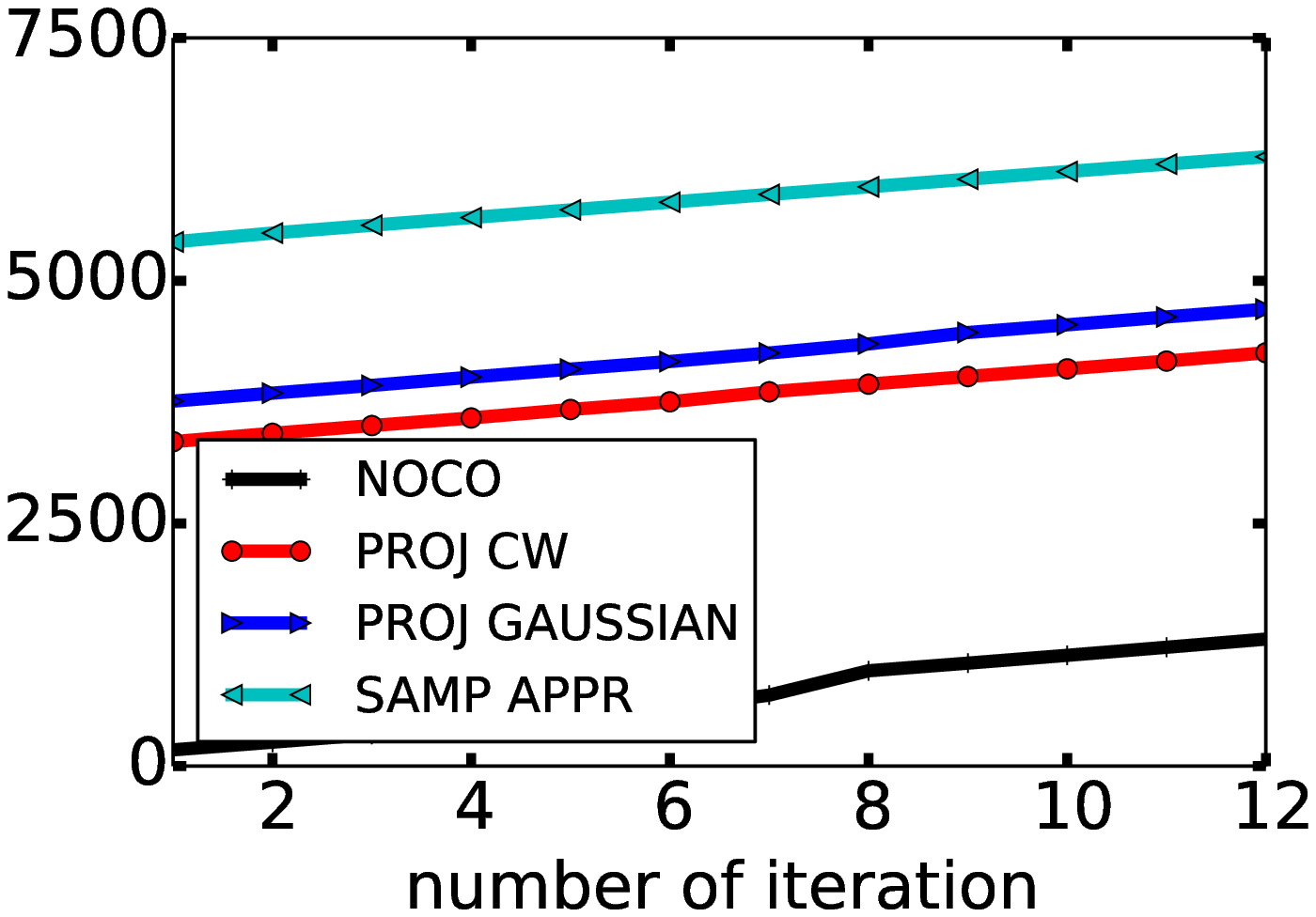}
}
\\
 \multicolumn{3}{c}{\bf small embedding dimension}
\\
\subfigure[$\|x - x^\ast\|_2/\|x^\ast\|_2$]{
\includegraphics[width=0.3\textwidth]{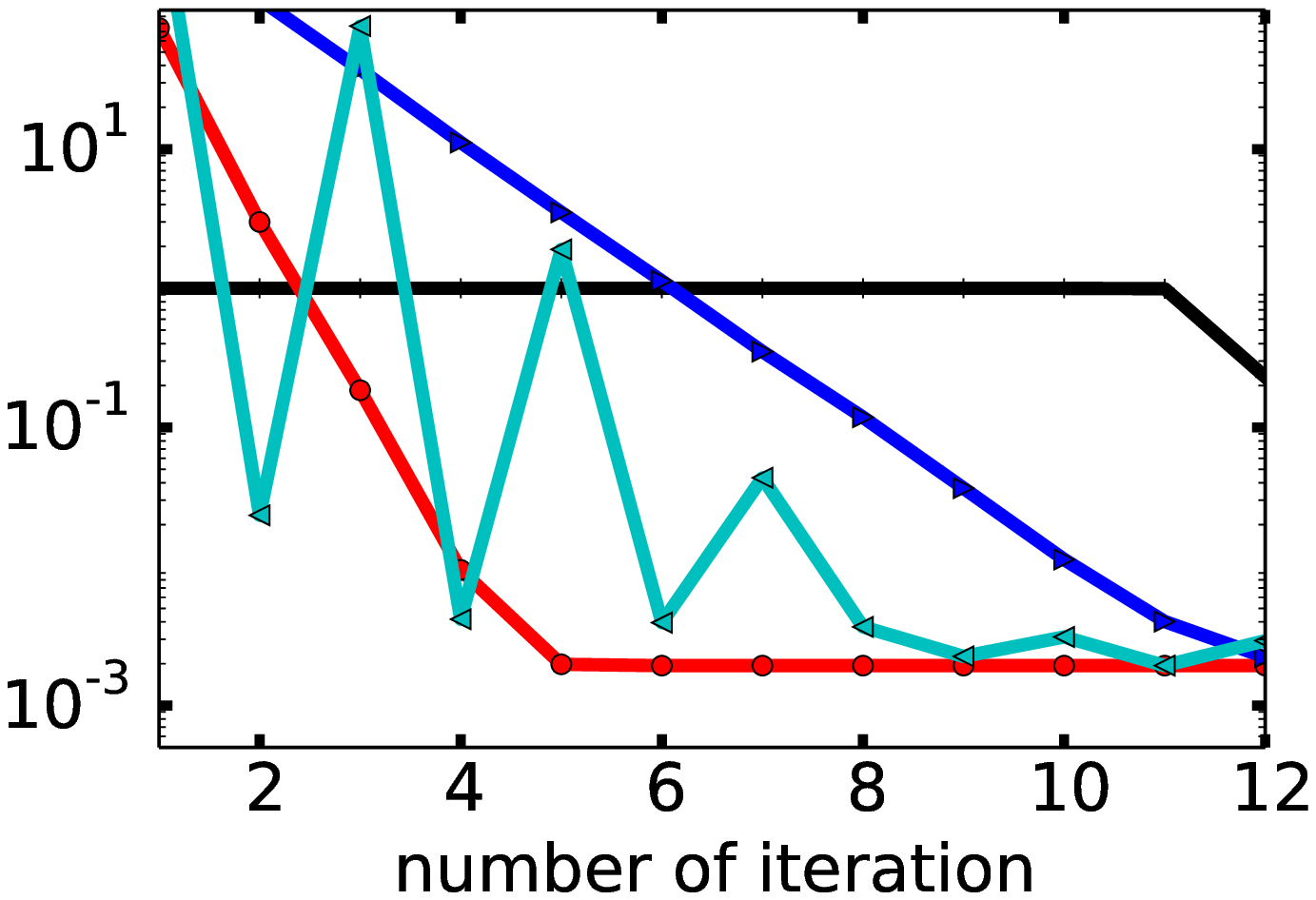}
}
&
\subfigure[$|f-f^\ast|/|f^\ast|$]{
\includegraphics[width=0.3\textwidth]{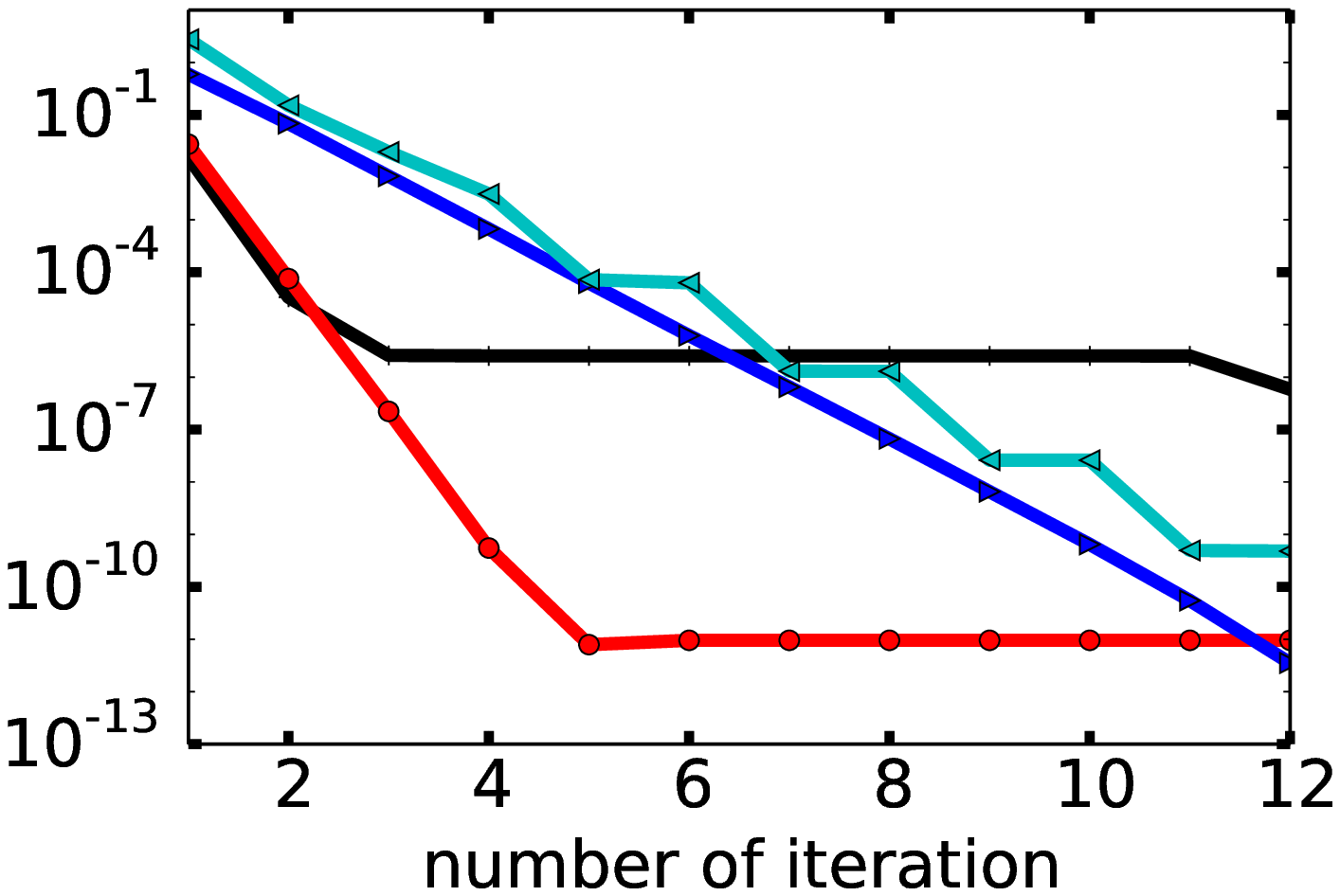}
}
&
\subfigure[Running time(sec)]{
\includegraphics[width=0.3\textwidth]{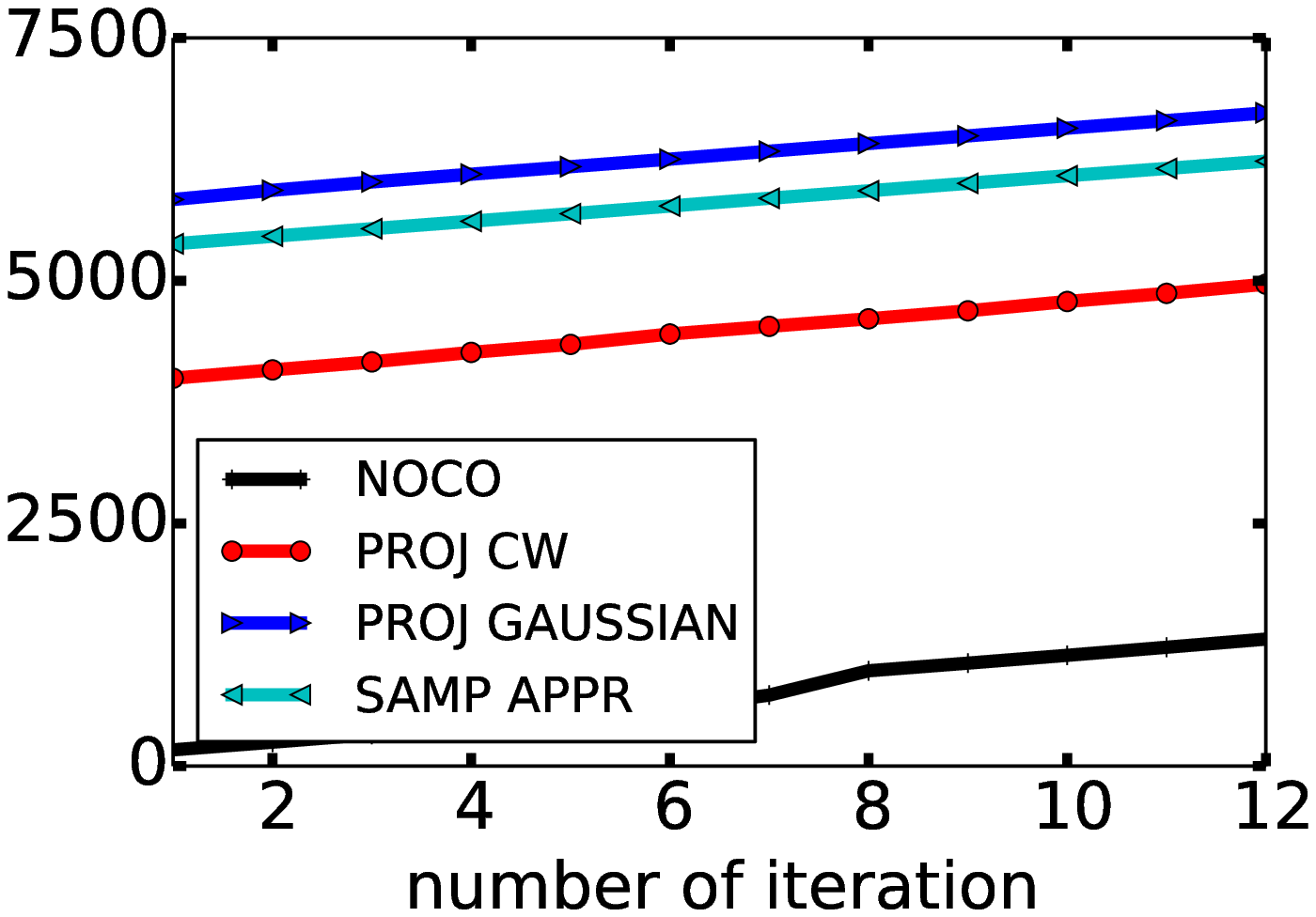}
}
\\
 \multicolumn{3}{c}{\bf large embedding dimension}
\end{tabular}
\end{centering}
\caption{ Evaluation of LSQR with randomized preconditioner on an NB matrix with size $1e8$ by $1000$ and condition number $1e6$.
Here, several ways for computing the embedding are implemented.
In \texttt{SAMP APPR}, the underlying random projection is \texttt{PROJ CW}
with projection dimension $3e5$.
For completeness, LSQR without preconditioner is evaluated, denoted by \texttt{NOCO}.
In above, by small embedding dimension, we mean $5e3$ for all the methods.
By large embedding dimension, we mean 3e5 for \texttt{PROJ CW}, 1e4 for \texttt{PROJ GAUSSIAN} and 5e4 for \texttt{SAMP APPR}.
For each method and embedding dimension, the following three quantities are computed: 
relative error of the objective $|f-f^\ast|/f^\ast$;
relative error of the certificate $\|x-x^\ast\|_2/\|x^\ast\|_2$;
and the running time to compute the approximate solution.
Each subplot shows one of the above quantities versus number of iteration, respectively.
For each setting, only one trial is performed.}
\label{fig:lsqr_nb}
\end{figure}

\begin{figure}[h]
\begin{centering}
\begin{tabular}{ccc}
\subfigure[$\|x - x^\ast\|_2/\|x^\ast\|_2$]{
\includegraphics[width=0.3\textwidth]{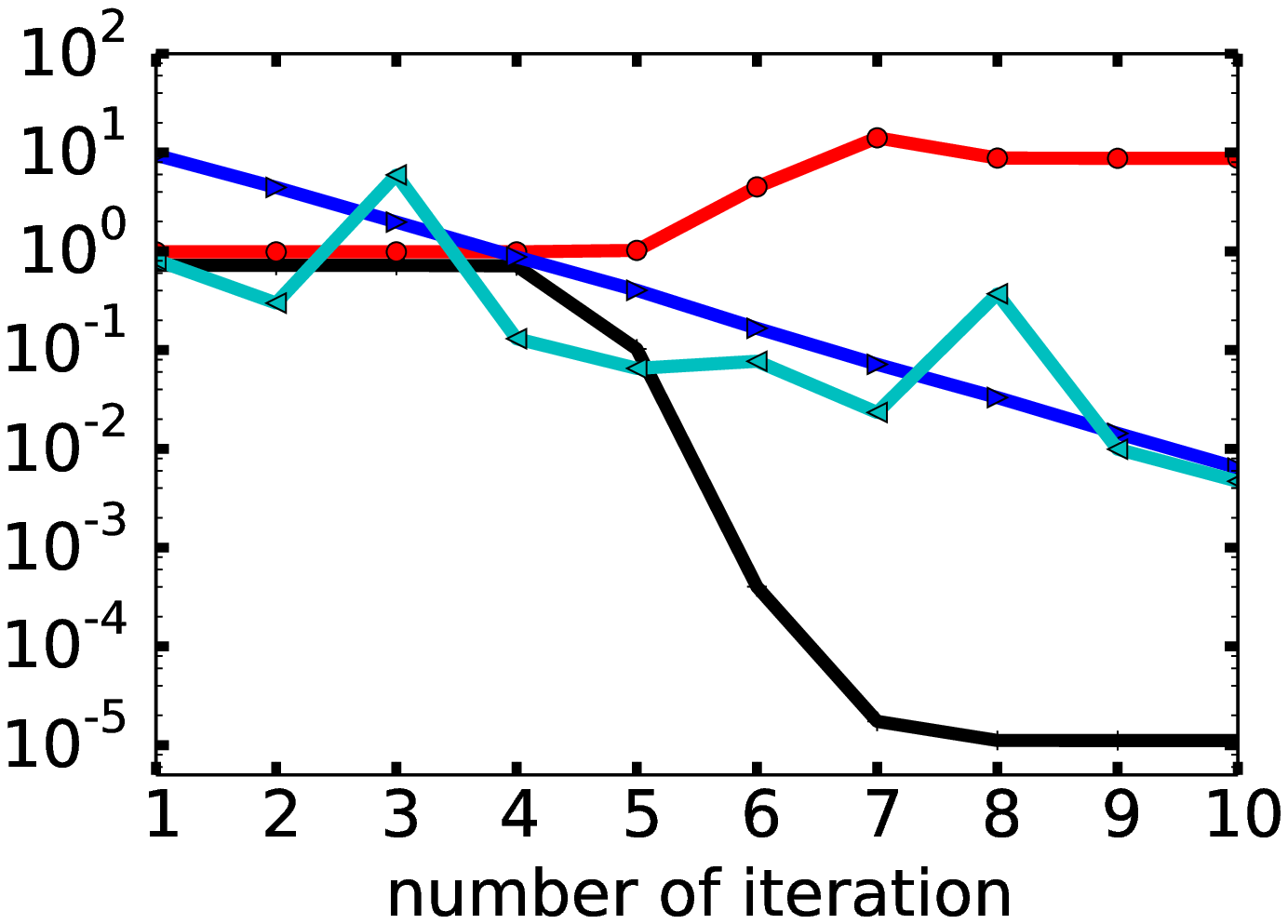}
}
&
\subfigure[$|f-f^\ast|/|f^\ast|$]{
\includegraphics[width=0.3\textwidth]{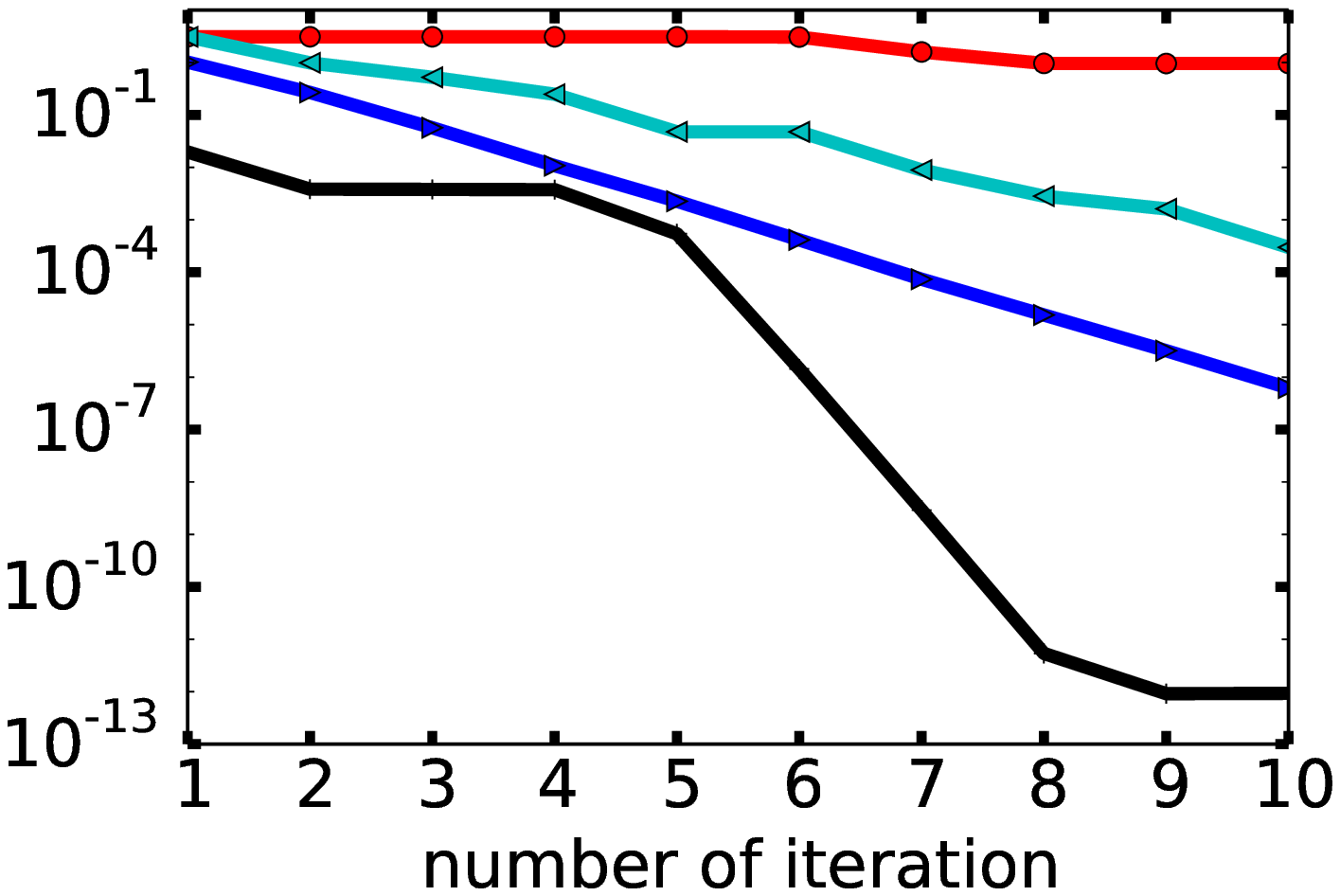}
}
&
\subfigure[Running time(sec)]{
\includegraphics[width=0.3\textwidth]{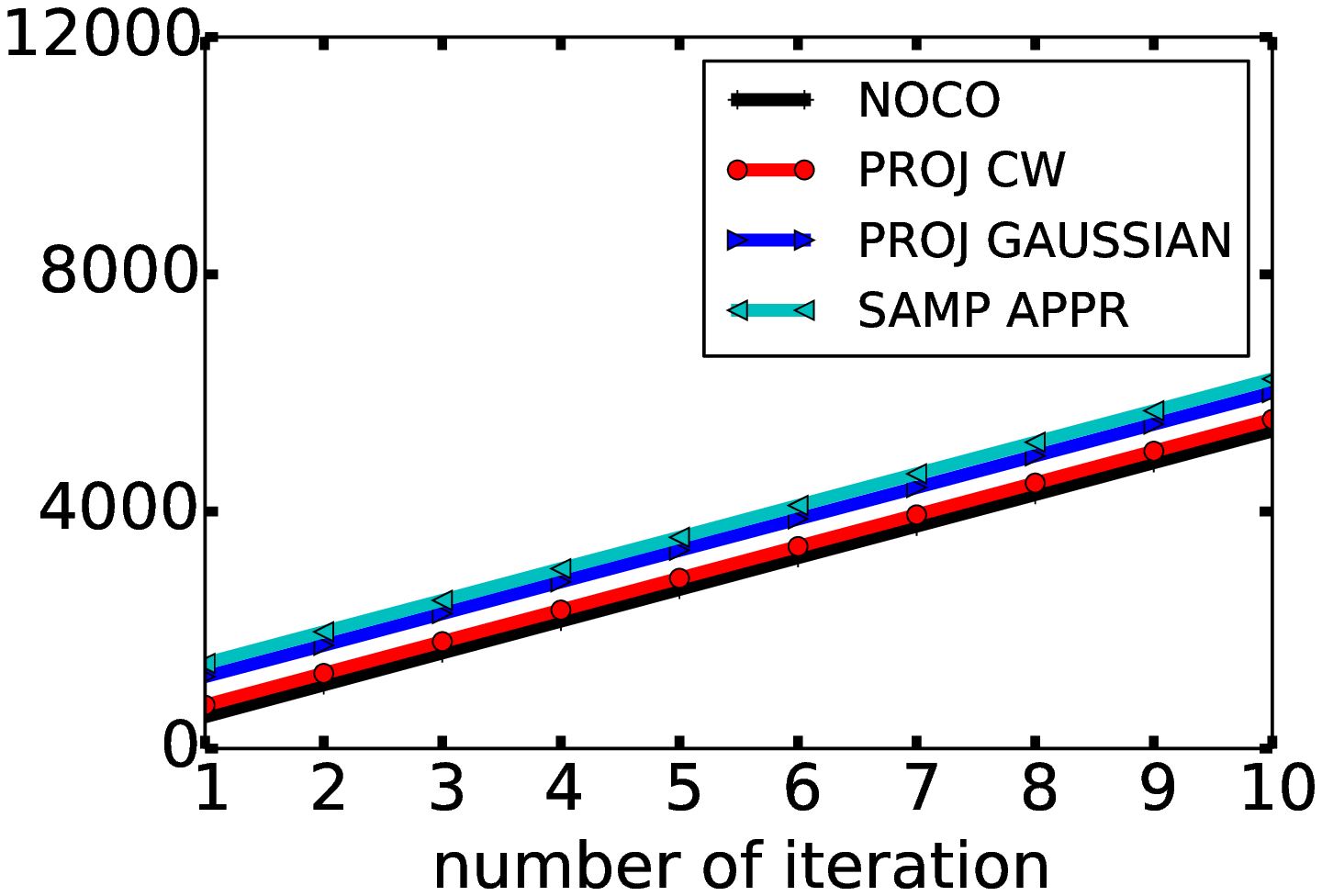}
}
\\
 \multicolumn{3}{c}{\bf small embedding dimension}
\\
\subfigure[$\|x - x^\ast\|_2/\|x^\ast\|_2$]{
\includegraphics[width=0.3\textwidth]{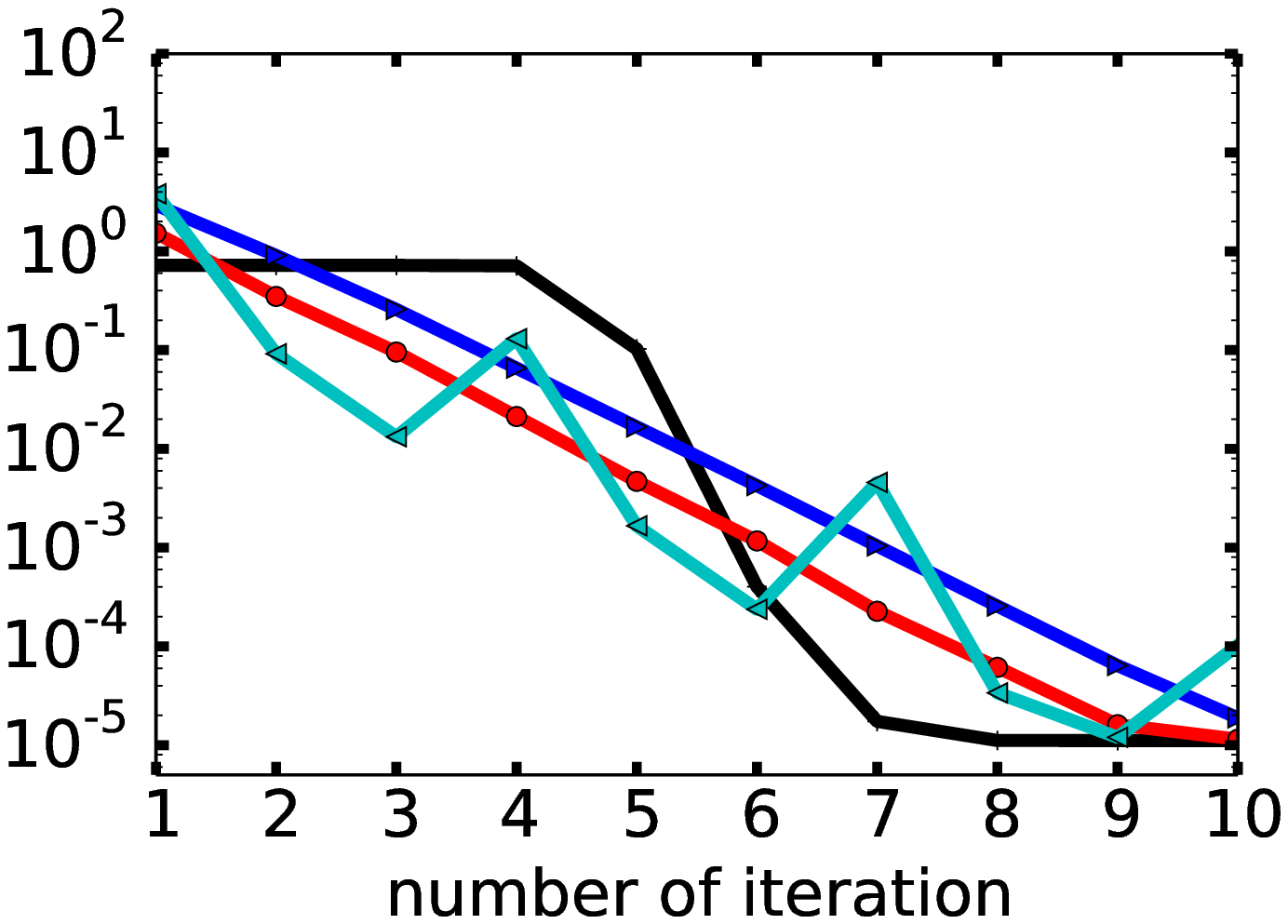}
}
&
\subfigure[$|f-f^\ast|/|f^\ast|$]{
\includegraphics[width=0.3\textwidth]{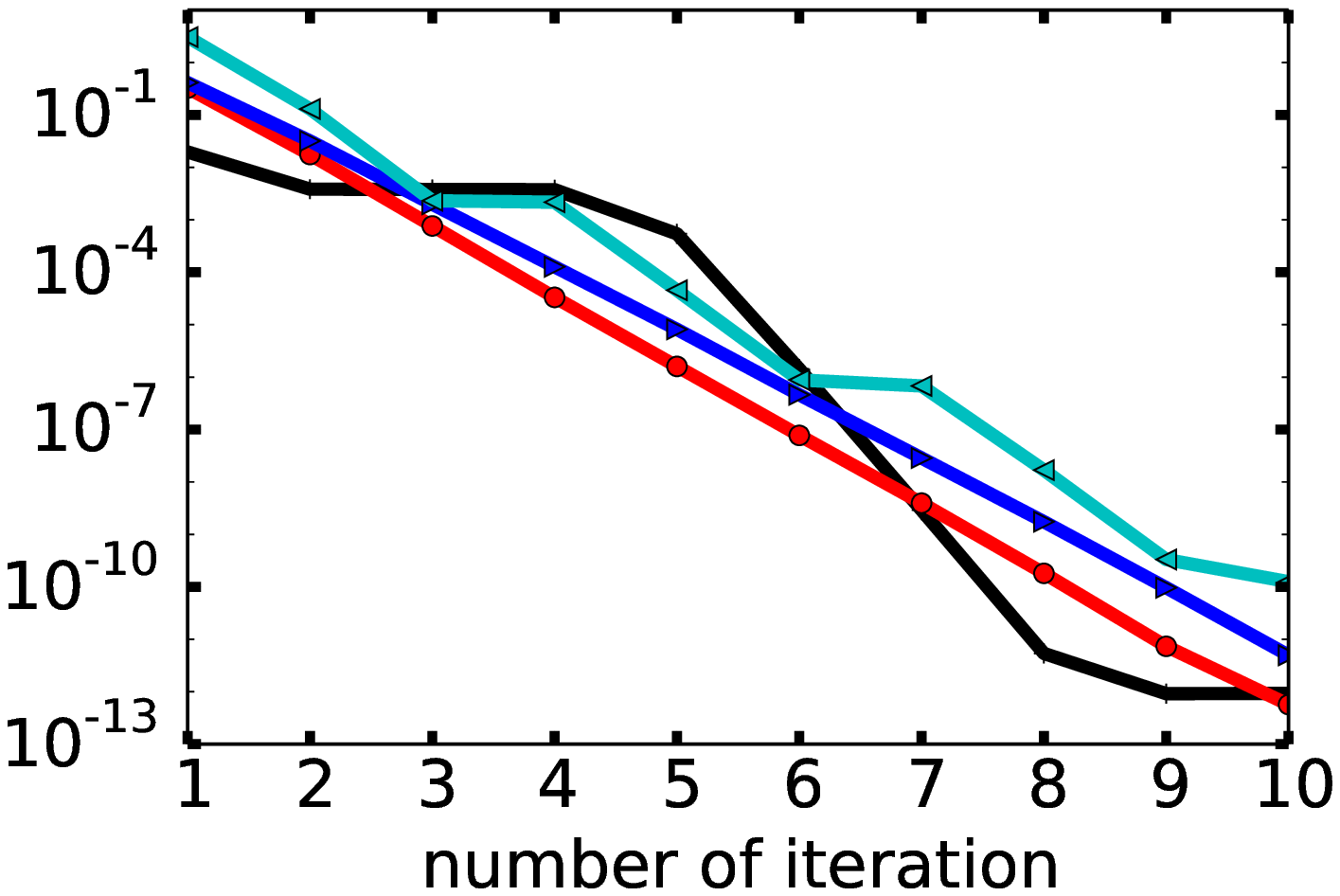}
}
&
\subfigure[Running time(sec)]{
\includegraphics[width=0.3\textwidth]{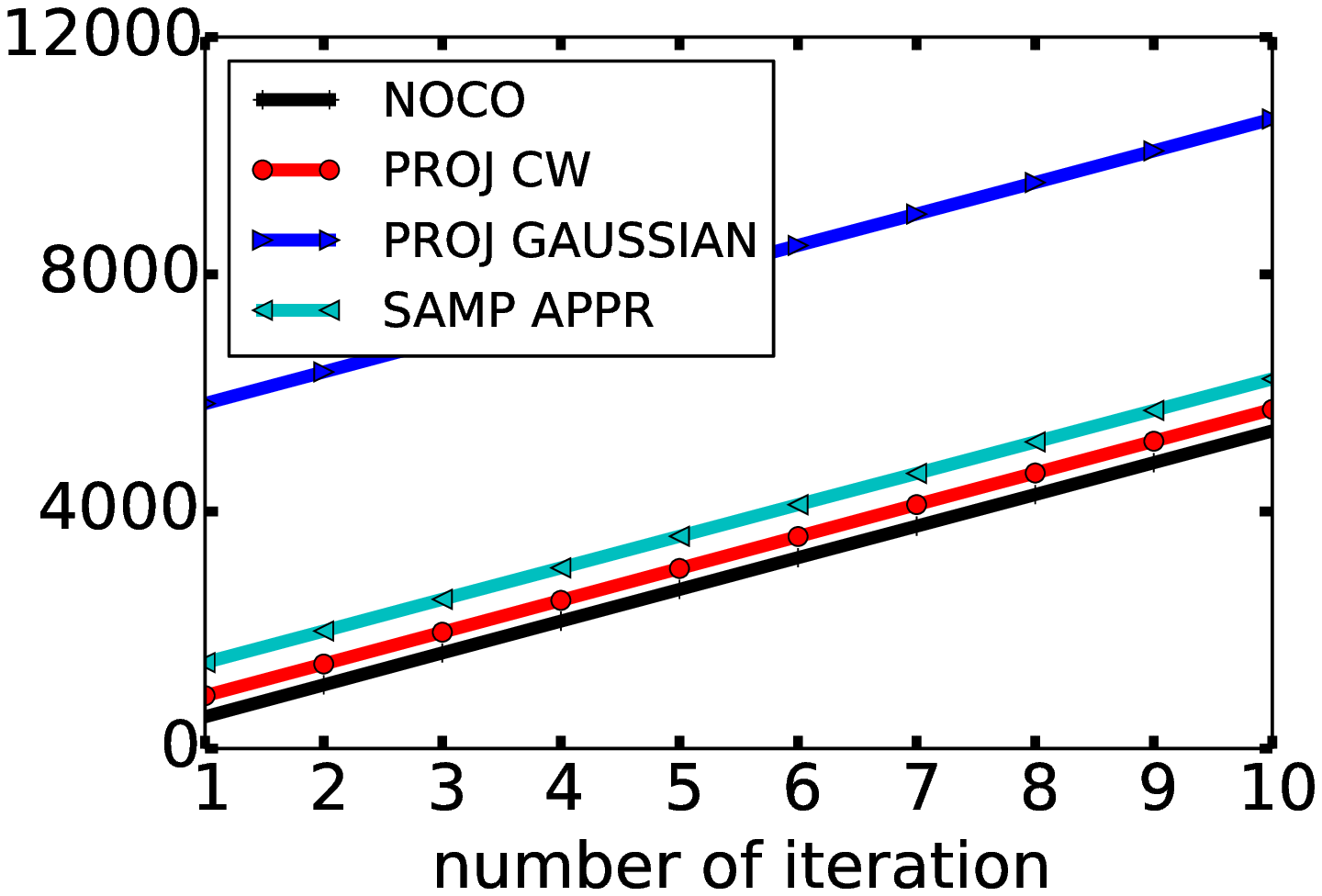}
}
\\
 \multicolumn{3}{c}{\bf large embedding dimension}
\end{tabular}
\end{centering}
\caption{ Evaluation of LSQR with randomized preconditioner on an NB matrix with size $1e7$ by $1000$ and condition number $5$.
Here, several ways for computing the embedding are implemented.
In \texttt{SAMP APPR}, the underlying random projection is \texttt{PROJ CW}
with projection dimension $3e5$.
For completeness, LSQR without preconditioner is evaluated, denoted by \texttt{NOCO}.
In above, by small embedding dimension, we mean $5e3$ for all the methods.
By large embedding dimension, we mean $3e5$ for \texttt{PROJ CW} and $5e4$ for the rest.
For each method and embedding dimension, the following three quantities are computed: 
relative error of the objective $|f-f^\ast|/f^\ast$;
relative error of the certificate $\|x-x^\ast\|_2/\|x^\ast\|_2$;
and the running time to compute the approximate solution.
Each subplot shows one of the above quantities versus number of iteration, respectively.
For each setting, 3 independent trials are performed and the median is reported.}
\label{fig:lsqr_ng}
\end{figure}

Here, we evaluate the use of these methods as preconditioners for 
high-precision iterative solvers.
Since the embedding can be used to compute a preconditioner for the original 
linear system, one can invoke iterative algorithms such as 
LSQR~\cite{paige1982lsqr} to solve the 
preconditioned least-squares problem.
Here, we will use LSQR.
We first evaluate the conditioning quality, \emph{i.e.}, $\kappa(AR^{-1})$, 
on an NB matrix with size $1e6$ by $500$ using several different ways for 
computing the embedding.
The results are presented in Table~\ref{table:cond}.
Then we test the performance of LSQR with these preconditioners on an NB matrix with
size $1e8$ by $1000$ and an NG matrix with size $1e7$ by $1000$.
For simplicity, for each method of computing the embedding, we try a small 
embedding dimension where some of the methods fail, and a large embedding 
dimension where most of the methods succeed.
See Figure~\ref{fig:lsqr_nb} and Figure~\ref{fig:lsqr_ng} for details.

The convergence rate of the LSQR phase depends on the preconditioning 
quality, \emph{i.e.}, $\kappa(AR^{-1})$ where $R$ is obtained by the QR 
decomposition of the embedding of $A$, $\Phi A$.
See Section~\ref{sxn:round_embed-embed} for more details.
Table~\ref{table:cond} implies that all the projection-based methods tend to 
yield preconditioners with similar condition numbers once the embedding 
dimension is large enough.
Among them, \texttt{PROJ CW} needs a much larger embedding dimension to be 
reliable (clearly consistent with its use in low-precision solvers).
In addition, overall, the conditioning quality of the sampling-based 
embedding method, \emph{i.e.}, \texttt{SAMP APPR} tends to be worse than 
that of projection-based methods.

As for the downstream performance, from Figure~\ref{fig:lsqr_nb} we can 
clearly see that, when a small embedding dimension is used, \emph{i.e.}, $s=5e3$,
\texttt{PROJ GAUSSIAN} yields the best preconditioner, as its better 
preconditioning quality translates immediately into fewer iterations for 
LSQR to converge.
This is followed by \texttt{SAMP APPR}.
This relative order is also suggested by Table~\ref{table:cond}.
As the embedding dimension is increased, i.e., using large embedding dimension, all the method yield significant improvements and produce much more accurate solutions compared to that of \texttt{NOCO} (LSQR without preconditioning), among which \texttt{PROJ CW} with embedding dimension $3e5$ converges to a nearly machine-precision solution within only $5$ iterations.
As for the running time, since each iteration of LSQR only involves with two matrix-vector multiplications (costs less than 2 minutes in our experiments), the overall running time is dominated by the time for computing the preconditioner.
As expected, \texttt{PROJ CW} runs the fastest and the running time of \texttt{PROJ GAUSSIAN} scales linearly in the embedding dimension. In \texttt{SAMP APPR}, the sampling process needs to make $1$-$2$ passes over the dataset but the running time is relatively stable regardless of the sampling size, as reflected in Figure~\ref{fig:lsqr_nb}(c)\&(f). 
Finally, note that the reason that the error does not drop monotonically in 
the solution vector is the following. 
With the preconditioners, we work on a transformed system, and the theory 
only guarantees monotonicity in the decreasing of the relative error of the 
certificate of the transformed system, not the original one.

Finally, a minor but potentially important point should be mentioned as a 
word of caution.
As expected, when the condition number of the linear system is large, 
vanilla LSQR does not converge at all.
On the other hand, when the condition number is very small, from 
Figure~\ref{fig:lsqr_ng}, there is no need to precondition.
If, in this latter case, a randomized preconditioning method is used, then 
the embedding dimension must be chosen to be sufficiently large: unless the 
embedding dimension is large enough such that the conditioning quality is 
sufficiently good, then preconditioned LSQR yields larger errors than even
vanilla LSQR.

\begin{table}[ht]
\begin{center}
\begin{sc}
\small
\begin{tabular}{c|ccccc}
  $c$ & PROJ CW & PROJ GAUSSIAN & PROJ RADEMACHER & PROJ SRDHT & SAMP APPR \\
\hline
  5e2 & 1.08e8 & 2.17e3 & 1.42e3 & 1.19e2 & 1.21e2\\
  1e3 & 1.1e6 & 5.7366 & 5.6006 & 7.1958 & 75.0290\\
  5e3 & 5.5e5 & 1.9059 & 1.9017 & 1.9857 & 25.8725\\
  1e4 & 5.1e5 & 1.5733 & 1.5656 & 1.6167 & 17.0679\\
  5e4 & 1.8e5 & 1.2214 & 1.2197 & 1.2293 & 6.9109\\   
  1e5 & 1.1376 & 1.1505 & 1.1502 & 1.1502 & 4.7573
\end{tabular}
\end{sc}
\end{center}
\caption{Quality of preconditioning on an NB matrix with size $1e6$ by $500$ using several kinds of embeddings.
For each setting, 5 independent trials are performed and the median is reported.}
\label{table:cond}
\end{table}

%% file: conclusion.tex
\section{Discussion and conclusion}
\label{sxn:conc}

Large-scale data analysis and machine learning problems present 
considerable challenges and opportunities to signal processing, 
electrical engineering, scientific computing, numerical linear algebra, 
and other research areas that have historically been developers of 
and/or consumers of high-quality matrix algorithms. 
RandNLA is an approach, originally from theoretical computer science, that 
uses randomization as a resource for the development of improved matrix 
algorithms, and it has had several remarkable successes in theory and in
practice in small to medium scale matrix computations in RAM.
The general design strategy of RandNLA algorithms (for problems such as 
$\ell_2$ regression and low-rank matrix approximation) in RAM is by now 
well known:
construct a sketch (either by performing a random projection or by random 
sampling according to a judiciously-chosen data-dependent importance 
sampling probability distribution), and then use that sketch to approximate 
the solution to the original problem (either by solving a subproblem on the 
sketch or using the sketch to construct a preconditioner for the original 
problem).

The work reviewed here highlights how, with appropriate modifications, 
similar design strategies can extend (for $\ell_2$-based regression 
problems as well as other problems such as $\ell_1$-based regression 
problems) to much larger-scale parallel and distributed environments 
that are increasingly common.
Importantly, though, the improved scalability often comes due to restricted 
communications, rather than improvements in FLOPS.
(For example, the use of Chebyshev semi-iterative method vs.\ LSQR in LSRN
on MPI; and the use of the MIE with multiple queries on MapReduce.) 
In these parallel/distributed settings, we can take advantage of the 
communication-avoiding nature of RandNLA algorithms to move beyond FLOPS to 
design matrix algorithms that use more computation than the traditional 
algorithms but that have much better communication profiles, and we can do 
this by mapping the RandNLA algorithms to the underlying architecture in 
very nontrivial ways.
(For example, using more computationally-expensive Gaussian projections to 
ensure stronger control on the condition number; and using the MIE with 
multiple initial queries to construct a very good initial search region.)
These examples of performing extra computation to develop algorithms with 
improved communication suggests revisiting other methods from numerical 
linear algebra, optimization, and scientific computing, looking in other 
novel ways beyond FLOPS for better communication properties for many 
large-scale matrix algorithms.